\documentclass[fleqn,usenatbib]{mnras}

\usepackage{newtxtext,newtxmath}

\usepackage[T1]{fontenc}

\DeclareRobustCommand{\VAN}[3]{#2}
\let\VANthebibliography\thebibliography
\def\thebibliography{\DeclareRobustCommand{\VAN}[3]{##3}\VANthebibliography}

\usepackage{graphicx}	
\usepackage{amsmath}	

\usepackage{amssymb}	
\usepackage{epstopdf}
\usepackage{longtable}
\usepackage{threeparttable, tablefootnote}
\graphicspath{{Images/}}

\title[Four bright $r$-process-enhanced stars]{Decoding the compositions of four bright $r$-process-enhanced stars }

\author[Saraf et al. 2023]{
Pallavi Saraf,$^{1,2}$\thanks{E-mail: pallavi.saraf@iiap.res.in (PS)}
Carlos Allende Prieto,$^{3,4}$
Thirupathi Sivarani,$^{1}$
Avrajit Bandyopadhyay,$^{5}$
\newauthor
Timothy C. Beers,$^{6,7}$
and A. Susmitha$^{1}$
\\
$^{1}$ Indian Institute of Astrophysics, Koramangala 2nd Block, Bangalore, 560034, India\\
$^{2}$ Pondicherry University, R.V. Nagar, Kalapet, 605014, Puducherry, India\\
$^{3}$ Instituto de Astrof\'isica de Canarias, E-38200 La Laguna, Tenerife, Spain\\
$^{4}$ Departamento de Astrof\'isica, Universidad de La Laguna, E-38205 La Laguna, Tenerife, Spain\\
$^{5}$ Department of Astronomy, University of Florida, Gainesville, FL 32601, United States of America\\
$^{6}$ Joint Institute for Nuclear Astrophysics-Center for the Evolution of the Elements (JINA-CEE), United States of America\\
$^{7}$ Department of Physics and Astronomy, University of Notre Dame, Notre Dame, IN 46556, United States of America
}

\date{Accepted XXX. Received YYY; in original form ZZZ}

\pubyear{2023}

\begin{document}
\label{firstpage}
\pagerange{\pageref{firstpage}--\pageref{lastpage}}
\maketitle

\begin{abstract}
There has been a concerted effort in recent years to identify the astrophysical sites of the $r$-process that can operate early in the Galaxy.  The discovery of many $r$-process-enhanced (RPE) stars (especially by the $R$-process Alliance collaboration) has significantly accelerated this effort. However, only limited data exist on the detailed elemental abundances covering the primary neutron-capture peaks. Subtle differences in the structure of the $r$-process pattern, such as the relative abundances of elements in the third peak, in particular, are expected to constrain the $r$-process sites further. Here, we present a detailed elemental-abundance analysis of four bright RPE stars selected from the HESP-GOMPA survey. Observations were carried out with the 10-m class telescope Gran Telescopio Canarias (GTC), Spain. The high spectral signal-to-noise ratios obtained allow us to derive abundances for 20 neutron-capture elements, including the third $r$-process peak element osmium (Os). We detect thorium (Th) in two stars, which we use to estimate their ages.
We discuss the metallicity evolution of Mg, Sr, Ba, Eu, Os, and Th in $r$-II and $r$-I stars,  based on a compilation of RPE stars from the literature. The strontium (Sr) abundance trend with respect to europium (Eu) suggests the need for an additional production site for Sr (similar to several earlier studies); this requirement could be milder for yttrium (Y) and zirconium (Zr). We also show that there could be some time delay between $r$-II and $r$-I star formation, based on the Mg/Th abundance ratios.
\end{abstract}




\section{Introduction}
\label{sec:introduction}
Understanding the formation of neutron-capture elements is one of the key topics in stellar astronomy. Since \cite{Burbidge.etal.1957} and \cite{Cameron.1957}, it has been recognized that the rapid neutron-capture process ($r$-process) is one of the prime mechanisms for producing heavy elements. However, we still do not know with certainty all of the major astrophysical sites of $r$-process-enrichment in the Galaxy.
Very metal-poor (VMP; [Fe/H] $\leq -2.0$) stars  are expected to preserve the nucleosynthesis signatures of the first generations of stars in the Galaxy \citep{Beers.Christlieb.2005, Frebel.etal.2005, Frebel.Norris.2015}. Hence, high-resolution spectroscopy of VMP stars provides the opportunity to study the detailed chemical abundances of elements present in these stars, and thus obtain information about the production and astrophysical sites of their formation. 

In the periodic table, elements up to zinc (Zn) can form in fusion reactions, but elements beyond Zn can only be formed via neutron-capture processes \citep{Burbidge.etal.1957, Cameron.1957, Clayton.1983book, Hansen.etal.2004book, Wanajo.etal.2018, Curtis.etal.2019}. According to the rate of neutron capture on seed nuclei, the neutron-capture processes are conventionally divided into three classes, (1) a rapid neutron-capture process ($r$-process), (2) a slow neutron-capture process ($s$-process), and (3) an intermediate neutron-capture process ($i$-process) \citep{Burbidge.etal.1957, Cameron.1957, Cowan.Rose.1977, Beers.Christlieb.2005, Sneden.etal.2008}. The sites of enrichment of $s$- and $i$-process elements  are thought to include asymptotic giant branch (AGB) stars through mass transfer from a  binary companion \citep{Herwig.2005, Campbell.etal.2008, Bisterzo.etal.2010, Doherty.etal.2015}. Other suggested sources for the $s$-process involve helium core-burning massive stars \citep{Truran.Iben.1977, Prantzos.etal.1990}, carbon core-burning massive stars \citep{Arnett.Thielemann.1985, Langer.etal.1986, Arcoragi.etal.1991}, and carbon shell-burning stars \citep{Raiteri.etal.1991}. 
Sources for the $i$-process include post-AGB stars \citep{Herwig.etal.2011}, rapidly accreting white dwarfs \citep{Denissenkov.etal.2017, Denissenkov.etal.2019}, and super-AGB stars \citep{Jones.etal.2016}.

\cite{Beers.Christlieb.2005} divided RPE stars into two categories, namely $r$-I (+0.3 $\leq$ [Eu/Fe] $\leq +1.0$, [Ba/Eu] $<$ 0), and $r$-II ([Eu/Fe] $> +1.0$, [Ba/Eu] $<$ 0). In the above definitions, europium (Eu) is taken as the primary indicator of $r$-process-enrichment because the abundance of Eu is dominated by the $r$-process, and is relatively easy to measure in high-resolution optical spectra. The barium (Ba) limits are introduced to remove likely $s$-process contamination. Another class of r-process-enhanced (RPE) stars is the so-called limited-$r$ ([Eu/Fe] $< +0.3$, [Sr/Ba] $> +0.5$, and [Sr/Eu] $> $0.0), which was introduced to account for the dispersion observed in the first peak of the $r$-process pattern \citep{Sneden.etal.2000, Travaglio.etal.2004, Frebel.2018}. Out of all the known metal-poor stars (up to [Fe/H] $\sim -1.0$), only about 2 percent are limited-$r$, 15 percent are $r$-I, and 5 percent are $r$-II stars \citep{Frebel.Jacobson.2016}. There are some $46$ limited-$r$, $426$ $r$-I, and $155$ $r$-II metal-poor stars known at present \citep{Gudin.etal.2021, Shank.etal.2023}.

Several possible sites for $r$-process-element production have been proposed. These include neutrino-driven winds in Type-II supernovae \citep{Woosley.Hoffman.1992, Takahashi.etal.1994, Arcones.etal.2007, Wanajo.etal.2018}, the prompt explosion of 
low-mass supernovae \citep{Wheeler.etal.1998, Sumiyoshi.etal.2001, Wanajo.etal.2003}, neutron star -- neutron star or neutron star -- black hole mergers \citep{Lattimer.Schramm.1974, Lattimer.Schramm.1976, Symbalisty.Schramm.1982, Meyer.1989, Freiburghaus.eta.1999, Goriely.etal.2011, Rosswog.etal.2014, Bovard.etal.2017}, and collapsars \citep{Woosley.1993, Nagataki.etal.2007, Fujimoto.etal.2008, Siegel.Metzger.2018, Siegel.etal.2019}.  The neutrino-driven winds in Type-II supernovae model over-produce elements near atomic mass A$\sim$90 and require a level of entropy to produce $r$-process elements higher than predicted in hydrodynamical simulations \citep{Takahashi.etal.1994}. The problem with the prompt explosions of low-mass supernovae model is whether it really explodes or not \citep{Bethe.1990, Sumiyoshi.etal.2001}. The collapse of a rapidly rotating massive star can lead to the formation of an accretion disc around the remnant black hole -- a collapsar -- and has been shown to be a potential candidate for $r$-process-element production \citep{Siegel.etal.2019}. All of these proposed sites are limited by various factors that remain to be verified by observations (and simulations). Given these uncertainties, the actual astrophysical site or sites of $r$-process enrichment is still a topic of debate.

Observations of the electromagnetic counterparts of binary neutron star mergers detected by the LIGO-Virgo collaboration exhibited a kilonova-like behaviour \citep{Arcavi.etal.2017, Smartt.etal.2017, Tanvir.etal.2017}. The excess red emission from the kilonova lightcurve and the detection of a strontium (Sr) spectral feature provide direct evidence for neutron star mergers as one of the robust sites of $r$-process element production \citep{Chornock.etal.2017, Drout.etal.2017,Metzger.2017, Shappee.etal.2017,Tanaka.etal.2017, Villar.etal.2017,Watson2019}.

A single neutron star merger (NSM) event can enrich a large number of stars with $r$-process-element ejecta, as observed in the $r$-process-rich ultra-faint dwarf galaxy Reticulum II (Ret II) \citep{Ji.etal.2016,Ji.etal.2022,Roederer.etal.2016}. We point out that \citet{Ji.etal.2022} reports that over 70 percent of the stars in Ret II with available high-resolution spectroscopic follow-up indicate strong enrichment by $r$-process elements.  However, there remain several difficulties in our understanding.  The long delay between the first supernovae explosions that produce Mg, Fe-peak elements, and ejection of $r$-process material from NSMs, for instance, does not support the observed $r$-process-enhancement by NSM in a pristine galaxy \citep{Argast.etal.2004, Wehmeyer.etal.2015, Tarumi.etal.2021}. The presence of large numbers of Eu-rich stars at low metallicity (see, e.g., \citealt{Holmbeck.etal.2018}) and the trend between [Eu/Fe] and  [Fe/H] for Milky Way stars also does not match with the simulations using NSMs as the primary site for $r$-process-element production \citep{Argast.etal.2004, Wehmeyer.etal.2015}. However, there have been efforts to explain $r$-process enrichment at low metallicity with accretion of stars from low-mass dwarf galaxies \citep{Hirai.etal.2015, Hirai.etal.2022} and neutron star natal kicks \citep{Banerjee.etal.2020}. 

Detailed abundances covering the full range of the $r$-process-element patterns for larger samples of stars clearly assist in efforts to constrain the sites of $r$-process-element production. In this work, we derive elemental abundances for 20 neutron-capture elements in four bright RPE stars.
We compare their elemental-abundance trends with a compiled list of RPE and ``normal" halo stars. We also discuss possibly interesting trends in the abundance ratios.

This paper is outlined as follows. In Section~\ref{sec:observation}, we discuss the observations, telescope, and the high-resolution spectrograph we employ. Data reduction and radial velocity determinations are provided in Section \ref{sec:data_reduction_radial_velocity}. Stellar-parameter estimation using different methods is discussed in Section~\ref{sec:line_list_stellar_paramters}. The detailed chemical abundances of our four program stars are briefly discussed in Section~\ref{sec:abundance_analysis}. Sections~\ref{sec:discussion} and \ref{sec:conclusions} are discussions of our main results and a summary of the key conclusions.

\section{Observations}
\label{sec:observation}
 In this study, we have selected four relatively bright ($V < $ 12) RPE VMP stars from \cite{Bandyopadhyay.etal.2020} [part of HESP-GOMPA survey \citep{Bandyopadhyay.etal.2020_HESP-GOMPA}] to obtain high signal-to-noise ratio (SNR) spectra, including the bluer spectral wavelengths that previously had too low SNRs for detailed study, to determine elemental abundances and upper limits of as many elements as possible, and to probe the rare-earth peak elements. These stars were initially selected from the SDSS-III MARVELS radial velocity survey \citep{Ge.etal.2015}.

Observations were carried out using the High Optical Resolution Spectrograph (HORuS) installed on the Gran Telescopio Canarias (GTC) \citep{HORUS.2020, HORUS.2021}. The GTC is a 10-m class telescope situated at the Roque de Los Muchachos Observatory on the island of La Palma, in the Canaries, Spain. HORuS is located at the Nasmyth focal plane of the telescope, and provides simultaneous wavelength coverage ($3800 - 6900$\,{\AA}) at a spectral resolution of $R \sim 25000$. The observations were obtained in the month of December, 2019. During the observations, we obtained the sets of bias and flat frames required for data-reduction procedure. We also obtained several Th-Ar lamp spectra for the wavelength calibration of the object spectra. All the useful information related to our program stars, including the name of the star, right ascension (RA), declination (DEC), $V$-band magnitude, with calculated SNR per pixel around the $4500$ \r{A} region, exposure time, and date of observation are listed in Table~\ref{tab:object_table}.

\begin{table*}
	\centering
	\caption{Basic information for our program stars. The spectral coverage is 3800$-$6900 \r{A}, and exposure times for each object were 1800 seconds, with two exposures of 900 seconds. The equinox and epoch for the coordinates is J2000.}
	\label{tab:object_table}
	\begin{tabular}{lccccccr} 
		\hline
		Star name & RA & DEC & $V$ mag & SNR & Date of Observation (UT)\\
		\hline
		TYC 3431-689-1 & 09:21:57.28 & +50:34:04.5 & 11.75 & 102 ($4500$ \r{A}) & 2019/12/22\\
		HD 263815 & 06:48:13.33 & +32:31:05.2 & 9.92 & 161 ($4500$ \r{A}) & 2019/12/23\\
		TYC 1191-918-1 & 00:43:05.28 & +19:48:59.1 & 9.90 & 107 ($4500$ \r{A}) & 2019/12/26\\
		TYC 1716-1548-1 & 23:19:23.84 & +19:17:15.4 & 11.57 & 126 ($4500$ \r{A}) & 2019/12/26\\
		\hline
	\end{tabular}
\end{table*}

\section{Data Reduction and Radial Velocity Determinations}
\label{sec:data_reduction_radial_velocity}
\subsection{Data Reduction}
We employ standard procedures for the spectroscopic data reduction of our program stars. First, all the spectra are corrected for bias using a median bias frame, followed by cosmic ray removal and scattered light correction. Then we perform a flat-fielding operation to correct the non-uniform pixel-to-pixel sensitivity of the detector. Wavelength calibration is accomplished using Th-Ar lamp spectra obtained with the same spectroscopic setup as that of the object. Finally, we normalized the continuum of the spectra to unity. All data-reduction operations were performed within the IMRED, CCDRED, and ECHELLE packages of NOAO's Image Reduction and Analysis Facility (IRAF)\footnote{https://iraf-community.github.io}.

\subsection{Radial Velocities}
Radial velocities (RVs) of our program stars with respect to the observer's frame of reference  are calculated by cross-correlating the observed spectra with a template synthetic spectrum with similar stellar parameters. We performed the cross-correlation task using the IDL routine CRSCOR for each spectral order in the entire spectral range from 3800 to 6900 \r{A}. We take the median of radial velocities of these orders as the geocentric radial velocity of the star and their standard deviation ($\sigma$) as the error in the radial velocity.

All the continuum-normalized spectra were corrected for the Doppler shift due to the radial velocities of the stars using the IRAF task DOPCOR. Table~\ref{tab:RVs_table} lists the geocentric RVs and the heliocentric RVs of our program stars corrected for the motion of the Earth around the Sun, the Gaia DR3 RVs \citep{Gaia.Collaboration.2022}, and the RVs reported by \cite{Bandyopadhyay.etal.2020}; the RVs reported by the latter source were geocentric. In this paper, we have corrected their RVs for the motion of earth around the Sun and report heliocentric RVs. Our RV calculations are in good agreement with the recent Gaia DR3 values. There is no apparent variation in RV that might indicate binaries among our targets.

\begin{table*}
	\centering
	\caption{Radial velocities of our program stars.}
	\label{tab:RVs_table}
    \begin{threeparttable}
	\begin{tabular}{lccccr} 
		\hline
	    Star name & Geocentric RV & Heliocentric RV & Error& RV GAIA DR3$^(a)$ & RV HESP$^{(b)}$\\
            & (km/s) & (km/s) & (km/s) & (km/s) & (km/s)\\
		\hline
		TYC 3431-689-1 & $-$127.98 & $-$112.95  & 0.28 & $-$110.75$\pm$0.44 	 & $-$111.96$\pm$0.50\\
		HD 263815 & 133.99 & 139.33 & 0.52 & 138.74$\pm$0.23 & 136.13$\pm$0.41\\
		TYC 1191-918-1 & $-$158.01 & $-$186.96 & 0.19 & $-$187.46$\pm$0.18 & $-$189.26$\pm$0.51\\
		TYC 1716-1548-1 & $-$250.00 & $-$278.27 & 0.43 & $-$277.10$\pm$0.20 & $-$277.89$\pm$0.47\\
		\hline
	\end{tabular}
    \begin{tablenotes}\footnotesize
    \item [a] \cite{Gaia.Collaboration.2022}
    \item [b] \cite{Bandyopadhyay.etal.2020} [we have listed here heliocentric RVs after correction.]
    \end{tablenotes}
    \end{threeparttable}
\end{table*}

\section{Line List and Stellar Parameters}
\label{sec:line_list_stellar_paramters}
\subsection{Line List}
\label{sec:line_list}
For an accurate determination of stellar parameters, we have restricted ourselves to spectral lines using the following selection criteria:

\noindent (i) Blended and asymmetric lines are avoided, due to potential contamination by other lines or species.
\newline (ii) Lines without clear continua are avoided, due to the difficulty of modeling the overly crowded regions of the spectrum.
\newline (iii) Lines with equivalent widths below 20 m\r{A} and above 120 m\r{A} are avoided;  below 20 m\r{A} the lines are often dominated by noise, and above 120 m\r{A} they can approach saturation.

We employ the above criteria for all of our program stars.  Appendix~\ref{sec:Fe1_Fe2_lines_list} provides a combined Fe line list for all the stars. The values of the lower excitation potential (LEP), the $\log gf$, and the measured equivalent widths are listed in this table.

\subsection{Stellar Parameters}
We employ both photometric and spectroscopic measurements for estimation of the stellar parameters (effective temperature: $T_{\rm eff}$, surface gravity: $\log g$, micro-turbulence velocity: $\xi$, and metallicity: [Fe/H]) of our program stars. Here we discuss in detail the approaches we use.

\subsubsection{Photometry}
\label{subsec:photometry}
\begin{figure}
	\centering
	\includegraphics[width=0.45\textwidth]{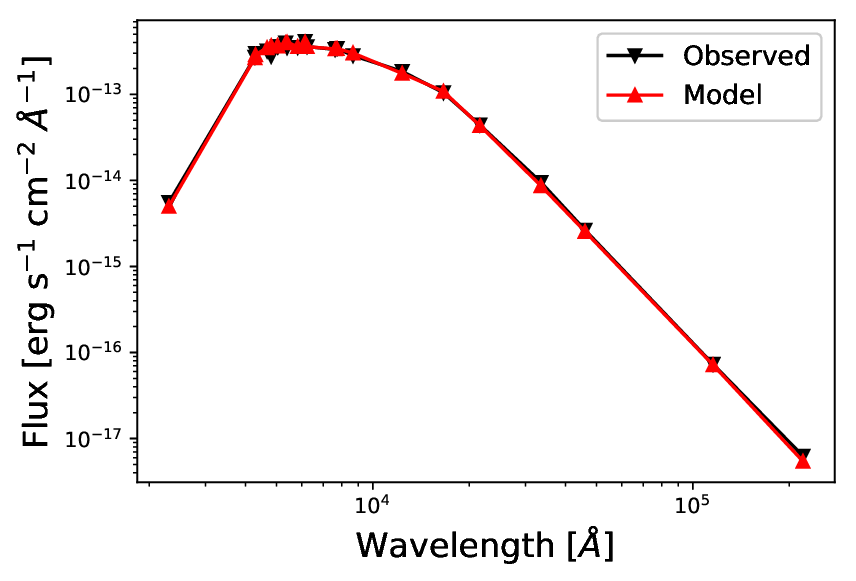}
    \caption{Temperature estimation using SED fitting for one of our program stars, HD 263815. The observed SED is shown with black downward triangles connected by a black line; the fitted SED is represented with red upward triangles, fitted with the red curve.}
    \label{fig:SED_HD263}
\end{figure}

Photometric colours, the spectral energy distribution (SED), and Gaia photometry \citep{Gaia.Collaboration.2021} are used to estimate the photometric temperatures. 
We use the extinction-corrected colours \citep{Schlafly.Finkbeiner.2011} with the temperature-colour relations described in \cite{Alonso.etal.1999} for temperature estimation. For the SED-based temperature estimation, we employ the SED-fitting web tool Virtual Observatory SED Analyzer (VOSA)\footnote{http://svo2.cab.inta-csic.es/theory/vosa/}. VOSA provides several theoretical stellar models to fit the SED,  obtained from various photometric surveys that are publicly available \citep{VOSA.2008}. For each star we apply a Bayesian analysis and chi-square fitting to obtain the temperature estimate. Fig.~\ref{fig:SED_HD263} shows an example of SED fitting for one of our program stars, HD 263815. Temperature estimates obtained from various photometric techniques are listed in different columns of Table~\ref{tab:temperatures}.

\begin{table*}
	\centering
	\caption{Estimate of effective temperature ($T_{\rm eff}$, in K) of our program stars, using different photometric and spectroscopic methods.}
	\label{tab:temperatures}
	\begin{tabular}{lccccccr} 
		\hline
		Star name & H$\alpha$ & V-K & J-K & J-H & Fe-I & SED & GAIA\\
		\hline
		TYC 3431-689-1 & 4900 & 5017 & 4938 & 4754 & 4850 & 5000 & 5450\\
		HD 263815 & 4700 & 4872 & 4809 & 4874 & 4700 & 4750 & 5016 \\
		TYC 1191-918-1 & 4700 & 4690 & 4613 & 4756 & 4650 & 4750 & 4877\\
		TYC 1716-1548-1 & 4550 & 4522 & 4286 & 4362 & 4500 & 4500 & 4644\\
		\hline
	\end{tabular}
\end{table*}

\subsubsection{Spectroscopy}
To estimate the stellar parameters using spectroscopic methods, we have identified Fe I and Fe II absorption lines in the spectra, taking into account the criteria mentioned in Section~\ref{sec:line_list}. Traditionally, Fe lines are used to estimate the stellar parameters due to their ubiquitous presence in optical spectra. Our selection criterion results in 110 Fe I and 17 Fe II lines for TYC 3431-689-1, 102 Fe-I and 18 Fe-II lines for HD 263815, 111 Fe-I and 20 Fe-II lines for TYC1716-1586-1, and 106 Fe-I and 21 Fe-II lines for TYC 1191-918-1. The equivalent width (EW) of each line is calculated interactively by Gaussian fitting of the line in the SPLOT task of IRAF. For a consistency check, we have also calculated the equivalent widths using an automated code, the Automatic Routine for line Equivalent widths in stellar Spectra (ARES; \citealt{ARES.2007}). Due to the good signal-to-noise of the spectra, both methods provide consistent results.

After line identification, we generated a grid of model stellar atmospheres using ATLAS9 \citep{ATLAS.1993, ATLAS.2003}.
We used the spectrum synthesis code TURBOSPECTRUM \citep{Turbospectrum.1998, Turbospectrum.2012} for abundance calculation of Fe I and Fe II lines, which were later used for the estimation of stellar parameters. The same codes are used for the abundance estimation of other elements in Section~\ref{sec:abundance_analysis}.

\begin{figure}
	\centering
	\includegraphics[width=0.45\textwidth]{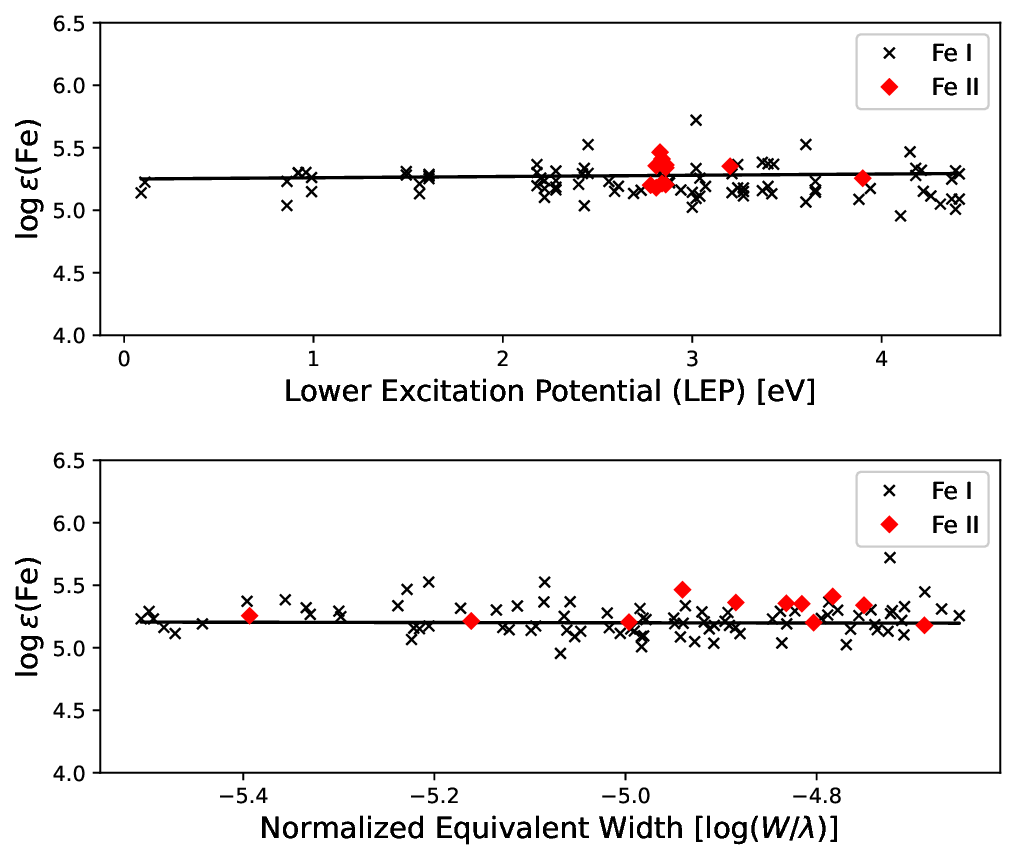}
    \caption{Top: Iron abundance, as a function of lower excitation potential (LEP). Bottom: Iron abundance, as a function of normalized equivalent width, for one of our program stars, HD 263815.}
    \label{fig:Fe_LEP_NEqW_HD263}
\end{figure}

\begin{figure}
	\centering
	\includegraphics[width=0.45\textwidth]{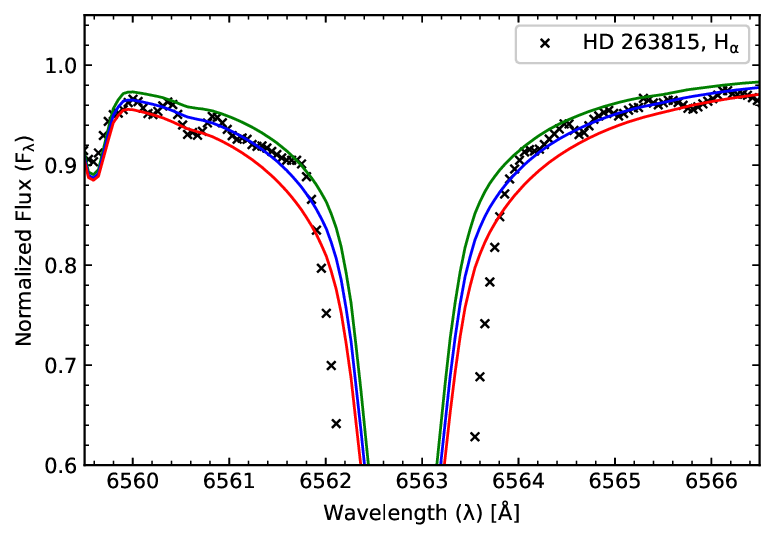}
    \caption{H$\alpha$ wing fitting for one of our program stars, HD 263815. The blue colour represents the best fit at a temperature of 4700\,K; 150\,K deviations in $T_{\rm eff}$ from the best fit are shown with green and red colours.}
    \label{fig:oplot_H1_6552.79}
\end{figure}

Spectroscopic temperature estimates are obtained by assuming excitation equilibrium among the neutral iron lines, i.e., by demanding that the abundance of Fe I lines is independent of the lower excitation potential (LEP), as shown in the top panel of Fig.~\ref{fig:Fe_LEP_NEqW_HD263}.  We have also fitted the wings of the H$\alpha$ absorption profile to obtain an independent estimate, as they are highly sensitive to small changes in temperature. As an illustration, the H$\alpha$ wing-fitting results are shown in Fig.~\ref{fig:oplot_H1_6552.79}. Note that the primary purpose of this exercise is to fit the wings of the H$\alpha$ line to check the consistency in temperature estimates. Thus, we have only shown the wings of H$\alpha$ line for clear visibility \citep[see,][]{Sahin.Lambert.2009, Matsuno.etal.2017}. Also, fitting of the H$_{\alpha}$ line requires full 3D non-LTE (NLTE) modeling, as the core part of the line forms in the upper atmosphere, whereas the wings form deep in the photosphere \citep{Leenaarts.etal.2012}.

\begin{figure*}
	\centering
	\includegraphics[width=\textwidth]{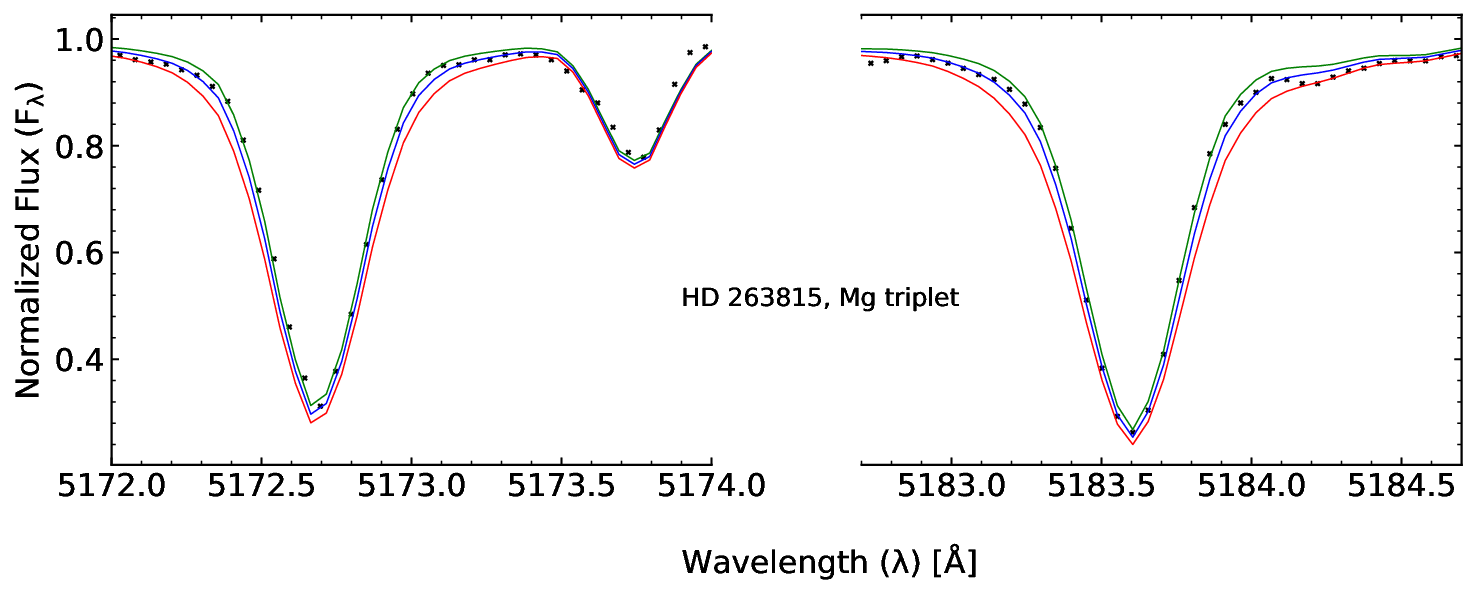}
    \caption{Mg-triplet wing fitting for one of our program stars, HD 263815. The best-fit line is shown with blue colours; 0.3 dex deviations from the best fit are shown with green and red lines.}
    \label{fig:Mg_triplet}
\end{figure*}

\begin{figure}
	\includegraphics[width=\columnwidth]{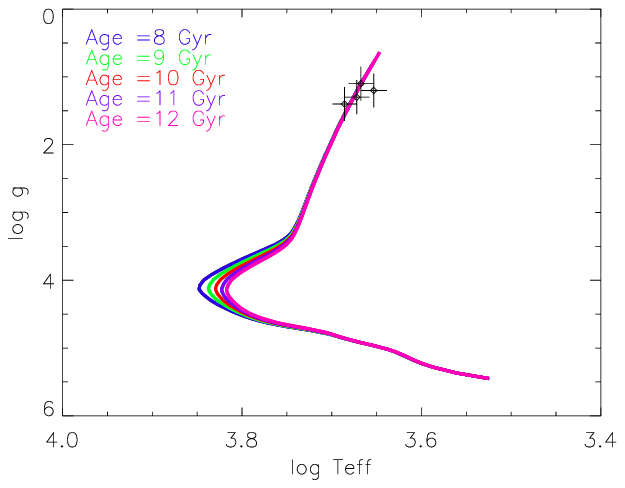}
    \caption{Surface gravity ($\log g$) vs. effective temperature ($\log T_{\rm eff}$) for our program stars. The black diamonds show the locations of the program stars along with the measurement uncertainties. Different colour curves shows isochrones ranging 8 to 12~Gyr in age.}
    \label{fig:isoc_temp_logg}
\end{figure}

Stellar surface gravity is estimated presuming the ionization balance of an element, i.e., by forcing the abundance of an element in its neutral state (Fe I) and ionized state (Fe II) to be the same by changing the surface gravity. 
We also fit the wings of the magnesium (Mg) triplet (5167\r{A}, 5172\r{A}, 5187\r{A}) to obtain an estimate of $\log g$, because the Mg-triplet wings are sensitive to small changes in surface gravity. As an illustration, Fig.~\ref{fig:Mg_triplet} shows fits for portions of the Mg-triplet region. For a consistency test, we place our program stars on stellar isochrones ranging from 8 to 12~Gyr in age. These isochrones are obtained from the CMD 3.6 website \footnote{http://stev.oapd.inaf.it/cgi-bin/cmd}, which uses PARSEC \citep{parsec2012} and COLIBRI \citep{marigo2013} tracks. Fig.~\ref{fig:isoc_temp_logg} shows the isochrones and our program stars.

It is  customary to use the abundance of iron (Fe) as an indicator of stellar metallicity. 
The absolute abundances are calculated on a log scale, with the number density of H assumed to be 12; relative abundances are calculated with respect to the solar ratio \footnote{Absolute abundance of element A: $\log \epsilon (A) = \log (N_{A}/N_{H}) + 12$ and the relative abundance of element A with respect to element B: $[A/B] = \log (N_{A}/N_{B}) - \log (N_{A}/N_{B})_{\odot}$, where $\odot$ represents values for the Sun.}. In this work, we have used solar abundances from \cite{Asplund.etal.2009}. Table \ref{tab:Adopted_stellar_parameters} lists metallicities for all of our program stars.

\begin{figure}
	\includegraphics[width=\columnwidth]{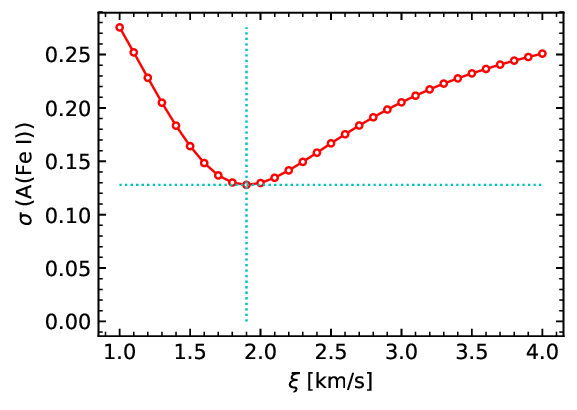}
    \caption{Micro-turbulence velocity estimation using dispersion in the abundances of 
    Fe I lines for one of our program stars, HD 263815. Here, the crossing point of vertical and horizontal lines represents the minimum of the distribution. The      micro-turbulence value that corresponds to this crossing point is taken as its estimate in the stellar atmosphere.}
    \label{fig:micro_turb_vel}
\end{figure}

To calculate the micro-turbulence velocity, we force the abundance of neutral iron lines (Fe I) to be independent of reduced equivalent width ($\log [EW/\lambda]$), as shown in the bottom panel of Fig.~\ref{fig:Fe_LEP_NEqW_HD263}. For an additional check, we also calculate an estimate of the micro-turbulence velocity using the method described in \citet{Sahin.Lambert.2009}, \citet{Reddy.etal.2012}, and \citet{Molina.etal.2014}, which makes use of the dispersion in the abundance of an element. The value of micro-turbulence where the dispersion becomes a minimum is taken as the micro-turbulence velocity of the atmosphere. For this analysis, we used the abundances of Ti I, Ti II, Cr I, Cr II, Fe I, Fe-II, V I, and V II, depending on the presence of significant numbers of clean lines in the spectra. For example, Fig.~\ref{fig:micro_turb_vel} shows the dispersion in the elemental abundance of Fe I lines for one of our program stars, HD 263815. The minimum of the distribution is the estimated micro-turbulence velocity in the stellar atmosphere. Table~\ref{tab:micro_turb_vel} lists micro-turbulence velocities of our program stars using the above methods. For further analysis, we have restricted ourselves to the stellar parameters obtained using the spectroscopic method, i.e., $T_{\rm eff}$ from excitation equilibrium, $\log g$ from ionization balance, micro-turbulence from zero trend of Fe I with normalized equivalent width, and metallicity from the mean Fe I abundance. The final adopted stellar parameters for our program stars are listed in Table~\ref{tab:Adopted_stellar_parameters}. Our parameters are in good agreement with those from \citet{Bandyopadhyay.etal.2020}, except for the metallicities. We identified the likely reason for this discrepancy as due to systematically lower values for equivalent widths obtained by the previous authors (used by them primarily for the Fe-peak elements).

\begin{table}
	\centering
	\caption{Estimates of micro-turbulence velocity ($\xi$) of our program stars using the methods described in the text.}
	\label{tab:micro_turb_vel}
	\begin{tabular}{lcr} 
		\hline
		Star name & using abundance of & using dispersion of\\
		& Fe I \& Fe II lines & Fe I abundance\\
		& (km/s) & (km/s) \\
		\hline
		TYC 3431-689-1 & 1.80 & 2.00 \\
		HD 263815 & 1.85 & 1.90\\
		TYC 1191-918-1 & 1.95 & 2.10\\
		TYC 1716-1548-1 & 1.95 & 2.00\\
		\hline
	\end{tabular}
\end{table}

\begin{table}
	\centering
	\caption{Final adopted stellar parameters for our program stars.}
	\label{tab:Adopted_stellar_parameters}
	\begin{tabular}{lcccc} 
		\hline
		Star name & $T_{\rm eff}$ & $\log g$ & [Fe/H] & $\xi$\\
		 & (K) &  &  & (km/s) \\
		\hline
		TYC 3431-689-1 & 4850 & 1.4 & $-$2.05 & 1.80 \\
		HD 263815 & 4700 & 1.3 & $-$2.22 & 1.85\\
		TYC 1191-918-1 & 4650 & 1.1 & $-$1.90 & 1.95\\
		TYC 1716-1548-1 & 4500 & 1.2 & $-$2.15 & 1.95\\
		\hline
	\end{tabular}
\end{table}

\section{Abundance analysis}
\label{sec:abundance_analysis}
In this study, we use a combination of equivalent-width analysis and spectrum
synthesis methods for the elemental-abundance estimates. For the lighter elements, up to Zn, one can easily find unblended lines in our spectra. Thus, we use the 
equivalent-width method, but also carried out spectrum synthesis for some of the lines to check the consistency. In the case of heavier elements, most of the absorption features contain contributions from other atomic transitions, thus spectrum synthesis is performed for them.

In the equivalent-width analysis we select lines with equivalent widths $<$ 120 m\r{A}, which are on the linear part of the Curve of Growth, and are not sensitive to the broadening of the wings. Note that we have dropped the lower-limit criterion of 20 m{\AA} to study more elements and lines. For each element, we have used the solar abundances from \cite{Asplund.etal.2009}. We have also incorporated the hyperfine-splitting information for Sc, V, Mn, Cu, Ba, La, Pr, Nd, Sm, and Eu. Solar isotopic ratios are employed.

\subsection{Light Elements: Molecular Bands (Carbon and Nitrogen)}
\begin{figure*}
	\includegraphics[width=\textwidth]{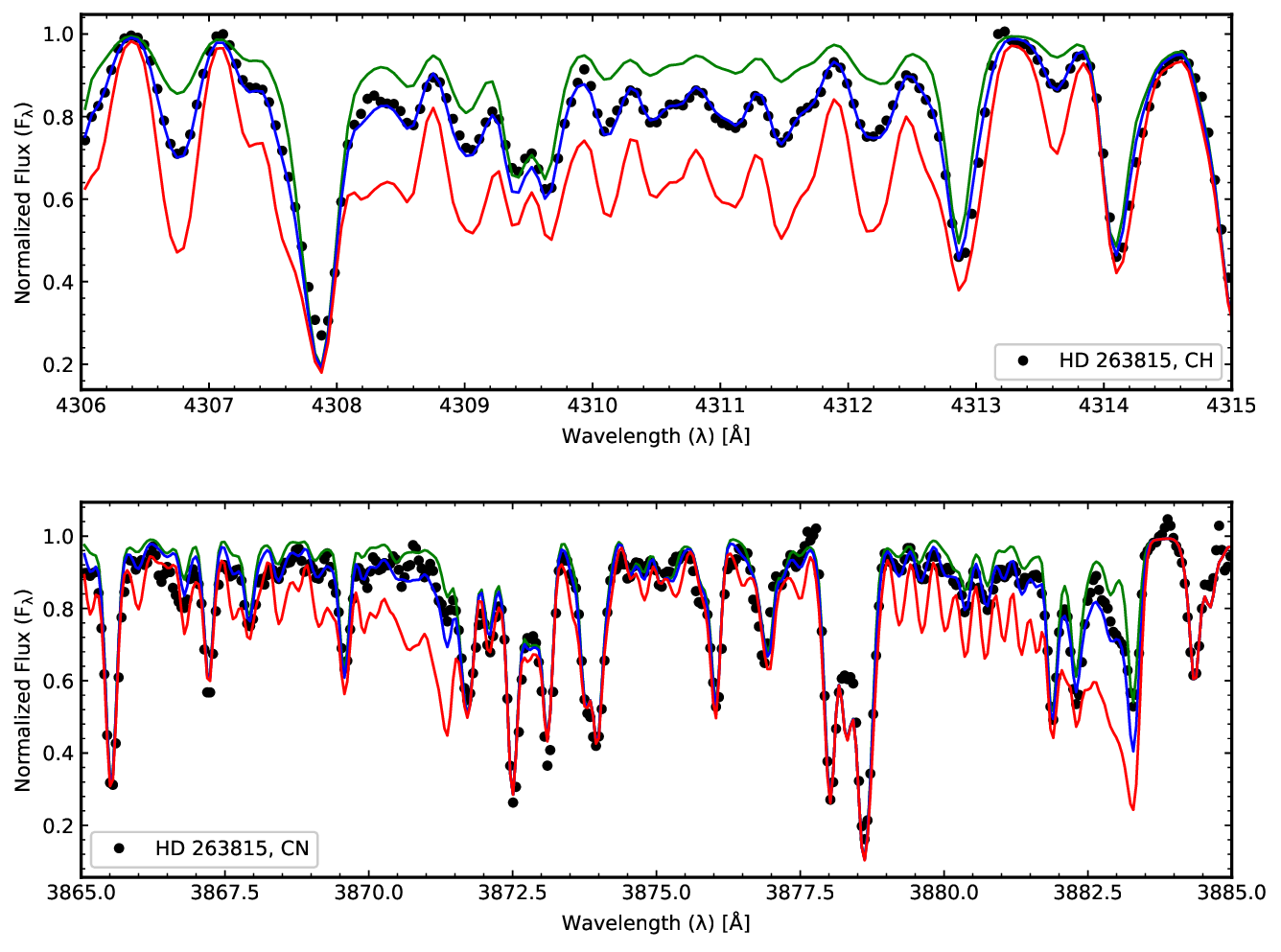}
    \caption{C and N abundance estimation for one of our program stars, HD 263815, using the spectrum-synthesis method. The top and bottom panels show the C and N synthesis, respectively. The black dotted points represent the observed spectra and the solid blue colours show the best fit. The green and red colours indicate $\pm$~0.5 dex deviations from the best fit.}
    \label{fig:CH_CN_band}
\end{figure*}

We estimate carbon (C) abundances from the CH $G$-band molecular region around 4313\,{\AA},  which could be obtained for all of our program stars using the spectrum-synthesis method. We have used the line list from \cite{Masseron_2014} for CH band synthesis. The top panel of  Fig.~\ref{fig:CH_CN_band} shows the observed spectrum using black dots and the best-fit synthetic spectrum using a blue colour line for one of our program stars, HD 263815. The green and red colours indicate $\pm~0.5$ dex deviations from the best-fit synthesis line. We have also attempted to constrain the $^{12}$C$/^{13}$C ratio. For this purpose, we performed spectral synthesis in 4211\,{\AA}, 4230.30\,{\AA} and 4231.45\,{\AA} regions. Our analysis shows that the average $^{12}$C$/^{13}$C is 3.0 for TYC 3431-689-1, 3.5 for HD 263815, 5.5 for TYC 1191-918-1, and 5.5 for TYC 1716-1548-1. Due to the high SNRs of our spectra in the blue region, we could also evaluate the nitrogen (N) abundances for all four of our program stars. The line list for CN band is adopted from \cite{Plez_and_cohen_2005}. We use the CN-band molecular region around 3883\,{\AA} for calculating the abundance of N. We performed the spectral synthesis of the CN band by keeping the carbon abundance fixed at the value obtained from the CH $G$-band. An example of the CN-band synthesis is shown in the bottom panel of Fig.~\ref{fig:CH_CN_band}.

Our program stars exhibit relatively low surface carbon abundances, as expected from their current evolutionary stages. All four stars fall in the red giant region, indicating that their C abundances have been depleted during evolution up the red giant branch. We use the method described in \cite{Placco.etal.2014} to correct the observed carbon abundances for the evolutionary states of the stars. Table~\ref{tab:r_process_classification} lists the LTE values of the observed C abundances from our analysis, along with their corrected C abundances.

\begin{table*}
	\centering
	\caption{Dilution of carbon corrected according to \protect \cite{Placco.etal.2014}, and $r$-process classification of our program stars from \protect \cite{Holmbeck.etal.2020}.}
	\label{tab:r_process_classification}
	\begin{threeparttable}
	\begin{tabular}{lcccccccccr} 
		\hline
		Star name & [Fe/H] & [C/Fe]$_{LTE}$ & $\Delta$[C/Fe]$^{a}_{cor}$ & [C/Fe]$_{cor}$ & [Sr/Fe] & [Ba/Fe] & [Eu/Fe] & [Ba/Eu] & [Sr/Ba] & Class\\
		\hline
		TYC 3431-689-1 & $-$2.05 & $-$0.60 & +0.55 & $-$0.05 & +0.15 & +0.25 & +0.80 & $-$0.55 & $-$0.10 & $r$-II \\
		HD 263815 & $-$2.22 & $-$0.58 & +0.63 & +0.05 & $-$0.18 & $-$0.13 & +0.52 & $-$0.65 & $-$0.05 & $r$-I\\
		TYC 1191-918-1 & $-$1.90 & $-$0.70 & +0.66 & $-$0.04 & $-$0.10 & $-$0.14 & +0.55 & $-$0.69 & +0.04 & $r$-I \\
		TYC 1716-1548-1 & $-$2.15 & $-$0.35 & +0.62 & +0.27 & +0.15 & +0.17 & +0.80 & $-$0.63 & $-$0.02 & $r$-II	\\
		\hline
	\end{tabular}
	\begin{tablenotes}\footnotesize
    \item [a] \href{https://vplacco.pythonanywhere.com/#}{https://vplacco.pythonanywhere.com/}.
\end{tablenotes}
\end{threeparttable}
\end{table*}

\subsection{Alpha-elements: Mg, Si, Ca, Ti}
We have identified the $\alpha$-elements Mg, Si, Ca, and Ti in our program stars. In all the stars, we identified numerous clean Ti I and Ti II lines, so the Ti abundance was calculated using the equivalent-width method. Although there are significant numbers of Ca I lines present in the spectra, we used both the equivalent-width and spectrum-synthesis methods for Ca abundance estimation. Abundances of Mg and Si were determined using the spectrum-synthesis method. We derive the Si abundance using the 4102\,{\AA} and 6155\,{\AA} lines.

\subsection{Odd-Z Elements: Na, Al}
\begin{figure}
	\includegraphics[width=\columnwidth]{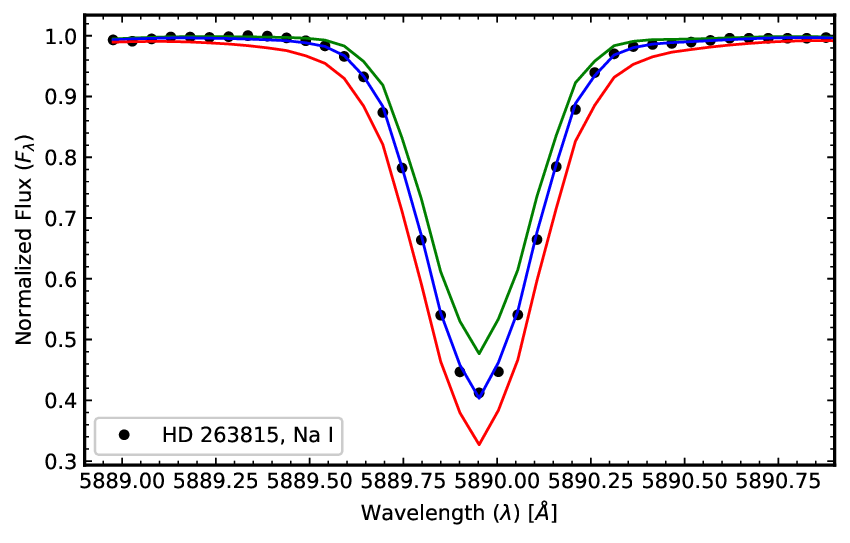}
    \caption{Synthesis of the Na-absorption feature. The black dotted points represent the observed spectra and the solid blue colour shows the best fit. Green and red colours indicate $\pm$~0.5 dex deviations from the best fit.}
    \label{fig:synthesis_Na}
\end{figure}

Among the odd-Z elements, we could calculate the abundances of sodium (Na) and Aluminium (Al). The sodium D1 and D2 lines at 5890\,{\AA} and 5896\,{\AA}, respectively, are used to calculate the abundance of Na. The abundance of Al is derived from the resonance lines present at 3941\,{\AA} and 3961\,{\AA}. These four lines are strongly affected by NLTE effects, so we have taken into account these corrections \citep{Baumueller.etal.1997, Gehren.etal.2004, Andrievsky.etal.2007, Andrievsky.etal.2008, Lind.etal.2011}. Proper care has been taken for the Na synthesis in accordance to \cite{Barklem.O'Mara.2001} for line broadening due to collisions. Fig.\ref{fig:synthesis_Na} shows the synthesis of the Na I line at 5890\,{\AA} to demonstrate the spectral synthesis considering collisional broadening. We have also verified the consistency of our estimated Na and Al abundances with those of other halo stars.

\subsection{Iron-peak Elements: Sc, V, Cr, Mn, Fe, Co, Ni, Cu, Zn }
We have calculated the abundances for some of the iron-peak elements Sc II, V I, V II, Cr I, Cr II, Mn I, Fe I, Fe II, Co I, Ni I, Cu I, and Zn I. A mix of both the equivalent-width method and the spectrum-synthesis method are used for abundance estimation. Hyperfine splitting is also considered wherever required. Iron (Fe I and Fe II) abundances are calculated using the equivalent-width method during the
stellar-parameter estimation procedure. 

Abundances of Co, Cu, and Zn are determined using the spectrum-synthesis method. We found only one line, at 5105.53\,{\AA}, useful for Cu abundance estimation. The Co abundance was evaluated using the 5020.82\,{\AA}, 4110.53\,{\AA}, and 4121.31\,{\AA} lines. We use the 4810.52\,{\AA}, and 4722.15\,{\AA} lines for the Zn-abundance calculation. 
Tables~\ref{tab:abundances_TYC3431}, \ref{tab:abundances_HD2638}, \ref{tab:abundances_TYC1716}, and \ref{tab:abundances_TYC1191} list the derived abundances of the iron-peak elements and the number of lines used for their estimation. All the abundances are consistent with previous estimates for halo stars.

\subsection{Neutron-capture Elements}
The absorption lines of neutron-capture elements are often blended with other elements. Sometimes they possess hyperfine structure as well. Thus, we use the spectrum-synthesis method for the abundance estimation of neutron-capture elements. We have identified a total of 20 neutron-capture elements: Sr, Y, Zr, Ba, La, Ce, Pr, Nd, Sm, Eu, Gd, Tb, Dy, Ho, Er, Tm, Lu, and Hf, and some important third $r$-process peak elements like Os and Th. Note that some elements may or may not be present in all of our program stars. In this study, the abundances of 10 elements (Pr, Gd, Tb, Ho, Er, Tm, Lu, Hf, Os, Th) of the 20 considered have been derived for the first time for our program stars. 

\subsubsection{First-peak elements}
We obtain measurements of the absorption lines for three first-peak $r$-process elements: Sr, Y, and Zr. The abundance of Sr is estimated from two lines at 4077\,{\AA}, and 4215\,{\AA}. Both of these lines yielded very similar abundances. We have also estimated Sr abundance using the Sr I line present at 4607{\AA}, and found that it yields significantly lower abundance because of NLTE effects \citep{Bergemann.etal.2012}. Therefore, we have not considered it for the average abundance calculation of Sr. The abundances of Y and Zr are estimated using 7 transitions for each element. Fig.~\ref{fig:synthesis_collage}(a) shows the synthesis of the Sr line in the 4215\,{\AA} wavelength region. Similar to many RPE stars, the first $r$-process peak does not match very well with the scaled-solar $r$-process pattern (see Fig.~\ref{fig:r_process_pattern_new.eps}).

\subsubsection{Second-peak elements}
In the second-peak region of the $r$-process pattern, we have detected a total of 15 elements: Ba, La, Ce, Pr, Nd, Sm, Eu, Gd, Tb, Dy, Ho, Er, Tm, Lu, and Hf. We were unable to measure Pr and Hf in TYC 1716-1548-1, and no Gd, Ho, and Er measurements were possible in TYC 1191-918-1. The element Tb was only measured in HD 263815 and TYC 3431-689-1, while Tm was available only in HD 263815. All the information related to the identified lines is listed in appendix~\ref{sec:other_lines_list}.
\begin{figure*}
	\includegraphics[width=\textwidth]{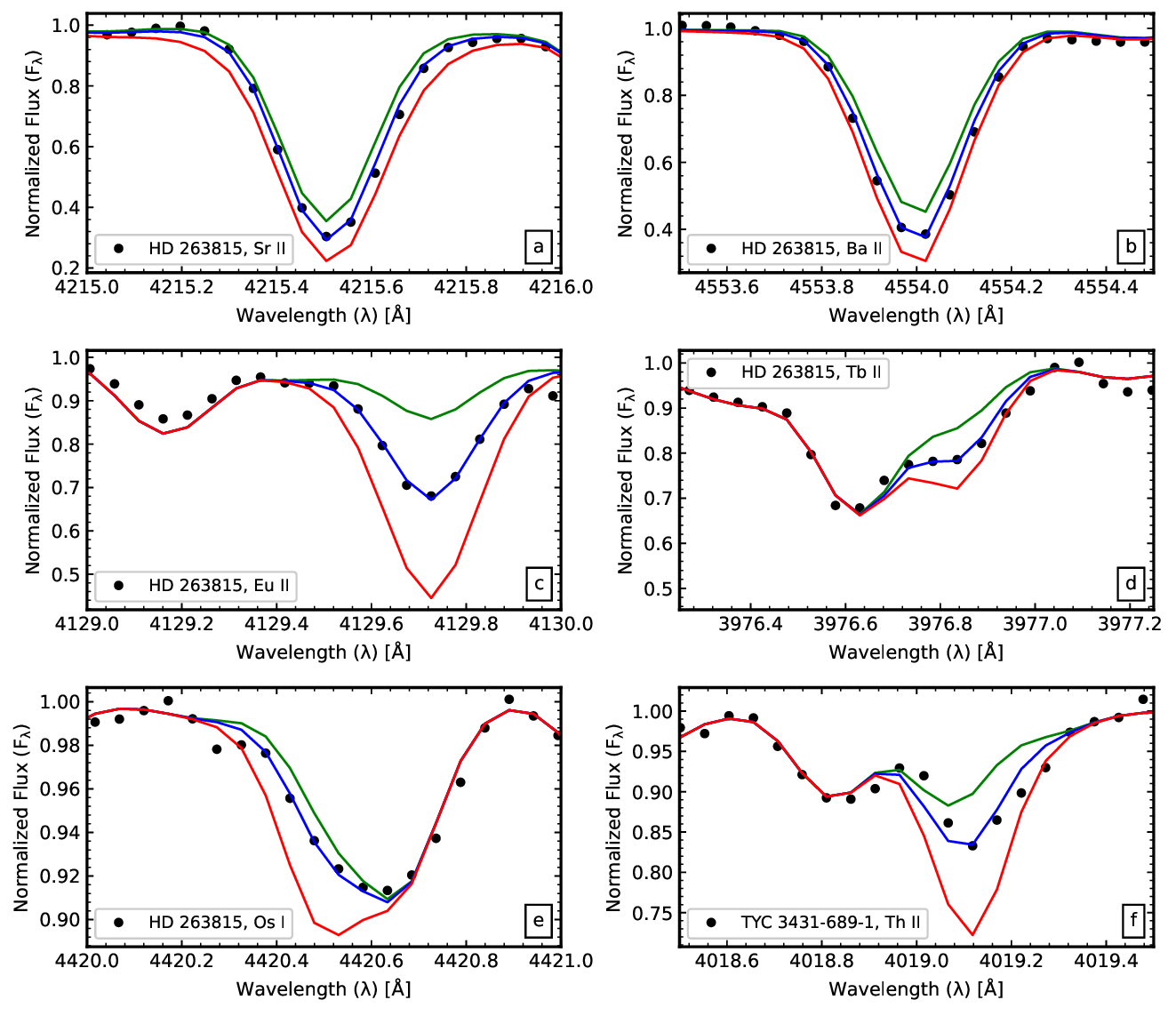}
    \caption{Synthesis of Sr, Ba, Eu, Tb, Os, and Th. The black dotted points represent the observed spectra and the solid blue colour shows the best fit. Green and red colours indicate $\pm$~0.5 dex deviations from the best fit.}
    \label{fig:synthesis_collage}
\end{figure*}

The abundance of Ba is estimated using the lines at 4554\,{\AA}, 4934\,{\AA}, 5853\,{\AA}, \,6142{\AA}, and 6496\,{\AA}. The line at 4934\,{\AA} is blended with an Fe I line whose $\log gf$ is not well-constrained \citep{Gallagher.etal.2012, Hansen.etal.2018}. We have corrected the abundance derived from the line at 4554\,{\AA} for NLTE effects from \cite{Short.Hauschildt.2006}, where the authors showed a maximum NLTE correction of +0.14 dex at [Fe/H] $= -4.0$. Fig.~\ref{fig:synthesis_collage}(b) shows the synthesis for one of the Ba lines at 
4554\,{\AA}. The Eu abundance is primarily derived from the two lines at 4129\,{\AA} and 4205\,{\AA}. Both these lines yield very similar abundance estimates. Fig.~\ref{fig:synthesis_collage}(c) shows the spectrum synthesis for the Eu line at 
4129\,{\AA}. We estimated Tb and Lu abundances using spectrum synthesis of lines at 3976\,{\AA} and 5476\,{\AA}, respectively. Fig.~\ref{fig:synthesis_collage}(d) and Fig.~\ref{fig:synthesis_Lu_Os}(b) show the synthesis of the Tb II and Lu II lines, respectively. Abundances of all the second-peak $r$-process elements are listed in 
Tables~\ref{tab:abundances_TYC3431}, \ref{tab:abundances_HD2638}, \ref{tab:abundances_TYC1716}, and \ref{tab:abundances_TYC1191}. These are in good agreement with the scaled-solar $r$-process pattern (see Fig.~\ref{fig:r_process_pattern_new.eps}).

\subsubsection{Third-peak elements}
\begin{figure}
	\includegraphics[width=\columnwidth]{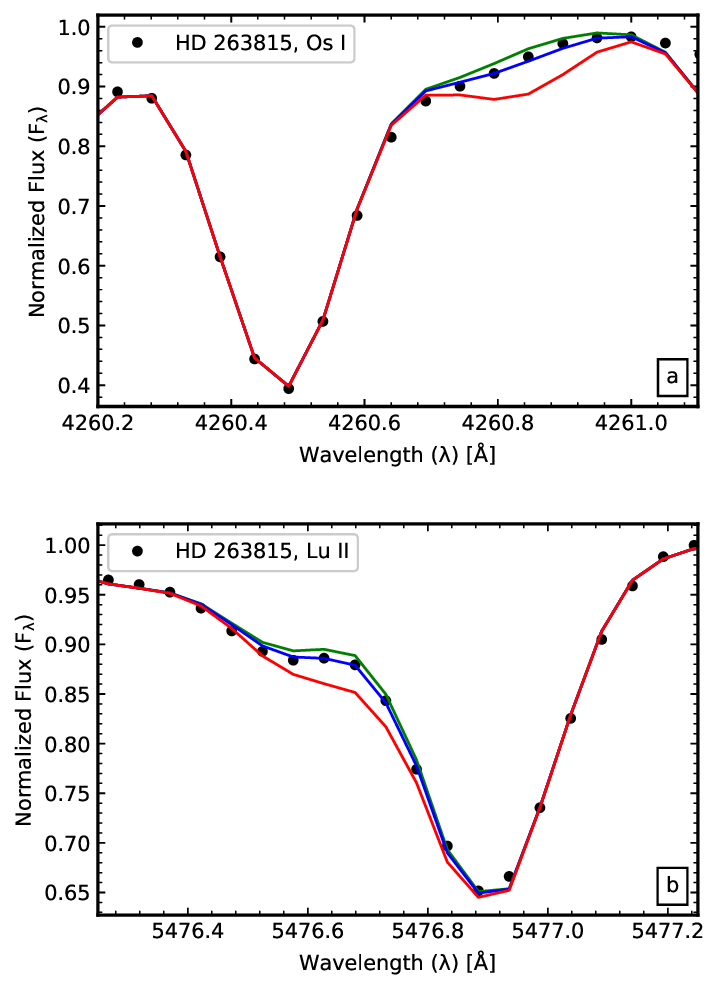}
    \caption{Synthesis of regions containing Os and Lu lines. The black dotted points represent the observed spectra and the solid blue colour shows the best fit. The green and red colours indicate $\pm$~0.5 dex deviations from the best fit.}
    \label{fig:synthesis_Lu_Os}
\end{figure}

We have estimated the abundances of two important elements (Os and Th) in the third $r$-process peak. The Os abundance is derived from two lines in the 4420\,{\AA} and 4265\,{\AA} regions. We have performed synthesis of the two Os lines shown in Fig.~\ref{fig:synthesis_collage}(e), and \ref{fig:synthesis_Lu_Os}(a). The Th abundance is estimated using one strong line present in the spectra in the 4019\,{\AA} region. It is highly blended with several other lines, such as V, Fe, Co, Ni, Ce, Nd, etc., along with the molecular CH $G$-band. We took care of all the blends while synthesizing Th. First, we estimated the abundances of blended elements using clean lines present in the other parts of the spectrum. Then we fixed the abundances of these blends and varied the Th abundance to match the observed spectrum with a synthetic spectrum. Table~\ref{tab:lines_list_for_Th_synthesis} lists all the lines used for the Th II 4019\,{\AA} region synthesis \cite[references,][]{Johnson.Bolte.2001, Hayek.etal.2009, Ren.etal.2012}. We could obtain an upper limit for lead (Pb) for three of our program stars. Fig.~\ref{fig:synthesis_collage}(f) shows the synthesis of Th in TYC 3431-689-1. To test the robustness of our Th determination in the spectra, we compute equivalent widths of all the contributing elements in the 4019.00$-$4019.44\,{\AA} range. We found the contributions of Th in this absorption line is 44 percent for TYC 3431-689-1 and 59 percent for TYC 1191-918-1. Due to the low signal-to-noise ratio (SNR < 15) in the 3959\,{\AA} region, we could not detect U in our program stars. The highly RPE stars with low [C/Fe] would be good candidates for U detection using GTC.

\begin{table}
	\centering
	\caption{Line list used for the spectrum synthesis of Th II lines in the 4019.129\,{\AA} region.
}
	\label{tab:lines_list_for_Th_synthesis}
	\begin{tabular}{lcccr} 
		\hline
		Species & wavelength ({\AA}) & log gf & LEP (eV)\\
		\hline
		Nd II & 4018.823 & $-$0.899 & 0.064\\
		Cr I  &  4018.826 & $-$2.629 & 3.648\\
		Cr I & 4018.863 & $-$2.822 & 4.440\\
		V I & 4018.929 & $-$0.651 & 2.581\\
		Mn I & 4018.999 & $-$1.497 & 4.354\\
		V II & 4019.036 & $-$2.704 & 3.753\\
		Fe I & 4019.042 & $-$2.780 & 2.608\\
		Fe I & 4019.052& $-$2.78 & 2.608\\
		Ce II & 4019.057 & $-$0.213 & 1.014\\
		Ni I & 4019.058 & $-$3.174 & 1.935\\
		Fe II & 4019.110  & $-$3.102 & 9.825\\
		Co I & 4019.126 & $-$2.270 & 2.280\\
		Th II & 4019.129 & $-$0.228 & 0.000\\
		V I & 4019.134 & $-$1.999 & 1.804\\
		Co I & 4019.163 & $-$3.136 & 2.871\\
		Fe II & 4019.181 & $-$3.532 & 7.653\\
		Co I & 4019.289 & $-$3.232 & 0.582\\
		Co I & 4019.299 & $-$3.769 & 0.629\\
		Cr II & 4019.289 & $-$5.604 & 5.330\\
		$^{12}$CH & 4019.440 & $-$7.971 & 1.172\\
		\hline
	\end{tabular}
\end{table}

\subsection{Uncertainties in Abundances}
The calculation of elemental abundances requires stellar parameters and the line list as inputs. The stellar parameters from spectroscopic methods also need an atomic line list as input. Each line has some intrinsic uncertainty in the atomic/molecular data and the line parameters, which propagates all the way to the estimation of elemental abundances. The SNR of the spectrum also contributes to this uncertainty \citep{Cayrel.1988}. In this study, we calculate errors in our abundance estimation assuming a 150\,K uncertainty in $T_{\rm eff}$, and a 0.25~dex uncertainty in $\log g$. Errors due to the SNR ratio are computed using the expression given in \citet{Cayrel.1988}. The total error in the elemental abundance is given as a quadratic sum of all the errors.

\section{Discussion}
\label{sec:discussion}
\subsection{Level of $r$-process Enhancement and Classification}
Traditionally, abundances of C, Fe, Sr, Ba, and Eu are used to classify metal-poor stars into several categories according to the enhancement level of these elements. 
Europium is often used as an indicator of the $r$-process, while Ba is often taken as an indicator of the $s$-process. Classifications of metal-poor stars were initially provided by \cite{Beers.Christlieb.2005}, and modifications were suggested by \citep{Frebel.2018}.  \cite{Cain.etal.2020} added a class of $r$-III stars to include extremely RPE stars with [Eu/Fe] $> + 2.0$. Recently, \cite{Holmbeck.etal.2020} updated the $r$-process classification using a large sample of RPE stars from  the RPA survey, most notably shifting the dividing line between $r$-I and $r$-II stars to [Eu/Fe] = +0.7 based on a statistical analysis, which we adopt for the rest of this paper. These classifications are listed in Table~\ref{tab:class_of_stars}. Some mildly RPE stars exhibit excess abundances of the first $r$-process peak elements that could be the result of a limited-$r$ process, defined as [Eu/Fe] $< +0.3$, [Sr/Ba] $> +0.5$, and [Sr/Eu] $ > 0.0$ \citep{Frebel.2018} and employed by \cite{Holmbeck.etal.2020}. The common presence of C-enhancement in VMP stars led to the classification of carbon-enhanced metal-poor (CEMP) stars, originally based on  [C/Fe] $> +1.0$ by \citet{Beers.Christlieb.2005}, but later revised to [C/Fe] $> +0.7$ by several authors, which we employ throughout this paper.

\begin{table*}
	\centering
	\caption{Classification of neutron-capture enhanced metal-poor stars.}
	\label{tab:class_of_stars}
	\begin{tabular}{llll} 
	\hline
		& \cite{Beers.Christlieb.2005} & \cite{Cain.etal.2020} & \cite{Holmbeck.etal.2020}\\
		\hline
		$r$-I & $0.3\leq{\rm [Eu/Fe]}\leq+1.0$ and ${\rm [Ba/Eu]}<0$ & $0.3\leq{\rm [Eu/Fe]}\leq+1.0$ and ${\rm [Ba/Eu]}<0$ & $0.3<{\rm [Eu/Fe]}\leq+0.7$ and ${\rm [Ba/Eu]}<0$\\
		$r$-II & ${\rm [Eu/Fe]}>+1.0$ and ${\rm [Ba/Eu]}<0$ & $1.0<{\rm [Eu/Fe]}\leq+2.0$ and ${\rm [Ba/Eu]}<0$ & ${\rm [Eu/Fe]}>+0.7$ and ${\rm [Ba/Eu]}<0$\\
		$r$-III & $-$ & ${\rm [Eu/Fe]}>+2.0$ and ${\rm [Ba/Eu]}<0$ & $-$\\
		s & ${\rm [Ba/Fe]}>+1.0$ and ${\rm [Ba/Eu]}>+0.5$ & $-$ & $-$\\
		r/s & $0.0<{\rm [Ba/Eu]}<+0.5$ & $-$ & $-$\\
        limited-r & $-$ & $-$ & [Eu/Fe] $< +0.3$, [Sr/Ba] $> +0.5$, and [Sr/Eu] $> 0.0$\\
		\hline
	\end{tabular}
\end{table*}

\begin{figure}
	\centering
	\includegraphics[width=\columnwidth]{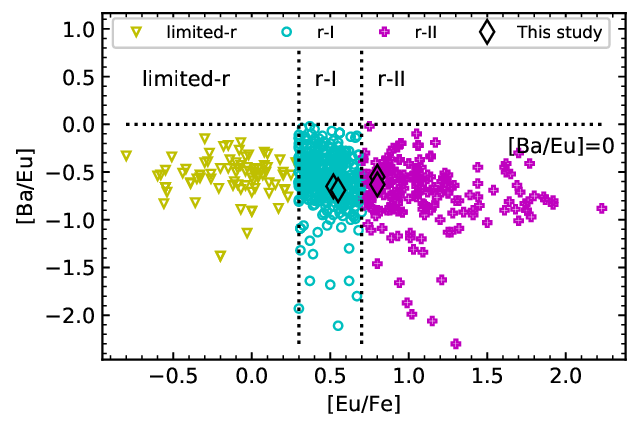}
    \caption{Distribution of [Ba/Eu] as a function of [Eu/Fe]. The limited-$r$ stars are shown in yellow colours, the $r$-I stars in cyan colours, and the $r$-II stars in magenta colours.  The four black diamonds indicate the $r$-I (cyan) and $r$-II (magenta) program stars in this study. References for the compiled data (including from the RPA): \protect\cite{McWilliam.etal.1995, Ryan.etal.1996, Burris.etal.2000, Fulbright.2000, Westin.etal.2000, Cowan.etal.2002, Cayrel.etal.2004, Christlieb.etal.2004, Honda.etal.2004, Aoki.etal.2005, Barklem.etal.2005, Ivans.etal.2006, Preston.etal.2006, Lai.etal.2008, Hayek.etal.2009, Aoki.etal.2010, Roederer.etal.2010, Cohen.etal.2013, Johnson.etal.2013, Hansen.etal.2012, Ishigaki.etal.2013, Mashonkina.etal.2014, Roederer.etal.2014, Jacobson.etal.2015, Howes.etal.2015, Li.etal.2015, T_Hansen.etal.2015, Howes.etal.2016, Placco.etal.2017, JINAbase2018, Cain.etal.2018, Frebel_2018, Hansen.etal.2018, Hawkins.etal.2018, Holmbeck.etal.2018, Roederer.etal.2018, Sakari.etal.2018, Mardini.etal.2019, Valentini.etal.2019, Xing.etal.2019, Cain.etal.2020, Holmbeck.etal.2020, Placco.etal.2020, Rasmussen.etal.2020}.}
    \label{fig:Ba_Eu_vs_Eu_Fe.eps}
\end{figure}

For the classification of our program stars, we first determined the carbonicity ([C/Fe]), and find that all of our program stars exhibit [C/Fe] < $+0.7$. We then calculated the [Eu/Fe] and [Ba/Eu] ratios to estimate the level of $r$-, and $s$-process enhancement. Fig.~\ref{fig:Ba_Eu_vs_Eu_Fe.eps} shows the location of our program stars in the [Ba/Eu] versus [Eu/Fe] plane. According to \cite{Beers.Christlieb.2005} classification, all of our target stars fall under the $r$-I category. However, according to the updated classification by \cite{Holmbeck.etal.2020}, which we use in this study, two stars (HD 263815 and TYC 1191-918-1) are $r$-I while the other two (TYC 3431-689-1 and TYC 1716-1548-1) are considered as $r$-II (see Fig.~\ref{fig:Ba_Eu_vs_Eu_Fe.eps}).

\subsection{The Light Elements (Z < 30)}
\begin{figure*}
	\centering
	\includegraphics[width=\textwidth]{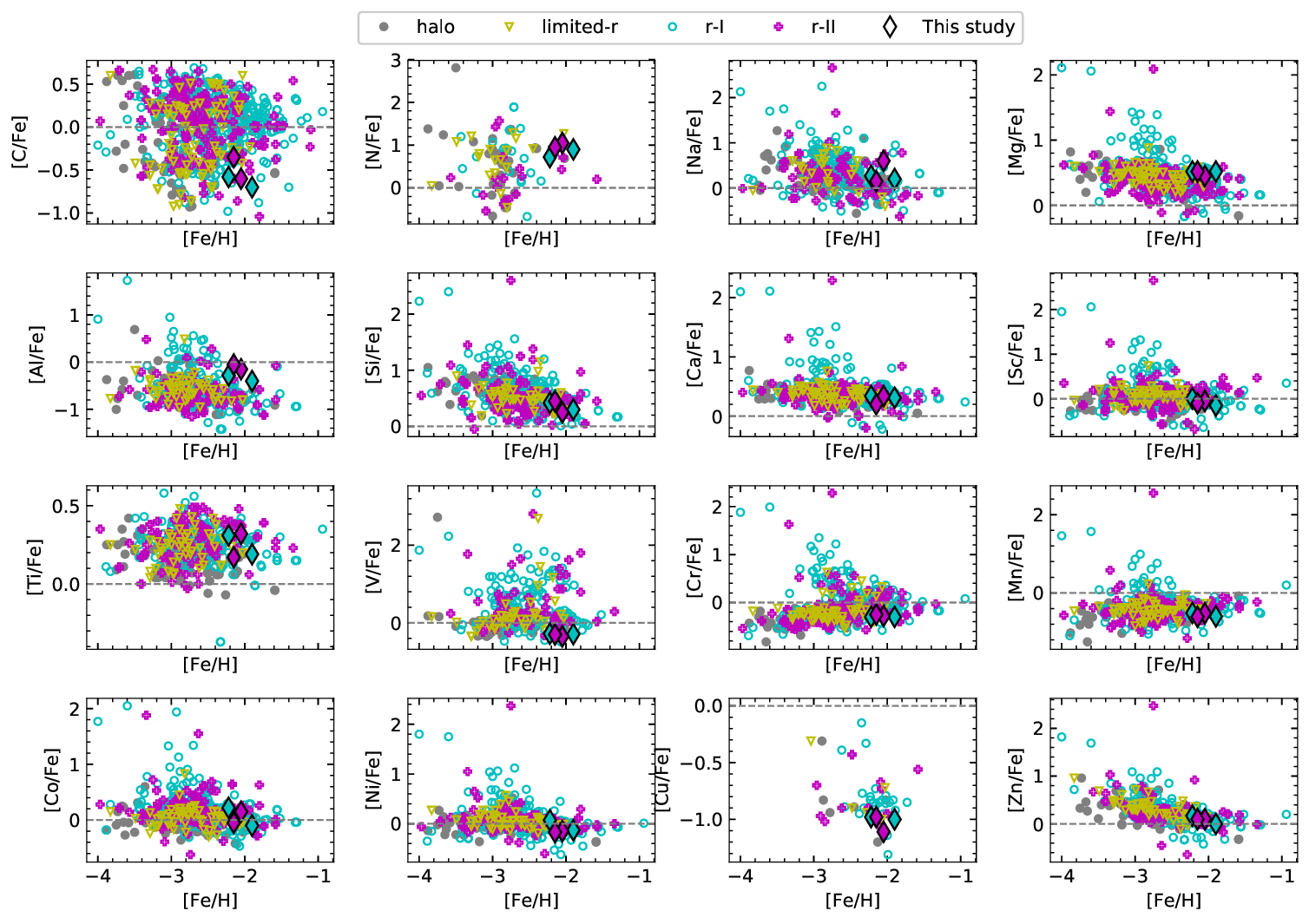}
    \caption{Abundances of the light elements as a function of metallicity. The limited-$r$ stars are shown in yellow colours, the $r$-I stars in cyan colours, and the $r$-II stars in magenta colours.  The four black diamonds indicate the $r$-I (cyan) and $r$-II (magenta) program stars in this study.  References for the $r$-process stars are given in the caption of Fig.~\ref{fig:Ba_Eu_vs_Eu_Fe.eps}. Normal metal-poor halo stars, indicated with grey filled circles, are taken from \protect\cite{SAGAbase.2008, Ryan.etal.1991, McWilliam.etal.1995, Burris.etal.2000, Fulbright.2000, Aoki.etal.2002, Johnson.2002, Cayrel.etal.2004, Honda.etal.2004, Aoki.etal.2005, Barklem.etal.2005, Jonsell.etal.2005, Preston.etal.2006, Lai.etal.2007, Zhang.etal.2009, Ishigaki.etal.2010, Roederer.etal.2010mar, Hollek.etal.2011, Allen.etal.2012, Aoki.etal.2013jan, Cohen.etal.2013, Ishigaki.etal.2013, Yong.etal.2013, Placco.etal.2014jan, Roederer.etal.2014jun, Hansen.etal.2015, Jacobson.etal.2015, Li.etal.2015jan}, and \protect\cite{JINAbase2018}. The grey horizontal lines represent the solar values.}
    \label{fig:Fe_H__X_Fe_light_elements}
\end{figure*}

Fig.~\ref{fig:Fe_H__X_Fe_light_elements} shows the abundances of all the light elements up to Zn for our program stars, along with the data collected from the literature. Our program stars fall on the lower [C/Fe] side, which is expected for their current evolutionary stage. Some of the $r$-II stars in the literature \citep{JINAbase2018} exhibit higher carbonicity ([C/Fe] $> +0.7$), which means those stars should fall into the CEMP-$r$ or CEMP-$r/s$ sub-class. We have not considered those stars in this study.

\begin{figure}
	\centering
	\includegraphics[width=\columnwidth]{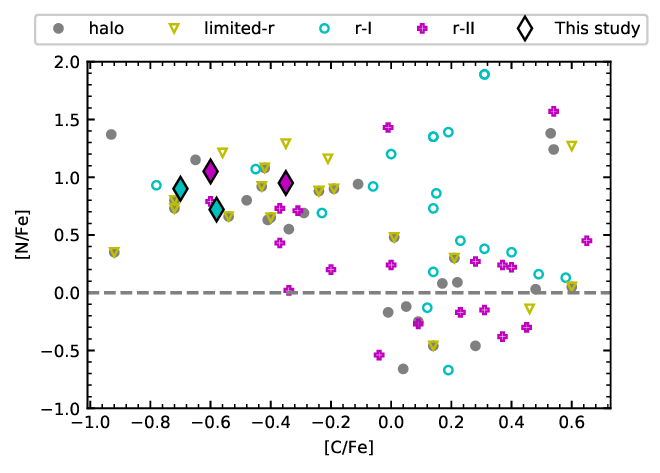}
    \caption{Relation between the N and C abundances. 
    The limited-$r$ stars are shown in yellow colours, the $r$-I stars in cyan colours, and the $r$-II stars in magenta colours. The four black diamonds indicate the $r$-I (cyan) and $r$-II (magenta) program stars in this study. References for the RPE stars are given in the caption of Fig.~\ref{fig:Ba_Eu_vs_Eu_Fe.eps}, and references for the normal metal-poor halo stars, indicated with grey filled circles, are listed in the caption of Fig.~\ref{fig:Fe_H__X_Fe_light_elements}. The grey horizontal line represents the solar value.}
    \label{fig:C_Fe__N_Fe}
\end{figure}

The mixed and unmixed stars are well-separated based on the C and N abundances, as shown in Fig.~\ref{fig:C_Fe__N_Fe}. Nearly two-thirds of the $r$-II stars fall in the unmixed region, defined with [N/Fe] $< +0.5$ \cite[see][]{Spite.etal.2005, Siqueira.etal.2014}. Hence, if the C and N abundances are from the natal cloud in which these stars formed, one would expect the presence of lithium in their atmospheres, which can help understand the amount of $r$-process ejecta mixed with the primordial cloud. Four of our program stars exhibit very low [C/N] ($ < -1.0$); they have likely undergone mixing processes. They also show $^{12}$C$/^{13}$C$ < 10$, which further supports mixing in the surface composition. The $^{12}$C$/^{13}$C values in our program stars are very close to the theoretically predicted CN cycle equilibrium value \citep{Caughlan.1965,Sneden.etal.1986}, which is consistent with the position of our target stars in Fig.~\ref{fig:C_Fe__N_Fe}. Similar values of $^{12}$C$/^{13}$C have been reported in other stars such as HD 122563 \citep{Lambert.Sneden.1977}.

Among the odd-Z elements, [Na/Fe] is slightly higher than the solar ratio for our program stars and [Al/Fe] is slightly sub-solar. These values are in good agreement with the known RPE stars. The Na and Al abundances in $r$-I and $r$-II stars exhibit a larger scatter compared to the normal halo stars. The enhanced N abundance along with Na could also indicate a very massive star progenitor for RPE stars. No Na-Mg anti-correlation is apparent. 

Halo stars usually exhibit a nearly constant enhancement in the abundance of $\alpha$-elements ($[\alpha$/Fe] $ \sim+0.4$). This level of enhancement in the stars comes from core-collapse supernovae. We found $[\alpha$ /Fe] =$~+0.35$ for TYC 3431-689-1, $[\alpha$ /Fe] =$~+0.35$ for HD 263815, $[\alpha$ /Fe] =$~+0.43$ for TYC 1716-1548-1, and $[\alpha$ /Fe] =$~+0.39$ for TYC 1191-918-1, consistent with the observed values for metal-poor halo stars \citep{McWilliam.1997review}.

\begin{figure}
	\centering
	\includegraphics[width=\columnwidth]{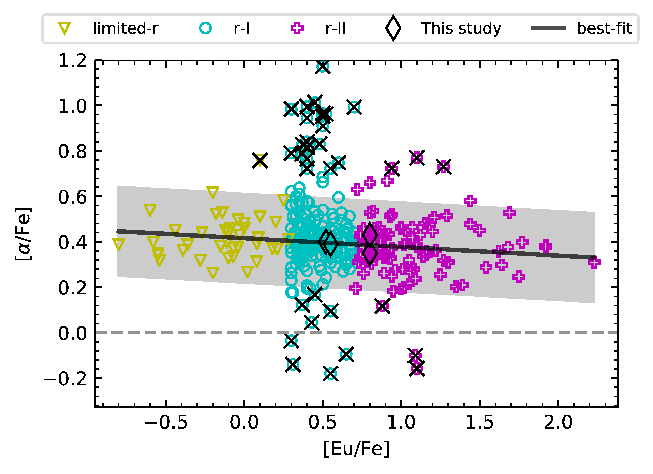}
    \caption{Alpha-element enhancements for RPE stars. The limited-$r$ stars are shown in yellow colours, the $r$-I stars in cyan colours, and the $r$-II stars in magenta colours. The four black diamonds indicate the $r$-I (cyan) and $r$-II (magenta) program stars in this study. References for the RPE stars are given in the caption of Fig.~\ref{fig:Ba_Eu_vs_Eu_Fe.eps}. The black line is the best fit after removing the 1$\sigma$ outliers denoted by overlaid crosses. The grey horizontal line represents the solar value.} 
    \label{fig:r_process_stars_alpha_Fe__Eu_Fe}
\end{figure}

To understand the connection between the $\alpha$-elements and $r$-process elements,  Fig.~\ref{fig:r_process_stars_alpha_Fe__Eu_Fe} shows the average $\alpha$-enhancement, i.e., $\frac{1}{4}$([Mg/Fe]+[Si/Fe]+[Ca/Fe]+[Ti/Fe]) for the RPE stars, as a function of [Eu/Fe]. In this figure, the black solid line shows the best fit after removing the 1$\sigma$ outliers denoted by overlaid crosses; the shaded region shows the dispersion around the best-fit line. There is only a very weak, statistically insignificant, negative correlation. Our program stars  nicely fall on the best-fit line.

The iron-peak elements Sc, Co, and Zn exhibit nearly solar ratios for our program stars, whereas V, Cr, Mn, and Cu fall in the sub-solar region. 
Abundances of all the iron-peak elements in RPE stars are in good agreement with normal metal-poor halo stars. On average, the halo stars show slightly sub-solar V abundances with a small dispersion, however, RPE stars exhibit significantly higher dispersions towards super-solar abundances. We find that the inclusion of the RPA data from \cite{Sakari.etal.2018} and \cite{Ezzeddine.etal.2020} results in a higher dispersion in the [V/Fe] ratio, ranging from $-0.8$ to $+3.3$. Otherwise, the [V/Fe] ratio remains in the range from $-0.8$ to $+1.0$, consistent with previous studies \citep{Sneden.etal.2016, Ou.etal.2020}. 
The [Zn/Fe] ratio increases towards the metal-poor end for $r$-I and $r$-II stars compared to the normal halo stars, indicating a possible additional contribution of Zn from the $r$-process-production site(s) at the metal-poor end. 
\begin{figure}
	\centering
	\includegraphics[width=\columnwidth]{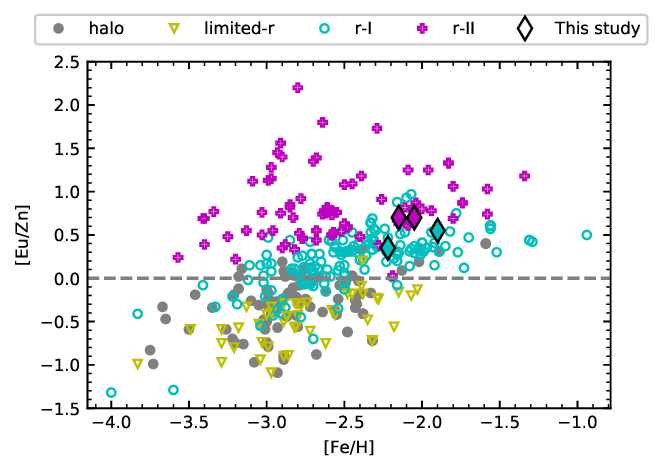}
    \caption{Evolution of the [Eu/Zn] ratio as a function of metallicity. The limited-$r$ stars are shown with yellow downward triangles,  $r$-I stars are displayed with open cyan circles, and $r$-II stars are represented by magenta crosses. The four black diamonds indicate the $r$-I (cyan) and $r$-II (magenta) program stars in this study. References for the RPE stars are given in the caption of Fig.~\ref{fig:Ba_Eu_vs_Eu_Fe.eps}, and references for normal metal-poor halo stars, indicated by grey filled circles, are listed in the caption of Fig.~\ref{fig:Fe_H__X_Fe_light_elements}. The grey horizontal line represents the solar value.}
    \label{fig:Fe_H__Eu_Zn}
\end{figure}
The [Eu/Zn] ratio also exhibits a value closer to normal halo stars at the metal-poor end, and increases with metallicity (see Fig.~\ref{fig:Fe_H__Eu_Zn}). This indicates that the metal-poor $r$-process-production sites manufactured Zn as well, and with increasing metallicity the $r$-process progenitors either produced less or no zinc. This is consistent with the recent claim for hypernovae or massive SNe being responsible for $r$-process production at the metal-poor end and NSMs contributing at higher metallicities \citep{Tarumi.etal.2021, Yong.etal.2021}. Four of our program stars show [Fe/H] in the range from $-2.22$ to $-1.90$ and exhibit near-solar [Zn/Fe] ratios. Also, the [Eu/Zn] ratio of our program stars lies towards the upper limit of [Eu/Zn] for the normal halo stars. It indicates that our program stars may have originated from gas that was primarily enriched by events occurring during the late evolution of the Galaxy, such as NSMs. The contribution from sites that were active in the early Universe may be negligible.

\subsection{Evolution of Heavy-element Abundance Ratios for RPE Stars}
\begin{figure*}
	\centering
	\includegraphics[width=\textwidth]{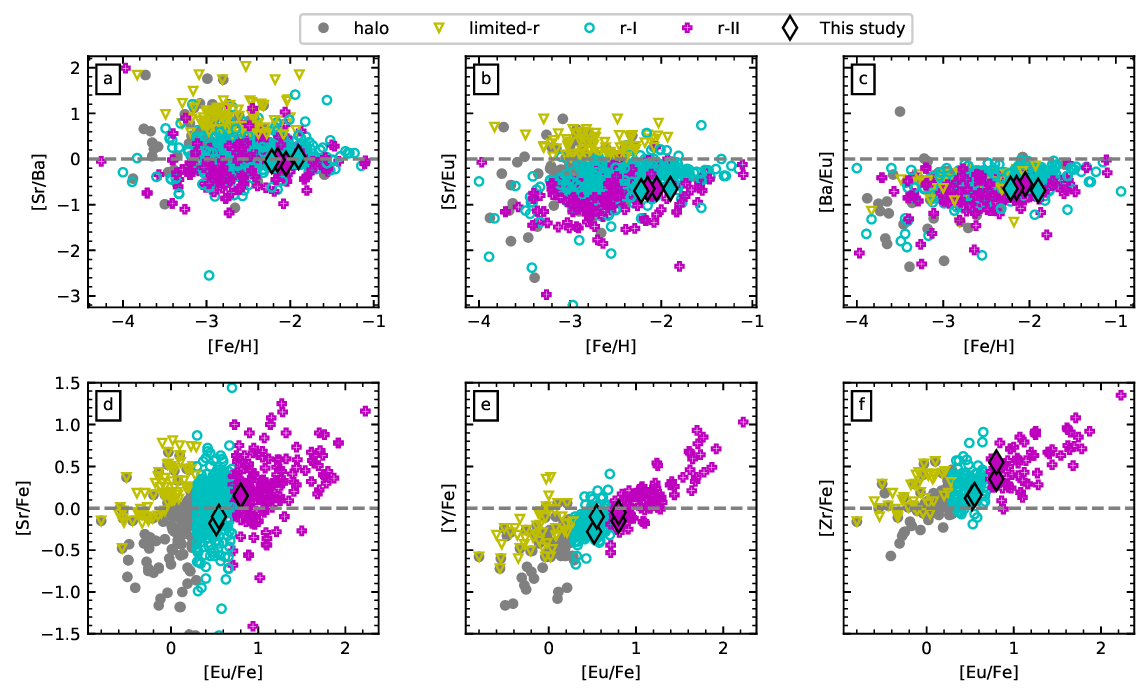}
    \caption{Evolution of neutron-capture abundance ratios as a function of [Fe/H] and [Eu/Fe]. The limited-$r$ stars are shown in yellow colours, the $r$-I stars in cyan colours, and the $r$-II stars in magenta colours.  The four black diamonds indicate the $r$-I (cyan) and $r$-II (magenta) program stars in this study. References for the RPE stars are given in the caption of Fig.~\ref{fig:Ba_Eu_vs_Eu_Fe.eps}, and references for normal metal-poor halo stars are listed in the caption of Fig.~\ref{fig:Fe_H__X_Fe_light_elements}.}
    \label{fig:r_process_stars_heavy_elements}
\end{figure*}

The [Sr/Ba] ratio reasonably separates the three sub-classes of RPE stars (see Fig.~\ref{fig:r_process_stars_heavy_elements}(a)). The limited-$r$ stars exhibit the highest [Sr/Ba] ratios, and the $r$-II stars show the lowest [Sr/Ba], whereas the $r$-I stars fall in between. However, some $r$-I and $r$-II stars also exhibit higher [Sr/Ba] over the metallicity span of the limited-$r$ stars.  At low metallicity, we expect the majority of Ba to have come from $r$-process event(s).  If Sr were produced from the same $r$-process event(s), we would expect a flat trend of [Sr/Ba] at low metallicity. However, the dispersion of [Sr/Ba] increases with decreasing metallicity (see Fig.~\ref{fig:r_process_stars_heavy_elements}(b)), indicating an additional formation site for Sr \citep{Travaglio.etal.2004, Aoki.etal.2013, F.Spite.etal.2018}.

We have also checked on the Sr ratio with respect to the $r$-process element Eu. The distribution of [Sr/Eu] exhibits a similar behaviour as [Sr/Ba], with higher dispersion at low metallicity. Fig.~\ref{fig:r_process_stars_heavy_elements}(d, e, and f) also exhibits correlations of [Sr/Fe], [Y/Fe], and [Zr/Fe] with respect to [Eu/Fe]. This suggests that some Sr, Y, and Zr are produced in an $r$-process event that synthesises Eu as well. A flatter trend and more scatter in [Sr/Fe] versus [Eu/Fe] could also imply  additional sites for the production of Sr as compared to Eu. The plots of [Y/Fe] and [Zr/Fe] versus [Eu/Fe] indicate clear trends, with much less scatter, as compared to the [Sr/Fe] ratio. This could also mean that the additional site that produces Sr does not produce Y and Zr in significant quantities.

Fig.~\ref{fig:r_process_stars_heavy_elements}(c) shows [Ba/Eu] as a function of metallicity. This ratio (excluding a few outliers) exhibits only a small dispersion for the $r$-I, $r$-II and limited-$r$ stars, suggesting a similar origin of Ba and Eu in RPE stars. It might be better understood as arising from different levels of dilution of the $r$-process-element yields from the same source. 

Four of our program stars exhibit solar [Sr/Ba] ratios. However, the [Sr/Eu] and [Ba/Eu] ratios are sub-solar. For our target stars, [Sr/Eu] ranges from $-0.65$ to $-0.70$ and [Ba/Eu] ranges from $-0.69$ to $-0.55$, indicating very little $s$-process contamination. The higher [Sr/Ba] ratio in our program stars as compared to [Sr/Eu] and [Ba/Eu] implies that the interstellar medium (ISM) where these objects were formed was already enriched with the events that produce first $r$-process peak elements and the main $r$-process elements.

Using the literature sample of RPE stars, we find that the limited-$r$ stars are poor in C, Fe, and Ba as compared to the $r$-I and $r$-II stars; see Fig.~\ref{fig:r_process_FeH_dist}(a, b, and d). However, the Sr abundance of the limited-$r$ stars spans a similar range as that of the $r$-I and $r$-II stars. All the limited-$r$ stars exhibit metallicities lower than [Fe/H] = $-2.0$, consistent with  previous findings \citep{Hansen.etal.2018, Sakari.etal.2018}. Only one star, 2MASS J14435196-2106283 from \cite{Holmbeck.etal.2020}, is found at [Fe/H] = $-1.93$. Also, the [C/H] and [Ba/H] abundance ratios in limited-$r$ stars are always smaller than $-2.0$. The evolution-corrected C abundance for all the limited-$r$ stars falls in the [C/H] $< -1.7$ range. The C and N abundances also indicate that the majority of limited-$r$ stars are in the mixed region, indicating that mixing of 
CN-processed material during their current RGB phase has taken place (see, Fig.~\ref{fig:C_Fe__N_Fe}). Mass transfer from an AGB companion that has gone through the 2nd dredge-up phase can also result in low carbon 
abundances.  The data on C and N abundances are limited, and we cannot rule out that the [C/N] ratio is a result of RGB mixing or due to massive-star wind ejecta. Lithium abundances, along with CN abundances, will provide valuable insights on in-situ/binary origin and an alternative primordial cloud mixed with low-carbon ejecta.
These conclusions are based on 46 limited-$r$, 314 $r$-I, and 128 $r$-II stars taken from \cite{Gudin.etal.2021}. The number of limited-$r$ stars is still not sufficiently large to comment further on their nature.

\subsection{Production Sites of the Rhird $r$-process Peak}
\begin{figure*}
	\centering
	\includegraphics[width=\textwidth]{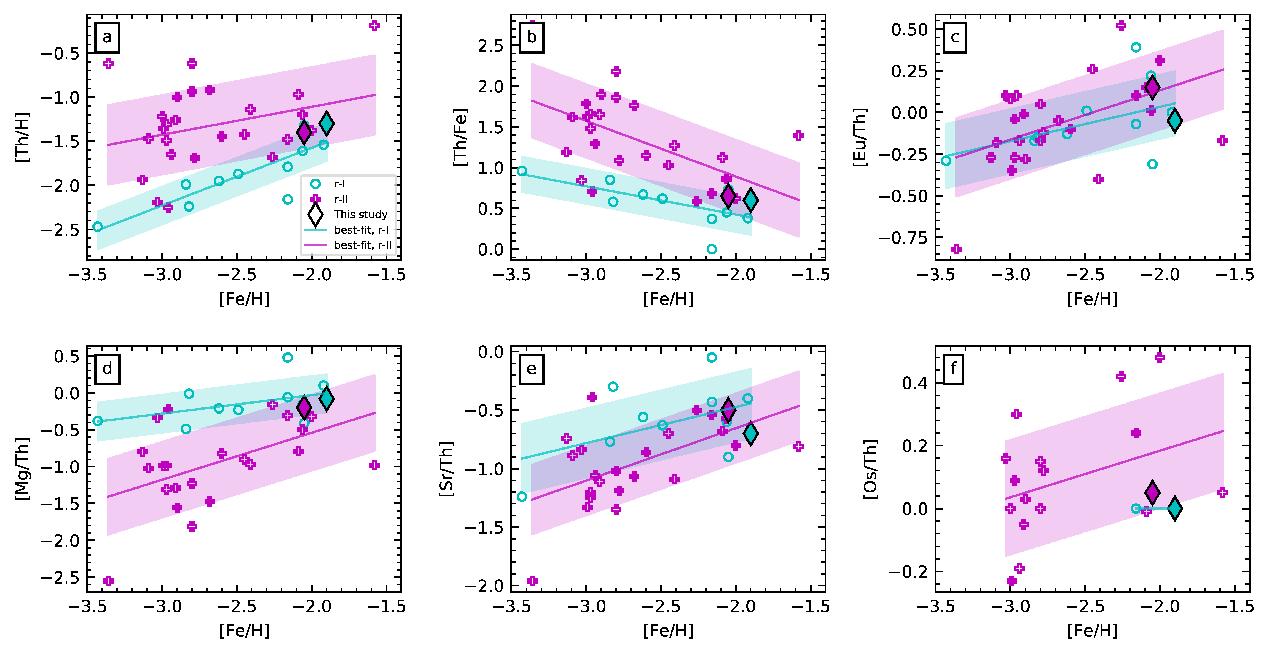}
    \caption{Evolution of ratios of Th with the light and heavy $r$-process elements as a function of metallicity. The $r$-I stars are displayed with open cyan circles, and $r$-II stars are represented by magenta crosses. The four black diamonds indicate the $r$-I (cyan) and $r$-II (magenta) program stars in this study. The cyan and magenta lines represent the best fits to the $r$-I and $r$-II stars, respectively. Their corresponding dispersions are shown with the shaded regions. We have compiled Th abundances from \protect\cite{Westin.etal.2000, Cowan.etal.2002, Johnson.etal.2002, Honda.etal.2004, Ivans.etal.2006, Aoki.etal.2005, Aoki.etal.2007, Hayek.etal.2009, Mashonkina.etal.2010, Mashonkina.etal.2014, Roederer.Preston.etal.2014, Roederer.etal.2014, Siqueira.etal.2014, Hill.etal.2017, Placco.etal.2017, Holmbeck.etal.2018, Sakari.Placco.etal.2018}}
    \label{fig:r_process_stars_with_Thorium}
\end{figure*}

Thorium is purely produced in $r$-process nucleosynthesis. To understand the evolution of Th with other light and heavy elements, we have shown various elemental ratios in Fig.~\ref{fig:r_process_stars_with_Thorium}. 
The [Th/H] ratio increases with increasing metallicity for both $r$-I and $r$-II stars, as can be seen in \ref{fig:r_process_stars_with_Thorium}(a). At a given metallicity, the Th abundance in $r$-II stars is higher than for the $r$-I stars, as evident from \ref{fig:r_process_stars_with_Thorium}(a and b). In contrast to [Th/H], the [Th/Fe] ratio decreases with metallicity, even among $r$-II stars. This could indicate that Fe-peak elements are enriched at a faster rate than the relatively slow radioactive decay of Th.

Fig.~\ref{fig:r_process_stars_with_Thorium}(c) shows an increasing [Eu/Th] ratio with [Fe/H]. Both $r$-I and $r$-II stars overlap with each other with small scatter. Equal [Eu/Th] ratios in $r$-I and $r$-II stars can be explained with the same production sites for both and more dilution of $r$-process ejecta in $r$-I stars.

The [Mg/Th] ratio, as a function of [Fe/H], in Fig.~\ref{fig:r_process_stars_with_Thorium}(d) shows interesting distributions of $r$-I and $r$-II stars. All the $r$-I stars have higher [Mg/Th] ratios than the $r$-II stars. With increasing metallicity, [Mg/Th] appears to converge to zero, where $r$-II stars have steeper slope than $r$-I stars. These trends indicate that, apart from the difference in the level of mixing of the $r$-process ejecta in the inherited composition of $r$-I and $r$-II stars, the $r$-II stars could have formed earlier than $r$-I stars.  Fig.~\ref{fig:r_process_stars_with_Thorium}(e) shows the [Sr/Th] evolution with [Fe/H]. It exhibits similar trends for $r$-I and $r$-II stars as that of [Eu/Th], but with larger dispersion. Finally, we have plotted the third-peak element [Os/Th] ratio with metallicity in Fig.~\ref{fig:r_process_stars_with_Thorium}(f). For $r$-II stars, it shows that the [Os/Th] ratio increases with metallicity, similar to the [Eu/Th] ratio. We have only two $r$-I stars with detected Os and Th, thus we cannot comment about their evolution. In the future, a larger number of RPE stars with detected Os and Th abundances should help reveal more about their evolution.

Two of our program stars with measured Th are at the larger [Fe/H] values of the sample. Also, both of them lying in the overlapping region of $r$-I and $r$-II trends, as can be seen in all the panels of Fig.~\ref{fig:r_process_stars_with_Thorium}. The separation of $r$-I and $r$-II sub-classes at the lower-metallicity end seems to disappear as we observe Th evolution with other elements from lighter to heavier elements. Being in overlapping regions and towards the larger [Fe/H] side indicates that our program stars are primarily enriched by single main $r$-process events, such as NSMs, with little contribution from a limited-$r$ process event.

\subsection[The r-process Pattern]{The $r$-process pattern}
The abundances of the elements produced in the $r$-process, as a function of their atomic numbers, exhibit a distinctive pattern known as the $r$-process pattern. This pattern shows three noticeable peaks, at atomic number $Z\sim$35 (first peak: best probed with Sr, Y, and Zr), $Z\sim$54 (second peak: best probed with Ba and La), and $Z\sim$78 (third peak: best probed with Os and Ir). These peaks are the result of three closed neutron shells with neutron numbers 50, 82, and 126, respectively \citep{Sneden.etal.2008, Ji.Frebel.2018}.

Numerous previous studies found that the abundance patterns of metal-poor RPE stars match very closely with the scaled-solar $r$-process pattern \citep{Sneden.etal.1998, Sneden.etal.2008}. This agreement is more pronounced for lanthanides, the elements from Ba to Hf. Lighter $r$-process elements exhibit scatter from the scaled-solar $r$-process pattern, which is attributed to the result of a different formation mechanism, called the weak $r$-process or limited-$r$ process. The scatter of the actinide elements was expected due to their radioactive nature. However, some stars exhibit Th in excess amounts as compared to the scaled-solar $r$-process pattern. These stars are known as actinide-boost stars \citep{Schatz.etal.2002,Mashonkina.etal.2014}.

\begin{figure*}
	\centering
	\includegraphics[width=\textwidth]{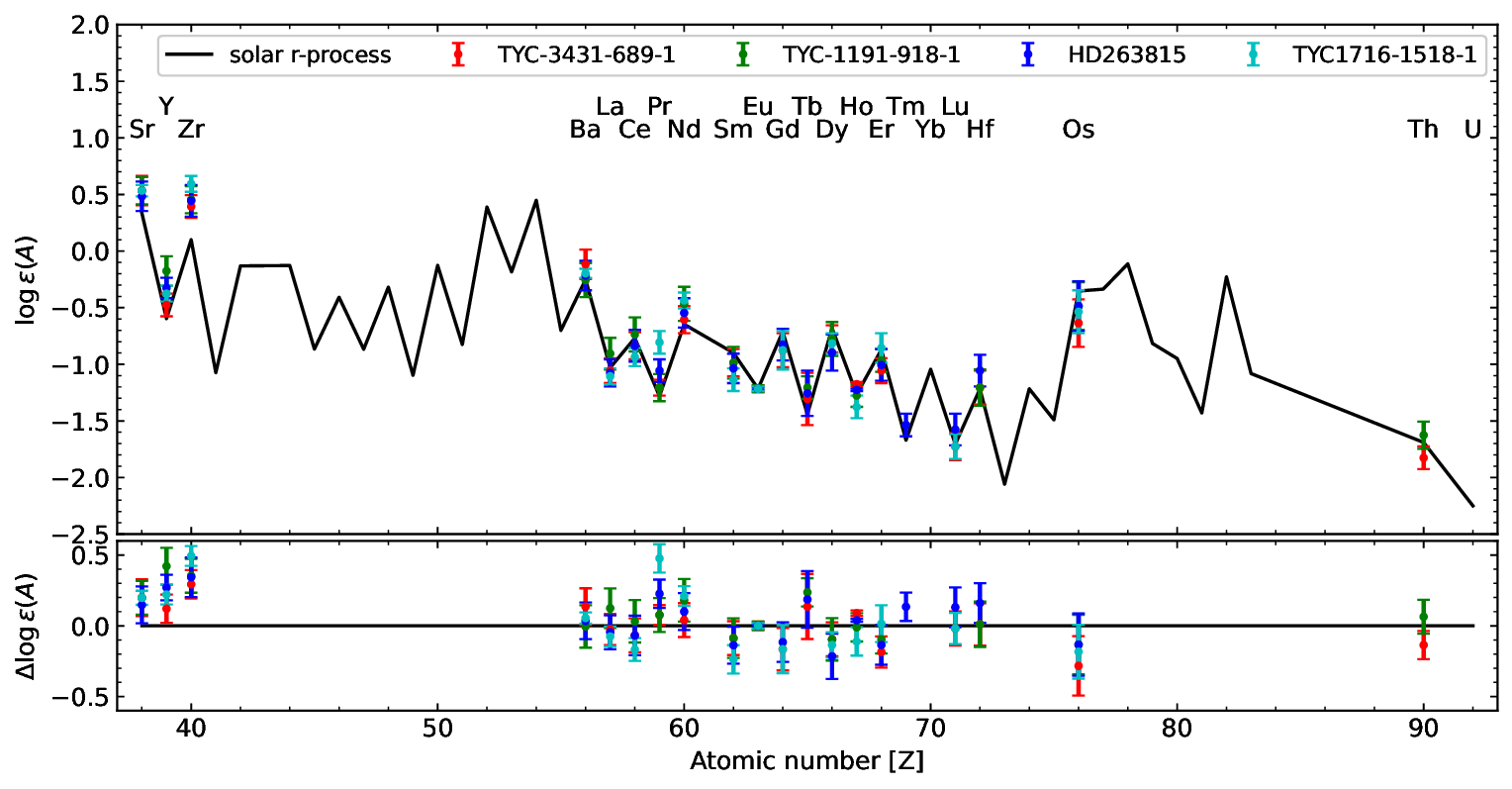}
    \caption{Top panel: The solid black line shows the solar $r$-process pattern, i.e., the absolute abundance ($\log \epsilon$) as a function of atomic number (Z). The elemental abundances of our program stars are scaled to the solar Eu abundance. Bottom panel: Residual abundances of our program stars are calculated as the difference of scaled stellar abundance and the solar abundance. The solar $r$-process pattern is adopted from \protect\cite{Arlandini.etal.1999}.}
    \label{fig:r_process_pattern_new.eps}
\end{figure*}

The top panel of Fig.~\ref{fig:r_process_pattern_new.eps} shows the abundance distribution of our program stars, as a function of their atomic number, scaled to the solar $r$-process pattern with respect to Eu. The bottom panel displays the residual abundances of the stars from the scaled-solar $r$-process pattern. The $r$-process-element abundances of our program stars also match with the scaled-solar $r$-process abundances, confirming the universality of the main $r$-process pattern (second-peak elements), along with the usual scatter for the first and third $r$-process elements. 

\subsection[effect of log gf values on actinide classification and age estimation]{Effect of $\log gf$ Values on Actinide Classification and Age estimation}
Among all of the actinide elements, we could only make estimates of the Th abundance for two of our program stars, TYC 1191-918-1 ($r$-I star) and TYC 3431-689-1 ($r$-II star). We have calculated an upper limit for a third star, HD 263815. Thorium is a radioactive element with a half-life of 14.05~Gyr. The production of Th is only possible in $r$-process nucleosynthesis \citep{Cowan.etal.1991}. 

There has been noticed a quite significant dispersion in the actinide-to-lanthanide abundance ratios among metal-poor stars (see \citealt{Holmbeck.etal.2018} and references therein). Generally, $\log \epsilon$(Th/Eu) is used as an indicator to measure the actinide-to-lanthanide abundance ratio. Around $30\%$ of RPE stars exhibit a 2 to 3 times higher Th/Eu ratio as compared to other RPE stars \citep{Hill.etal.2002, Mashonkina.etal.2014}. These stars, with $\log \epsilon$ (Th/Eu) $> -0.5 \pm 0.15$, are known as actinide-boost stars \citep{Holmbeck.etal.2018}. Recently, \cite{Holmbeck.etal.2019} used $\log \epsilon$(Th/Dy) to classify stars in different actinide categories as follows:
\vspace{0.5cm}
\newline actinide-deficient: \hspace{0.5cm} $\log \epsilon$(Th/Dy) $< -1.20$
\newline actinide-normal: \hspace{0.7cm} $-1.20 \leq$ $\log \epsilon$(Th/Dy) $\leq -0.90$
\newline actinide-boost: \hspace{1.0cm} $\log \epsilon$(Th/Dy) $> -0.90$
\vspace{0.5cm}

The radioactive nature of Th is often used as chronometer to estimate stellar ages (more precisely, the time elapsed since its production, hence a lower limit on the stellar age). The thorium chronometer is based on the universality of the $r$-process pattern. The depletion of Th with respect to non-radioactive elements heavier than Ba, such as Ce, Eu, Dy, Os, Ir, etc., are used for stellar age calculation. The age of a star can be estimated using the following relation \citep{Cayrel.etal.2001, Cain.etal.2018, Hansen.El-Souri.etal.2018}:
\begin{equation}
    \Delta t = 46.7[\log \epsilon{\rm (Th/X)}_{i} - \log \epsilon{\rm (Th/X)}_{f}],
    \label{eqn:age_calc}
\end{equation}
where $X$ is a non-radioactive element above Ba, $\log \epsilon {\rm (Th/X)}_{i}$ is the initial abundance ratio at the time of star's birth, also known as the production ratio (PR), and $\log \epsilon
{\rm (Th/X)}_{f}$ is the final abundance ratio, i.e., now. The error in age is calculated from:
\begin{equation}
    \Delta t_{err} = 46.7\sqrt{\sigma^{2}_{\log \epsilon{\rm (Th)}} + \sigma^{2}_{\log \epsilon{\rm (X)}}},
    \label{eqn:age_calc_err}
\end{equation}
where $\sigma_{\log \epsilon {\rm (Th)}}$, and $\sigma_{\log \epsilon {\rm (X)}}$ are the errors in the abundance of Th and element X, respectively. Since the age determination depends only on 
this abundance ratio, it is necessary to precisely estimate the Th abundance -- a small error of 0.2 dex in $\log$(Th/Eu) can produce a 9.3~Gyr error in the age estimate. 

\begin{figure}
	\centering
	\includegraphics[width=0.45\textwidth]{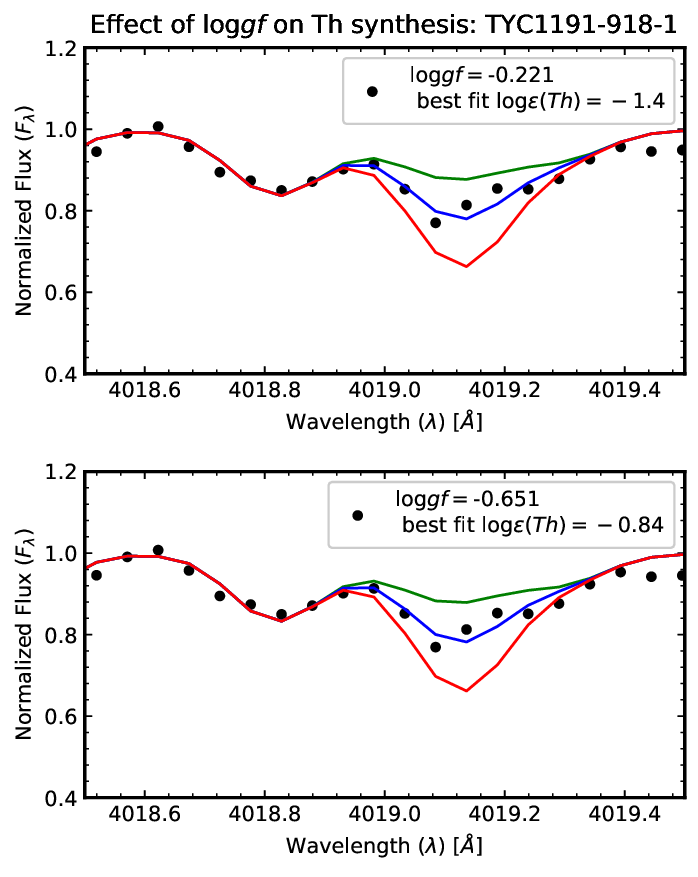}
    \caption{Synthesis of the Th II 4019\,{\AA} region in TYC 1191-918-1 for two different values of $\log gf$, $-0.228$ and $-0.651$. The dotted line represents the observed spectrum and the solid blue line is the best-fit synthetic spectrum. The green and red lines show 0.5 dex deviations from the best-fit line.}
    \label{fig:effect_of_log_gf}
\end{figure}

Even if we carefully consider effects of the blending in the abundance estimation of Th, the choice of $\log gf$ can significantly affect its abundance determination. In the literature, the value of $\log gf$ for this species is continuously being updated; the current value of $\log gf$ may differ from previously employed values, resulting in different abundance estimates. For example, Fig.~\ref{fig:effect_of_log_gf} compares the spectrum synthesis of the Th II line at 4019\,{\AA} in TYC 1191-918-1 for two different $\log gf$ values, $-0.228$ and $-0.651$. Here, the dotted line corresponds to the observed spectrum and the solid blue line is the best fit, while the green and red lines correspond to 0.5 dex deviations from the best fit. As expected, different $\log gf$ values yield different Th abundances, leading to  significant errors in the age estimation of the star. We adopt an updated value of $\log gf = -0.228$ for Th abundance estimation, and use the resulting abundance for the age calculation of our program stars. The PR of elements in equation~\eqref{eqn:age_calc} is taken from \cite{Schatz.etal.2002} and \cite{Hill.etal.2017}. Estimated ages from individual species and their medians for Th-detected stars are listed in Tables~\ref{tab:Ages_TYC3431} and \ref{tab:Ages_TYC1191}. Following recent PR values from \citep{Hill.etal.2017}, the median age of TYC 3431-689-1 is 14.36$\pm7.3$~Gyr, which can be expected for a VMP star. However, the median age of TYC 1191-918-1 is 9.19$\pm$8.98~Gyr, which is lower than expected, due to the higher Th abundance.

\begin{table*}
	\caption{Age estimation of TYC 3431-689-1.}
	\label{tab:Ages_TYC3431}
	\begin{tabular}{lccccr} 
		\hline
		& & TYC 3431-689-1 &  \\
		$Th/X$ & PR \citep{Schatz.etal.2002} & Age (Gyr) & PR \citep{Hill.etal.2017} & Age (Gyr) & $\sigma$ (Gyr)\\
		\hline
		Th/Ba & $-$ & $-$ & $-$1.058 & 30.50 & 7.67 \\
		Th/La & $-$0.60 & 7.95 & $-$0.362 & 19.08 & 6.95 \\
		Th/Ce & $-$0.79 & 9.35 & $-$0.724 & 12.44 & 7.30 \\
		Th/Pr & $-$0.30 & 14.96 & $-$0.313 & 14.36 & 5.71 \\
		Th/Nd & $-$0.91 & 14.50 & $-$0.928 & 13.65 & 7.30 \\
		Th/Sm & $-$0.61 & 10.75 & $-$0.796 & 2.05 & 7.30 \\
		Th/Eu & $-$0.33 & 13.09 & $-$0.240 & 17.3 & 4.88 \\
		Th/Gd & $-$0.81 & 6.54 & $-$0.569 & 17.8 & 8.43 \\
		Th/Dy & $-$0.89 & 7.48 & $-$0.827 & 10.43 & 7.30 \\
		Th/Ho & $-$ & $-$ & $-$0.071 & 27.08 & 4.77 \\
		Th/Er & $-$0.68 & 4.21 & $-$0.592 & 8.32 & 6.95 \\
		Th/Hf & $-$0.20 & 19.64 & $-$0.036 & 27.31 & 8.4 \\
		Th/Os & $-$1.15 & 1.87 & $-$0.917 & 12.77 & 10.88 \\
		\hline
		Median & & 9.35 & & 14.36  &  7.30\\
		\hline
	\end{tabular}
\end{table*}

\begin{table*}
	\caption{Age estimation of TYC 1191-918-1.}
	\label{tab:Ages_TYC1191}
	\begin{tabular}{lccccr} 
		\hline
		& & TYC 1191-918-1 &  \\
		$Th/X$ & PR \citep{Schatz.etal.2002} & Age (Gyr) & PR \citep{Hill.etal.2017} & Age (Gyr) & $\sigma$ (Gyr)\\
		\hline
		Th/Ba & $-$ & $-$ & $-$1.058 & 14.59 & 8.98 \\
		Th/La & $-$0.60 & 5.61 & $-$0.362 & 16.74 & 8.62 \\
		Th/Ce & $-$0.79 & 4.67 & $-$0.724 & 7.76 & 8.98\\
		Th/Pr & $-$0.30 & 5.61 & $-$0.313 & 5.00 & 7.93 \\
		Th/Nd & $-$0.91 & 11.69 & $-$0.928 & 10.85 & 8.98 \\
		Th/Sm & $-$0.61 & 1.40 & $-$0.796 & $-$7.29 & 8.62 \\
		Th/Eu & $-$0.33 & 3.74 & $-$0.240 & 7.95 & 5.78 \\
		Th/Gd & $-$0.81 & $-$ & $-$0.569 & $-$ & $-$ \\
		Th/Dy & $-$0.89 & $-$1.87 & $-$0.827 & 1.07 & 8.98 \\
		Th/Ho & $-$ & $-$ & $-$0.071 & $-$ & $-$ \\
		Th/Er & $-$0.68 & $-$ & $-$0.592 & $-$ & $-$ \\
		Th/Hf & $-$0.20 & 10.29 & $-$0.036 & 17.96 & 9.35 \\
		Th/Os & $-$1.15 & $-$0.467 & $-$0.917 & 10.43 & 11.31 \\
		\hline
		Median & & 4.67 & & 9.19  & 8.98\\
		\hline
	\end{tabular}
\end{table*}

\begin{figure}
	\centering
	\includegraphics[width=\columnwidth]{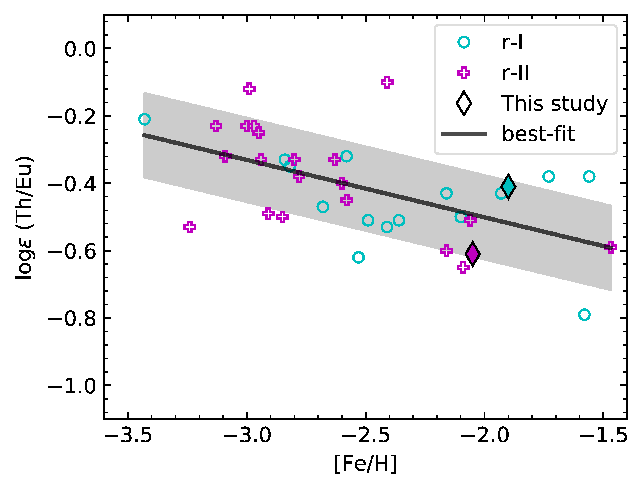}
    \caption{The $\log \epsilon$ (Th/Eu) value, as a function of metallicity, for RPE stars with measured Th. The $r$-I stars are shown with cyan-coloured open circles and the $r$-II stars with 
    magenta-coloured crosses. The black diamonds indicate the $r$-I (cyan) and $r$-II (magenta) program stars in this study. The black line represents the best fit to the data and the shaded region shows its dispersion. The references for stars with detected Th are given in the caption of Fig.~\ref{fig:r_process_stars_with_Thorium}.}
    \label{fig:Th_r-process}
\end{figure}

Th and U are the most useful elements for estimating ages for $\sim
10$ Gyr-old stars. The primary sources of uncertainty in the Th chronometer are its long half-life of 14.05~Gyr and production ratio of the second and third $r$-process peak elements. The detection of U along with Th would have been of much importance to constrain the ages of our program stars independent from stellar evolution models, and thus the minimum age of the 
Universe because U has a smaller half-life of 4.5~Gyr, both the elements (Th and U) belongs to same (third) $r$-process peak, and are only produced in $r$-process nucleosynthesis \citep[see, ][]{Cayrel.etal.2001, Barbuy.etal.2011}.

We have also calculated the actinide-to-lanthanide abundance ratio for our program stars with detected Th. Following \cite{Holmbeck.etal.2018}, both of our program stars with Th belong to the actinide-normal category. On the other hand, according to the \cite{Holmbeck.etal.2019} criterion, TYC 3431-689-1 again falls in the actinide-normal class, whereas TYC 1191-918-1 moves into the actinide-boost region, close to the separation boundary. We believe this determination of an actinide boost may be due to the uncertainties in our abundance estimation. TYC 1191-918-1 was earlier reported to be an actinide-boost star by \cite{Bandyopadhyay.etal.2020}, with $\log \epsilon$ (Th/Eu) $= -0.20$. There could be three possible reasons for their higher actinide-to-lanthanide ratio: (1) differences in the adopted stellar parameters, (2) incomplete accounting of the contribution from blended elements, and/or (3) their choice of a lower $\log gf$ value.

Fig.~\ref{fig:Th_r-process} shows $\log \epsilon$ (Th/Eu), as a function of [Fe/H], for our two Th-detected stars, overlaid with RPE stars from the literature. Our program stars are shown with black diamonds, whereas $r$-I and $r$-II stars are represented by cyan circles and magenta crosses, respectively. The black solid line represents the best-fit linear relationship, and the grey colour region shows the dispersion around the best-fit line. From inspection, there is a clear trend of a decreasing actinide-to-lanthanide ratio with increasing metallicity. If we take metallicity as a proxy for age, this implies that older stars are more actinide rich than younger stars, contrary to the universality of the $r$-process pattern. According to this paradigm, older stars should be more actinide poor than the younger stars. This opposite trend could be the result of two possibilities for non-universal production ratio: (1) the late universe is enriched by lanthanides or (2) the early universe is enriched by actinides. The first possibility appears more likely in terms of the production site(s) of these elements, although the production sites of the actinides are less-well understood.

The assumption of a linear trend of the actinide-to-lanthanide production ratio with metallicity may be one of the possible reasons for higher dispersion in this ratio, leading to the actinide-boost phenomenon. Assuming a constant actinide-to-lanthanide production ratio yields negative ages for actinide-boost stars and large uncertainties in the age estimation using this method. If the $r$-process pattern is truly universal, the production ratio needs to be calibrated for the chemical evolution of the Galaxy. And of course, a much larger sample of RPE stars is needed to provide better constraints on the astrophysical formation sites of such stars.

\section{Conclusions}
\label{sec:conclusions}

In this study, we estimate a more complete set of elemental abundances for four relatively bright ($V < 12$) VMP RPE stars observed with the 10-m class GTC telescope. Due to the excellent SNR of the spectra in the blue region, we could estimate the abundances of ten neutron-capture elements for these stars that were not reported by \cite{Bandyopadhyay.etal.2020}: Pr, Gd, Tb, Ho, Hf, Er, Tm, Lu, Os, and Th, and improved estimates for others. We find that two of our targets fall under the $r$-I classification, while the other two are $r$-II stars. We could measure Th in  TYC 1191-918-1($r$-I star) and TYC 3431-689-1 ($r$-II star). In our study, we also obtain upper limits of for the Pb abundances of HD 263815, TYC 3431-689-1, and TYC 1716-1548-1.The elemental-abundance patterns of our stars are well-matched with the scaled-solar $r$-process patterns in the main $r$-process region (Ba to Hf). We find [C/N] $< -1.0$ and $^{12}$C$/^{13}$C$ < 10$, which indicates internal mixing of CN-cycle nucleosynthesis products altering the surface composition.

We also study the elemental-abundance trends among the RPE stars; the main conclusions are listed below.
\begin{itemize}
    \item
    The RPE stars exhibit a similar trend in [$\alpha$/Fe] and [Fe/H] as compared to the normal metal-poor halo stars. However, r-II stars exhibit a declining trend with respect to [Eu/Fe], which may indicate efficient mixing of the ISM at higher metallicities.  Four of our program stars also follow the expected trends of other RPE stars.
    \item
    The compiled RPE stars exhibit similar trends in the Fe-peak elements as that of the halo stars, however there is a large scatter noticed in the trends of Sc,V,Cr, especially at low metallicities.  A homogeneous analysis might be useful to confirm various trends. The targets of the current study  follow the same trend as that of the other halo stars at the same metallicities.
    Also, the Fe-peak elements have flat trends with respect to [Eu/Fe], excepting Mn, Cr, and Zn. We found that metal-poor stars exhibit high [Zn/Fe] and low [Eu/Zn], hinting at Zn production with $r$-process events in the early Galaxy. Four of our program stars show [Fe/H] in the range from $-2.22$ to $-1.90$, and exhibit near-solar [Zn/Fe] ratios, suggesting insignificant contribution of Zn from the r-process source.
    \item
    The [Sr/Fe] ratio exhibits a relatively flat trend with respect to [Eu/Fe] and large scatter, however [Y/Fe] and [Zr/Fe] show steeper trends and less scatter with respect to [Eu/Fe]. This indicates that the site that made Eu is also a significant contributor of Y and Zr as compared to Sr.
    \item
    The abundance trend in Th and Mg with respect to [Fe/H] indicates that there could be some delay between the formation of $r$-II stars and $r$-I stars.
    \item 
    Due to the radioactive nature of Th, we could calculate age estimates for TYC 3431-689-1 and TYC 1191-918-1, as 14.36$\pm7.3$ and 9.19$\pm8.98$~Gyr, respectively. We emphasize the role that production ratios, nuclear physics, and astrophysical sites of the $r$-process play a role on the uncertainty in age estimation.
\end{itemize}

Some of the interesting signatures we find are suggestive, but firm conclusions require additional data (more stars, more abundances) based on a homogeneous analysis, continuing on the path pioneered by the RPA. Along with the discovery of a larger number of RPE stars,
the abundances of many elements across the neutron-capture peaks over a larger metallicity range can help identify the astrophysical sites of the $r$-process and their relative contribution to the chemical history of the Galaxy.

\section*{Acknowledgements}
We thank the anonymous reviewer for valuable comments and suggestions which improved our paper. PS thanks ABSR, AK, and RS for their useful discussions during this study. We thank Vinicius Placco for providing carbon correction values. CAP is thankful for funding from the Spanish government through grants AYA2014-56359-P, AYA2017-86389-P and PID2020-117493GB-100. TCB acknowledges partial support for this work from grant PHY 14-30152; Physics Frontier Center/JINA Center for the Evolution of the Elements (JINA-CEE), and OISE-1927130: The International Research Network for Nuclear Astrophysics (IReNA), awarded by the US National Science Foundation. This work makes use of data from the European Space Agency (ESA) mission {\it Gaia} (\url{https://www.cosmos.esa.int/gaia}), processed by the {\it Gaia} Data Processing and Analysis Consortium (DPAC, \url{https://www.cosmos.esa.int/web/gaia/dpac/consortium}). Funding for the DPAC has been provided by national institutions, in particular the institutions participating in the {\it Gaia} Multilateral Agreement. This work makes use of Astropy:\footnote{http://www.astropy.org} a community-developed core Python package and an ecosystem of tools and resources for astronomy \citep{astropy:2013, astropy:2018, astropy:2022}. This publication makes use of VOSA, developed under the Spanish Virtual Observatory (https://svo.cab.inta-csic.es) project funded by MCIN/AEI/10.13039/501100011033/ through grant PID2020-112949GB-I00. This work makes use of Matplotlib \citep{Matplotlib.2007}, NumPy \citep{Numpy.2020}, and Pandas \citep{Pandass.paper.2010, Pandas.software.2020}.

\section*{DATA AVAILABILITY}
The data used in this research will be shared on reasonable request to the corresponding author.



\bibliographystyle{mnras}
\bibliography{R_process_paper_arxiv} 


\newpage
\appendix

\section{Elemental abundances of elements}
In Tables~\ref{tab:abundances_TYC3431}, \ref{tab:abundances_HD2638}, \ref{tab:abundances_TYC1716},  and \ref{tab:abundances_TYC1191}, we list the elemental abundances derived for TYC 3431-689-1, HD 263815, TYC 1716-1548-1 and TYC 1191-918-1 respectively. Each table contains the name of the element, the atomic/molecular species used for abundance estimation, the atomic number, the number of lines used for abundance estimation, the solar abundances of the corresponding element, the abundance of the program star from the spectral-synthesis method (except for Ti, for which we have plenty of clear lines available), the relative abundance of program star with respect to the solar value and with respect to Fe ([X/Fe]), and the uncertainty in the abundance estimation. Whenever possible, for the relative abundance, we have consider absolute abundance obtained from the 
spectral-synthesis method.

\begin{table*}
	\centering
	\caption{Detailed abundance determination for TYC 3431-689-1.}
	\label{tab:abundances_TYC3431}
	\begin{tabular}{lccccccccr} 
		\hline
		Element & species & atomic no. & N-lines & Solar & A(X)$_{\rm syn}$ & [X/H] & [X/Fe] & $\sigma$\\
		\hline
	    C & CH & 6 & 1 & 8.39 & 5.74 & $-$2.65 & $-$0.60 & 0.23\\
		N & CN & 7 & 1 & 7.78 & 6.78 & $-$1.00 & +1.05 & 0.18\\
		Na & Na I & 11 & 2 & 6.17 & 4.72 & $-$1.45 & +0.60 & 0.25\\
		Mg & Mg I & 12 & 7 & 7.53 & 5.93 & $-$1.60 & +0.45 & 0.17\\
		Al & Al I & 13 & 2 & 6.37 & 4.17 & $-$2.20 & $-$0.15 & 0.14\\
		Si & Si I & 14 & 2 & 7.51 & 5.71 & $-$1.80 & +0.25 & 0.13\\
		Ca & Ca I & 20 & 14 & 6.31 & 4.6 & $-$1.71 & +0.34 & 0.12\\
		Sc & Sc I & 21 & 1 & 3.17 & 0.67 & $-$2.50 & $-$0.45 & 0.31\\
		Sc & Sc II & 21 & 9 & 3.17 & 1.07 & $-$2.10 & $-$0.05 & 0.10\\
		Ti & Ti I & 22 & 18 & 4.90 & 3.07 & $-$1.83 & +0.22 & 0.19\\
		Ti & Ti II & 22 & 31 & 4.90 & 3.18 & $-$1.72 & +0.33 & 0.08\\
		V & V I & 23 & 4 & 4.00 & 1.6 & $-$2.40 & $-$0.35 & 0.23\\
		V & V II & 23 & 2 & 4.00 & 1.9 & $-$2.10 & $-$0.05 & 0.11\\
		Cr & Cr I & 24 & 13 & 5.64 & 3.31 & $-$2.33 & $-$0.28 & 0.19\\
		Cr & Cr II & 24 & 5 & 5.64 & 3.54 & $-$2.10 & $-$0.05 & 0.08\\
		Mn & Mn I & 25 & 8 & 5.39 & 2.82 & $-$2.57 & $-$0.52 & 0.19\\
		Co & Co I & 27 & 3 & 4.92 & 3.02 & $-$1.90 & +0.15 & 0.24\\
		Ni & Ni I & 28 & 16 & 6.23 & 4.03 & $-$2.20 & $-$0.15 & 0.13\\
		Cu & Cu I & 29 & 1 & 4.21 & 1.05 & $-$3.16 & $-$1.11 & 0.19\\
		Zn & Zn I & 30 & 2 & 4.60 & 2.65 & $-$1.95 & +0.10 & 0.08\\
		Sr & Sr II & 38 & 2 & 2.92 & 1.02 & $-$1.90 & +0.15 & 0.13\\
		Y & Y II & 39 & 7 & 2.21 & 0.01 & $-$2.20 & $-$0.15 & 0.10\\
		Zr & Zr II & 40 & 7 & 2.58 & 0.88 & $-$1.70 & +0.35 & 0.10\\
		Ba & Ba II & 56 & 5 & 2.17 & 0.37 & $-$1.80 & +0.25 & 0.13\\
		La & La II & 57 & 4 & 1.13 & $-$0.57 & $-$1.70 & +0.35 & 0.11\\
		Ce & Ce II & 58 & 3 & 1.70 & $-$0.35 & $-$2.05 & +0.00 & 0.12\\
		Pr & Pr II & 59 & 1 & 0.58 & $-$0.72 & $-$1.30 & +0.75 & 0.07\\
		Nd & Nd II & 60 & 6 & 1.45 & $-$0.12 & $-$1.57 & +0.48 & 0.12\\
		Sm & Sm II & 62 & 3 & 1.00 & $-$0.5 & $-$1.50 & +0.55 & 0.12\\
		Eu & Eu II & 63 & 2 & 0.52 & $-$0.73 & $-$1.25 & +0.80 & 0.03\\
	    Gd & Gd II & 64 & 2 & 1.11 & $-$0.39 & $-$1.50 & +0.55 & 0.15\\
	    Tb & Tb II & 65 & 1 & 0.28 & $-$0.82 & $-$1.10 & +0.95 & 0.23\\
        Dy & Dy II & 66 & 3 & 1.14 & $-$0.29 & $-$1.43 & +0.62 & 0.12\\
	    Ho & Ho II & 67 & 1 & 0.51 & $-$0.69 & $-$1.20 & +0.85 & 0.02\\
	    Er & Er II & 68 & 2 & 0.93 & $-$0.57 & $-$1.50 & +0.55 & 0.11\\
	    Lu & Lu II & 71 & 1 & 0.06 & $-$1.24 & $-$1.30 & +0.75 & 0.12\\
	    Hf & Hf II & 72 & 2 & 0.88 & $-$0.72 & $-$1.60 & +0.45 & 0.15\\
	    Os & Os I & 76 & 2 & 1.25 & $-$0.15 & $-$1.40 & +0.65 & 0.21\\
		Pb & Pb II & 82 & 1 & 2.00 & <0.1 & $-$ & $-$ & $-$\\
		Th & Th II & 90 & 1 & 0.06 & $-$1.34 & $-$1.40 & +0.65 & 0.1\\	
	   \hline
	\end{tabular}
\end{table*}

\begin{table*}
	\centering
	\caption{Detailed abundance determination for HD 263815.}
	\label{tab:abundances_HD2638}
	\begin{tabular}{lccccccccr} 
		\hline
	      Element & species & atomic no. & N-lines & Solar & A(X)$_{\rm syn}$ & [X/H] & [X/Fe] & $\sigma$\\
		\hline
	    C & CH & 6 & 1 & 8.39 & 5.59 & $-$2.80 & $-$0.58 & 0.25\\
		N & CN & 7 & 1 & 7.78 & 6.28 & $-$1.50 & +0.72 & 0.18\\
		Na & Na I & 11 & 2 & 6.17 & 4.22 & $-$1.95 & +0.27 & 0.26\\
		Mg & Mg I & 12 & 7 & 7.53 & 5.83 & $-$1.70 & +0.52 & 0.17\\
		Al & Al I & 13 & 2 & 6.37 & 3.87 & $-$2.50 & $-$0.28 & 0.15\\
		Si & Si I & 14 & 2 & 7.51 & 5.71 & $-$1.80 & +0.42 & 0.11\\
		Ca & Ca I & 20 & 14 & 6.31 & 4.43 & $-$1.88 & +0.34 & 0.14\\
		Sc & Sc I & 21 & 1 & 3.17 & 0.39 & $-$2.78 & $-$0.56 & 0.31\\
		Sc & Sc II & 21 & 9 & 3.17 & 0.96 & $-$2.21 & +0.01 & 0.09\\
		Ti & Ti I & 22 & 14 & 4.90 & 2.9 & $-$2.00 & +0.22 & 0.19\\
		Ti & Ti II & 22 & 33 & 4.90 & 3.0 & $-$1.90 & +0.32 & 0.08\\
		V & V I & 23 & 4 & 4.00 & 1.50 & $-$2.50 & $-$0.28 & 0.24\\
		V & V II & 23 & 3 & 4.00 & 1.90 & $-$2.10 & +0.12 & 0.08\\
		Cr & Cr I & 24 & 13 & 5.64 & 3.12 & $-$2.52 & $-$0.30 & 0.22\\
		Cr & Cr II & 24 & 5 & 5.64 & 3.44 & $-$2.20 & +0.02 & 0.06\\
		Mn & Mn I & 25 & 8 & 5.39 & 2.66 & $-$2.73 & $-$0.51 & 0.21\\
		Co & Co I & 27 & 3 & 4.92 & 2.92 & $-$2.00 & +0.22 & 0.27\\
		Ni & Ni I & 28 & 16 & 6.23 & 4.08 & $-$2.15 & +0.07 & 0.14\\
		Cu & Cu I & 29 & 1 & 4.21 & 1.01 & $-$3.20 & $-$0.98 & 0.21\\
		Zn & Zn I & 30 & 2 & 4.60 & 2.55 & $-$2.05 & +0.17 & 0.05\\
		Sr & Sr II & 38 & 2 & 2.92 & 0.52 & $-$2.40 & $-$0.18 & 0.13\\
		Y & Y II & 39 & 7 & 2.21 & $-$0.29 & $-$2.50 & $-$0.28 & 0.09\\
		Zr & Zr II & 40 & 7 & 2.58 & 0.48 & $-$2.10 & +0.12 & 0.14\\
		Ba & Ba II & 56 & 5 & 2.17 & $-$0.18 & $-$2.35 & $-$0.13 & 0.13\\
		La & La II & 57 & 4 & 1.13 & $-$1.04 & $-$2.17 & +0.05 & 0.12\\
		Ce & Ce II & 58 & 3 & 1.70 & $-$0.8 & $-$2.50 & $-$0.28 & 0.14\\
		Pr & Pr II & 59 & 2 & 0.58 & $-$1.02 & $-$1.60 & +0.62 & 0.1\\
		Nd & Nd II & 60 & 6 & 1.45 & $-$0.51 & $-$1.96 & +0.26 & 0.13\\
		Sm & Sm II & 62 & 3 & 1.00 & $-$1.00 & $-$2.00 & +0.22 & 0.13\\
		Eu & Eu II & 63 & 2 & 0.52 & $-$1.18 & $-$1.70 & +0.52 & 0.01\\
	    Gd & Gd II & 64 & 2 & 1.11 & $-$0.79 & $-$1.90 & +0.32 & 0.14\\
	    Tb & Tb II & 65 & 1 & 0.28 & $-$1.22 & $-$1.50 & +0.72 & 0.2\\
        Dy & Dy II & 66 & 3 & 1.14 & $-$0.86 & $-$2.00 & +0.22 & 0.16\\
	    Ho & Ho II & 67 & 2 & 0.51 & $-$1.19 & $-$1.70 & +0.52 & 0.01\\
	    Er & Er II & 68 & 2 & 0.93 & $-$0.97 & $-$1.90 & +0.32 & 0.14\\
	    Tm & Tm II & 69 & 1 & 0.0 & $-$1.5 & $-$1.50 & +0.72 & 0.1\\
	    Lu & Lu II & 71 & 1 & 0.06 & $-$1.54 & $-$1.60 & +0.62 & 0.14\\
	    Hf & Hf II & 72 & 2 & 0.88 & $-$1.02 & $-$1.90 & +0.32 & 0.14\\
	    Os & Os I & 76 & 1 & 1.25 & $-$0.45 & $-$1.70 & +0.52 & 0.22\\
		Pb & Pb II & 82 & 1 & 2.00 & <$-$0.1 & <$-$2.1 & <0.12 & $-$\\
		Th & Th II & 90 & 1 & 0.06 & <$-$2.04 & $-$ & $-$ & $-$\\
	    \hline
	\end{tabular}
\end{table*}

\begin{table*}
	\centering
	\caption{Detailed abundance determination for TYC 1716-1548-1.}
	\label{tab:abundances_TYC1716}
	\begin{tabular}{lccccccccr} 
		\hline
		Element & species & atomic no. & N-lines & Solar & A(X)$_{\rm syn}$ & [X/H] & [X/Fe] & $\sigma$\\
		\hline
	    C & CH & 6 & 1 & 8.39 & 5.89 & $-$2.50 & $-$0.35 & 0.27\\
		N & CN & 7 & 1 & 7.78 & 6.58 & $-$1.20 & +0.95 & 0.20\\
		Na & Na I & 11 & 2 & 6.17 & 4.17 & $-$2.00 & +0.15 & 0.32\\
		Mg & Mg I & 12 & 7 & 7.53 & 5.9 & $-$1.63 & +0.52 & 0.23\\
		Al & Al I & 13 & 2 & 6.37 & 4.17 & $-$2.20 & $-$0.05 & 0.17\\
		Si & Si I & 14 & 2 & 7.51 & 5.81 & $-$1.70 & +0.45 & 0.15\\
		Ca & Ca I & 20 & 14 & 6.31 & 4.36 & $-$1.95 & +0.20 & 0.19\\
		Sc & Sc I & 21 & 1 & 3.17 & 0.46 & $-$2.71 & $-$0.56 & 0.3\\
		Sc & Sc II & 21 & 9 & 3.17 & 0.92 & $-$2.25 & $-$0.1 & 0.05\\
		Ti & Ti I & 22 & 15 & 4.90 & 2.92 & $-$1.98 & +0.17 & 0.28\\
		Ti & Ti II & 22 & 29 & 4.90 & 3.27 & $-$1.63 & +0.52 & 0.06\\
		V & V I & 23 & 4 & 4.00 & 1.55 & $-$2.45 & $-$0.30 & 0.28\\
		Cr & Cr I & 24 & 13 & 5.64 & 3.24 & $-$2.40 & $-$0.25 & 0.28\\
		Cr & Cr II & 24 & 5 & 5.64 & 3.49 & $-$2.15 & +0.00 & 0.09\\
		Mn & Mn I & 25 & 8 & 5.39 & 2.63 & $-$2.76 & $-$0.61 & 0.26\\
		Co & Co I & 27 & 3 & 4.92 & 2.72 & $-$2.20 & $-$0.05 & 0.29\\
		Ni & Ni I & 28 & 16 & 6.23 & 3.91 & $-$2.32 & $-$0.17 & 0.17\\
		Cu & Cu I & 29 & 1 & 4.21 & 1.08 & $-$3.13 & $-$0.98 & 0.26\\
		Zn & Zn I & 30 & 2 & 4.60 & 2.55 & $-$2.05 & +0.10 & 0.04\\
		Ga & Ga I & 31 & 1 & 2.88 & 0.48 & $-$2.40 & $-$0.25 & $-$\\
		Sr & Sr II & 38 & 2 & 2.92 & 0.92 & $-$2.00 & +0.15 & 0.05\\
		Y & Y II & 39 & 7 & 2.21 & 0.01 & $-$2.20 & $-$0.05 & 0.07\\
		Zr & Zr II & 40 & 7 & 2.58 & 0.98 & $-$1.60 & +0.55 & 0.07\\
		Ba & Ba II & 56 & 7 & 2.17 & 0.19 & $-$1.98 & +0.17 & 0.04\\
		La & La II & 57 & 4 & 1.13 & $-$0.72 & $-$1.85 & +0.30 & 0.07\\
		Ce & Ce II & 58 & 3 & 1.70 & $-$0.55 & $-$2.25 & $-$0.10 & 0.08\\
        Pr & Pr II & 59 & 2 & 0.58 & $-$0.42 & $-$1.00 & +1.15 & 0.1\\		
        Nd & Nd II & 60 & 3 & 1.45 & $-$0.05 & $-$1.50 & +0.65 & 0.07\\
		Sm & Sm II & 62 & 2 & 1.00 & $-$0.75 & $-$1.75 & +0.40 & 0.1\\
		Eu & Eu II & 63 & 2 & 0.52 & $-$0.83 & $-$1.35 & +0.80 & 0.02\\
	    Gd & Gd II & 64 & 2 & 1.11 & $-$0.49 & $-$1.60 & +0.55 & 0.17\\
	    Dy & Dy II & 66 & 3 & 1.14 & $-$0.43 & $-$1.57 & +0.58 & 0.09\\
	    Ho & Ho II & 67 & 2 & 0.51 & $-$0.99 & $-$1.50 & +0.65 & 0.1\\
        Er & Er II & 68 & 2 & 0.93 & $-$0.47 & $-$1.40 & +0.75 & 0.13\\
	    Lu & Lu II & 71 & 1 & 0.06 & $-$1.33 & $-$1.39 & +0.76 & 0.11\\
	    Os & Os I & 76 & 2 & 1.25 & $-$0.15 & $-$1.40 & +0.75 & 0.19\\
		Pb & Pb II & 82 & 1 & 2.00 & <0.1 & <$-$1.9 & <0.25 & $-$\\
		\hline
	\end{tabular}
\end{table*}

\begin{table*}
	\centering
	\caption{Detailed abundance determination for TYC 1191-918-1.}
	\label{tab:abundances_TYC1191}
	\begin{tabular}{lcccccccccr} 
		\hline
		Element & species & atomic no. & N-lines & Solar & A(X)$_{\rm syn}$ & [X/H] & [X/Fe] & $\sigma$\\
		\hline
        C & CH & 6 & 1 & 8.39 & 5.79 & $-$2.60 & $-$0.70 & 0.24\\
		N & CN & 7 & 1 & 7.78 & 6.78 & $-$1.00 & +0.90 & 0.19\\
		Na & Na I & 11 & 2 & 6.17 & 4.47 & $-$1.70 & +0.20 & 0.26\\
		Mg & Mg I & 12 & 7 & 7.53 & 6.15 & $-$1.38 & +0.52 & 0.18\\
		Al & Al I & 13 & 2 & 6.37 & 4.07 & $-$2.30 & $-$0.40 & 0.13\\
		Si & Si I & 14 & 2 & 7.51 & 5.91 & $-$1.60 & +0.30 & 0.13\\
		Ca & Ca I & 20 & 14 & 6.31 & 4.72 & $-$1.59 & +0.31 & 0.15\\
		Sc & Sc II & 21 & 9 & 3.17 & 1.12 & $-$2.05 & $-$0.15 & 0.12\\
		Ti & Ti I & 22 & 17 & 4.90 & 3.19 & $-$1.71 & +0.19 & 0.23\\
		Ti & Ti II & 22 & 22 & 4.90 & 3.4 & $-$1.50 & +0.40 & 0.10\\
		V & V I & 23 & 4 & 4.00 & 1.81 & $-$2.19 & $-$0.29 & 0.24\\
		V & V II & 23 & 1 & 4.00 & 2.1 & $-$1.90 & +0.00 & 0.13\\
		Cr & Cr I & 24 & 13 & 5.64 & 3.44 & $-$2.20 & $-$0.30 & 0.24\\
		Cr & Cr II & 24 & 5 & 5.64 & 3.72 & $-$1.92 & $-$0.02 & 0.1\\
		Mn & Mn I & 25 & 8 & 5.39 & 2.89 & $-$2.50 & $-$0.60 & 0.17\\
		Co & Co I & 27 & 3 & 4.92 & 2.92 & $-$2.00 & $-$0.10 & 0.29\\
		Ni & Ni I & 28 & 16 & 6.23 & 4.20 & $-$2.03 & $-$0.13 & 0.15\\
		Cu & Cu I & 29 & 1 & 4.21 & 1.31 & $-$2.90 & $-$1.00 & 0.21\\
		Zn & Zn I & 30 & 2 & 4.60 & 2.70 & $-$1.90 & +0.00 & 0.07\\
		Sr & Sr II & 38 & 2 & 2.92 & 0.92 & $-$2.00 & $-$0.10 & 0.12\\
		Y & Y II & 39 & 7 & 2.21 & 0.21 & $-$2.00 & $-$0.10 & 0.13\\
		Zr & Zr II & 40 & 5 & 2.58 & 0.84 & $-$1.74 & +0.16 & 0.12\\
		Ba & Ba II & 56 & 3 & 2.17 & 0.13 & $-$2.04 & $-$0.14 & 0.15\\
		La & La II & 57 & 4 & 1.13 & $-$0.52 & $-$1.65 & +0.25 & 0.14\\
		Ce & Ce II & 58 & 3 & 1.70 & $-$0.35 & $-$2.05 & $-$0.15 & 0.15\\
		Pr & Pr II & 59 & 2 & 0.58 & $-$0.82 & $-$1.40 & +0.50 & 0.12\\
		Nd & Nd II & 60 & 6 & 1.45 & $-$0.08 & $-$1.53 & +0.37 & 0.15\\
		Sm & Sm II & 62 & 3 & 1.00 & $-$0.6 & $-$1.60 & +0.30 & 0.14\\
		Eu & Eu II & 63 & 2 & 0.52 & $-$0.83 & $-$1.35 & +0.55 & 0.03\\
	    Tb & Tb II & 65 & 1 & 0.28 & $-$1.82 & $-$1.10 & +0.80 & 0.12\\
        Dy & Dy II & 66 & 3 & 1.14 & $-$0.39 & $-$1.53 & +0.37 & 0.15\\
	    Ho & Ho II & 67 & 2 & 0.51 & $-$0.89 & $-$1.40 & +0.50 & 0.13\\
        Er & Er II & 68 & 2 & 0.93 & $-$0.58 & $-$1.51 & +0.39 & 0.12\\
        Hf & Hf II & 72 & 1 & 0.88 & $-$0.82 & $-$1.70 & +0.20 & 0.16\\
	    Os & Os I & 76 & 2 & 1.25 & $-$0.1 & $-$1.35 & +0.55 & 0.21\\
		Th & Th II & 90 & 1 & 0.06 & $-$1.24 & $-$1.30 & +0.60 & 0.12\\
		\hline
	\end{tabular} 
\end{table*}

\section[Fe-I and Fe-II lines information]{F\MakeLowercase{e}-I and F\MakeLowercase{e}-II lines information}
\label{sec:Fe1_Fe2_lines_list}
Table~\ref{tab:Fe1_Fe2_lines_list} provides a complete Fe I and Fe II line list for all the program stars. It includes the elemental ionization, wavelength, lower excitation potential (LEP), and $\log gf$. The equivalent widths of lines identified in the spectra are listed under corresponding program star. References of the $\log gf$ values for these lines are written below the Table~\ref{tab:Fe1_Fe2_lines_list}.

\clearpage
\onecolumn
\begin{longtable}{lccccccccr}
	\caption{Information about the Fe line list.}
	\label{tab:Fe1_Fe2_lines_list}\\
	\hline
	Element & wavelength ({\AA}) & EP (eV)& $\log gf$ & TYC 3431-689-1 & HD 263815 & TYC 1716-1548-1 & TYC 1191-918-1 \\
	\hline
Fe I  & 4107.488 & 2.831 & $-$0.879 &  84.9 &  84.5  &  106.9  &  111.2 \\
Fe I  & 4170.902 & 3.017 & $-$1.086 &  81.7 &  78.9  &  86.5  &  111.9 \\
Fe I  & 4174.913 & 0.915 & $-$2.969 &  80.1 &  100.8  &  134.1  &  119.3 \\
Fe I  & 4175.636 & 2.845 & $-$0.827 &  74.9 &  81.8  &  96.6  &  104.1 \\
Fe I  & 4182.383 & 3.017 & $-$1.18 &  51.0 &  43.7  &  52.3  &  $-$ \\
Fe I  & 4206.697 & 0.052 & $-$3.96 &  110.3 &  $-$  &  156.8  &  153.6\\
Fe I  & 4216.22 & 2.453 & $-$5.376 &  133.9 &  109.1  &  185.2  &  252.2\\
Fe I  & 4225.454 & 3.417 & $-$0.51 &  104.2 &  $-$  &  131.5  &  $-$ \\
Fe I  & 4233.603 & 2.482 & $-$0.604 &  85.8 &  $-$  &  99.4  &  97.2\\
Fe I  & 4433.219 & 3.654 & $-$0.7 &  34.5 &  33.5  &  51.3  &  54.7 \\
Fe I  & 4433.782 & 3.602 & $-$1.267 &  40.6 &  27.6  &  52.3  &  53.0 \\
Fe I  & 4442.339 & 2.198 & $-$1.255 &  98.4 &  99.3  &  119.3 &  90.0\\
Fe I  & 4447.717 & 2.223 & $-$1.342 &  91.6 &  87.0  &  113.6  &  110.3\\
Fe I  & 4494.563 & 2.198 & $-$1.136 &  101.2 &  103.6  &  121.4  &  128.2\\
Fe I  & 4517.525 & 3.071 & $-$1.858 &  20.8 &  16.3  &  29.7  &  32.2 \\
Fe I  & 4602.001 & 1.608 & $-$3.154 &  35.7 &  39.7  &  69.5 &  60.9 \\
Fe I  & 4602.941 & 1.485 & $-$2.209 &  91.5 &  98.6 &  114.2  &  115.1\\
Fe I  & 4625.045 & 3.241 & $-$1.34 &  35.3 &  38.0  &  53.3  &  57.2 \\
Fe I  & 4632.912 & 1.608 & $-$2.913 &  55.5 &  55.7  &  94.6  &  83.7 \\
Fe I  & 4710.283 & 3.018 & $-$1.612 &  33.9 &  36.3  &  46.5  &  58.5 \\
Fe I  & 4733.592 & 1.485 & $-$2.988 &  61.9 &  60.6  &  82.0  &  85.0\\
Fe I  & 4736.773 & 3.211 & $-$0.752 &  67.4 &  68.8  &  90.3  &  90.9\\
Fe I  & 4871.318 & 2.865 & $-$0.363 &  101.4 &  103.2  &  123.6  &  122.7\\
Fe I  & 4872.138 & 2.882 & $-$0.567 &  95.5 &  94.8  &  119.5  &  93.8\\
Fe I  & 4890.755 & 2.875 & $-$0.394 &  101.1 &  101.0  &  123.9  &  120.5\\
Fe I  & 4891.492 & 2.851 & $-$0.112 &  107.8 &  116.1  &  131.2  &  127.2\\
Fe I  & 4903.31 & 2.882 & $-$0.926 & 63.1 &  78.6  &  98.4  &  95.2 \\
Fe I  & 4924.77 & 2.279 & $-$2.241 &  46.2 &  51.0  &  69.8  &  70.8 \\
Fe I  & 4939.687 & 0.859 & $-$3.34 &  73.2 &  72.0  &  103.7  &  95.5\\
Fe I  & 4946.388 & 3.368 & $-$1.17 &  36.6 &  29.7  &  63.9 &  53.8 \\
Fe I  & 4982.524 & 4.103 & 0.144 & 43.3 &  42.6  &  49.4  &  54.7 \\
Fe I  & 4994.13 & 0.915 & $-$3.08 &  76.1 &  95.0  &  122.1  &  106.7 \\
Fe I  & 5001.864 & 3.881 & 0.01 &  57.6 &  57.1  &  75.1  &  74.5 \\
Fe I  & 5002.793 & 3.396 & $-$1.58 &  21.8 &  20.1  &  34.6  &  29.5 \\
Fe I  & 5014.943 & 3.943 & $-$0.303 &  45.8 &  40.4  &  54.5  &  58.1 \\
Fe I  & 5027.12 & 4.154 & $-$0.559 &  34.8 &  29.7  &  46.3 &  50.5 \\
Fe I  & 5039.252 & 3.368 & $-$1.573 &  22.2 &  22.2  & 38.3 &  37.6 \\
Fe I  & 5044.211 & 2.851 & $-$2.038 &  32.1 &  23.6 &  38.6  &  39.9 \\
Fe I  & 5049.82 & 2.279 & $-$1.355 &  86.0 &  92.1  &  118.5  &  112.9 \\
Fe I  & 5051.635 & 0.915 & $-$2.795 &  108.1 &  111.7  &  140.2 &  129.2\\
Fe I  & 5060.079 & 0 & $-$5.46 &  24.1 &  $-$  &  65.3 &  $-$ \\
Fe I  & 5068.766 & 2.94 & $-$1.042 &  68.0 &  66.1  &  113.3  &  87.0 \\
Fe I  & 5079.74 & 0.99 & $-$3.22 &  88.7 &  82.7  &  132.6  &  120.2\\
Fe I  & 5083.339 & 0.958 & $-$2.958 &  94.1 & 101.9  &  133.9 &  121.7\\
Fe I  & 5090.774 & 4.256 & $-$0.4 &  25.3 &  17.2  &  34.7  &  33.5 \\
Fe I  & 5125.117 & 4.22 & $-$0.14 &  43.5 &  31.2  &  55.6  &  60.4 \\
Fe I  & 5127.359 & 0.915 & $-$3.307 &  77.5 &  85.5 &  113.8  &  106.5\\
Fe I  & 5133.689 & 4.178 & 0.14 &   60.3 &  59.3  &  76.9  &  78.7 \\
Fe I  & 5150.84 & 0.99 & $-$3.003 &  80.9 & 88.6  &  114.5  &  108.5 \\
Fe I  & 5162.273 & 4.178 & 0.02 &  55.8 &  49.4  &  71.9 &  69.5 \\
Fe I  & 5166.282 & 0 & $-$4.195 &  91.3 &  101.1  &  132.4  &  119.7 \\
Fe I  & 5171.596 & 1.485 & $-$1.793 &  113.9 &  105.9  &  150.9  & 133.9\\
Fe I  & 5191.455 & 3.038 & $-$0.551 &  100.8 &  91.1  &  148.5 &  133.2 \\
Fe I  & 5192.344 & 2.998 & $-$0.421 &  90.2 &  88.4  &  122.9  &  114.9\\
Fe I  & 5194.942 & 1.557 & $-$2.09 &  100.6 &  97.8  &  140.2  &  119.8\\
Fe I  & 5198.711 & 2.223 & $-$2.135 &  59.5 &  58.4  &  89.7  &  85.4\\
Fe I  & 5202.336 & 2.176 & $-$1.838 &  89.6 &  84.9 &  120.0  &  113.7\\
Fe I  & 5215.181 & 3.266 & $-$0.871 &  48.7 &  51.5  &  66.8 &  68.8\\
Fe I  & 5216.274 & 1.608 & $-$2.15 &  95.2 &  98.8  &  119.7  &  116.6 \\
Fe I  & 5217.389 & 3.211 & $-$1.07 &  44.2 &  45.3  &  69.0 &  66.6 \\
Fe I  & 5225.526 & 0.11 & $-$4.789 &  49.8 &  54.9  &  91.4  &  78.7 \\
Fe I  & 5232.94 & 2.94 & $-$0.058 &  110.4 &  115.5  &  131.1  &  131.8\\
Fe I  & 5247.05 & 0.087 & $-$4.946 &  35.3 &  41.8 &  81.4  &  77.3\\
Fe I  & 5250.209 & 0.121 & $-$4.938 &  30.0 &  $-$  &  60.6  &  57.9\\
Fe I  & 5250.646 & 2.198 & $-$2.181 &  49.2 &  $-$  &  65.1 &  73.5\\
Fe I  & 5254.97 & 4.294 & $-$4.035 &  63.5 &  65.8  &  100.4  &  97.2 \\
Fe I  & 5263.306 & 3.266 & $-$0.879 &  47.9 &  53.3  &  68.8 &  72.2\\
Fe I  & 5266.555 & 2.998 & $-$0.386 &  91.4 &  96.6 &  116.3  &  112.3 \\
Fe I  & 5281.79 & 3.038 & $-$0.834 &  72.8 &  69.6 &  87.5  &  82.2\\
Fe I  & 5283.621 & 3.241 & $-$0.432 &  67.9 &  $-$  &  89.4  &  84.8\\
Fe I  & 5322.041 & 2.279 & $-$2.803 &  14.6 &  17.5  &  35.4 &  31.4 \\
Fe I  & 5324.179 & 3.211 & $-$0.103 &  105.5 &  100.8  &  126.5 &  122.9\\
Fe I  & 5332.9 & 1.557 & $-$2.777 &  58.7 &  67.8  &  102.1  &  91.0\\
Fe I  & 5339.929 & 3.266 & $-$0.647 &  62.2 &  68.6  &  86.9 &  87.5 \\
Fe I  & 5367.467 & 4.415 & 0.443 &  50.0 &  47.5  &  60.4  &  64.0\\
Fe I  & 5369.962 & 4.371 & 0.536 &  59.6 &  55.8  &  70.5 &  75.4 \\
Fe I  & 5383.369 & 4.312 & 0.645 &  62.0 &  63.6  &  72.3  &  79.4\\
Fe I  & 5389.479 & 4.415 & $-$0.41 &  21.4 &  17.1  &  27.5 &  29.6 \\
Fe I  & 5393.168 & 3.241 & $-$0.715 &  65.0 & 66.8  &  88.8  &  83.8 \\
Fe I  & 5397.128 & 0.915 & $-$1.993 &  134.4 &  $-$  &  164.8  &  158.1 \\
Fe I  & 5400.502 & 4.371 & $-$0.16 &  34.6 &  27.2  &  42.5  &  42.1\\
Fe I  & 5415.199 & 4.386 & 0.642 &  66.6 &  56.3  &  74.0  &  76.7\\
Fe I  & 5445.042 & 4.386 & $-$0.02 &  41.8 &  36.6  &  49.2  &  58.4\\
Fe I  & 5497.516 & 1.011 & $-$2.849 &  101.7 &  110.5  &  133.6  &  $-$ \\
Fe I  & 5501.465 & 0.958 & $-$3.047 &  89.1 &  99.4 &  129.6  &  116.8 \\
Fe I  & 5506.779 & 0.99 & $-$2.797 &  105.2 &  109.8  &  131.9  &  127.9 \\
Fe I  & 5572.842 & 3.396 & $-$0.275 &  75.4 &  82.2  &  115.1  &  103.2\\
Fe I  & 5576.089 & 3.43 & $-$1 &  57.8 &  48.8  &  67.8  &  72.2 \\
Fe I  & 5586.822 & 3.695 & $-$5.961 &  91.5 &  85.6  &  114.4  &  109.1 \\
Fe I  & 5624.542 & 3.417 & $-$0.755 &  57.0  &  50.5  &  68.6  &  73.8\\
Fe I  & 5762.992 & 4.209 & $-$0.45 &  33.2  &  26.7   &  40.1 &  41.7 \\
Fe I  & 5956.694 & 0.859 & $-$4.605 &  22.5  &  19.1  &  56.5  &  38.8\\
Fe I  & 6065.482 & 2.608 & $-$1.53 &  68.8  &  68.3  &  97.4  &  89.9 \\
Fe I  & 6136.615 & 2.453 & $-$1.4 &  91.8 &  92.3  &  144.4  &  134.9 \\
Fe I  & 6137.692 & 2.588 & $-$1.403 &  77.4 &  75.1 &  102.0  &  97.8 \\
Fe I  & 6213.43 & 2.223 & $-$2.482 &  45.4  &  38.7 &  70.7  &  58.5\\
Fe I  & 6230.723 & 2.559 & $-$1.281 &  79.0  &  88.8  &  111.1  &  108.6 \\
Fe I  & 6232.641 & 3.654 & $-$1.223 &  23.6  &  19.4  &  25.8  &  32.7 \\
Fe I  & 6246.319 & 3.602 & $-$0.733 &  35.1  &  37.3  &  60.7  &  56.9 \\
Fe I  & 6252.555 & 2.404 & $-$1.687 &  79.2  &  75.6  &  96.3 &  97.1 \\
Fe I  & 6265.134 & 2.176 & $-$2.55 &  29.4  &  45.9 &  71.4 &  65.3 \\
Fe I  & 6318.018 & 2.453 & $-$2.338 &  51.3 &  52.0  &  74.6  &  83.7 \\
Fe I  & 6393.601 & 2.433 & $-$1.432 &  76.8  &  79.1  &  104.1  &  107.9 \\
Fe I  & 6411.649 & 3.654 & $-$0.595 &  43.1 &  47.7  &  70.0 &  69.9 \\
Fe I  & 6421.351 & 2.279 & $-$2.027 &  67.1  &  67.0  &  101.5  &  90.8\\
Fe I  & 6430.846 & 2.176 & $-$2.006 &  75.8 &  73.9 &  105.7  &  95.4\\
Fe I  & 6592.914 & 2.727 & $-$1.473 &  53.8  &  63.4 &  82.3 &  92.7\\
Fe I  & 6593.87 & 2.433 & $-$2.422 & 35.3&  38.1  &  69.9  &  58.1 \\
Fe I  & 6663.442 & 2.424 & $-$2.479 &  40.1  &  33.4  &  59.4  &  52.6\\
Fe I  & 6677.987 & 2.692 & $-$1.418 &  74.4 &  68.2 &  114.7  &  94.0 \\
Fe II  & 4233.172 & 2.583 & $-$1.9 &  91.8 &  $-$  & 79.1   &  97.2 \\
Fe II  & 4416.83 & 2.778 & $-$2.41 &  70.8 &  69.5 &  72.1  &  85.8\\
Fe II  & 4489.183 & 2.828 & $-$2.97 &  61.0 &  51.5 &  68.7  &  61.5\\
Fe II  & 4491.405 & 2.856 & $-$2.7 &  60.7 &  58.6 &  69.1  &  73.6\\
Fe II  & 4508.288 & 2.856 & $-$2.25 &  81.6	 &  80.1 &  82.3  & 91.3\\
Fe II  & 4515.339 & 2.844 & $-$2.45 &  68.2	 &  74.4 &  71.5  &  88.2\\
Fe II  & 4520.224 & 2.807 & $-$2.6 &  68.8	 &  66.6 &  69.1  &  85.5\\
Fe II  & 4541.524 & 2.856 & $-$2.79 &  44.7	 &  45.8 &  38.5  & 65.5\\
Fe II  & 4582.835 & 2.844 & $-$3.09 &  34.9	 &  31.6 &  32.0 &  45.2\\
Fe II  & 4583.837 & 2.807 & $-$1.86 &  92.9	 &  94.2  &  98.3  &  111.9 \\
Fe II  & 4620.521 & 2.828 & $-$3.24 &  32.7	 &  $-$  &  30.8 &  43.1\\
Fe II  & 4666.797 & 7.94 & $-$3.798 &  30.6	 &  27.7  &  36.5  &  46.0\\
Fe II  & 5197.568 & 10.398 & $-$2.116 &  72.2	 &  64.5 &  77.6 &  79.7\\
Fe II  & 5276.002 & 3.199 & $-$1.94 &  87.4 &  80.7 &  90.4  &  94.0 \\
Fe II  & 5362.869 & 3.199 & $-$2.739 &  56.0	 &  55.3 &  56.5  &  71.9 \\
Fe II  & 6247.57 & 5.956 & $-$4.011 &  30.23 &  23.0  &  28.7 &  30.0 \\
Fe II  & 6456.383 & 3.903 & $-$2.1 &  30.9 &  26.1  &  36.1  &  44.3 \\
\hline
\end{longtable}
References for $\log gf$ values for Fe I and Fe II lines: \cite{VALD_linelist.1999} [VALD], \cite{NIST_database}  [NIST], \href{http://kurucz.harvard.edu/linelists.html}{Kurucz Linelist [http://kurucz.harvard.edu/linelists.html]}.


\section{Other Lines List}
\label{sec:other_lines_list}
In table Table~\ref{tab:line_list_of_other_elements}, we have listed a complete line list for elements other than Fe I and Fe II. The values of the lower excitation potential (LEP), and the $\log gf$ values listed in this table are adopted from the recent literature. Again, a checkmark represents the presence of a line in the spectra.

\begin{longtable}{lcccccccr}
	\caption{Information about the line list.}
	\label{tab:line_list_of_other_elements}\\
	\hline
	Element & wavelength ({\AA}) & EP (eV) & $\log gf$ & TYC 3431-689-1 & HD 263815 & TYC 1716-1548-1 & TYC 1191-918-1 & Reference\\
	\hline
Na I & 5889.951 & 0.000 & 0.117 & \checkmark & \checkmark & \checkmark & \checkmark & (1)\\
Na I & 5895.924 & 0.000 & $-$0.184 & \checkmark & \checkmark & \checkmark & \checkmark & (1)\\
Mg I & 5711.088 & 4.346 & $-$1.833 &\checkmark & \checkmark & \checkmark & \checkmark & (2)\\
Mg I & 5528.405 & 4.346 & $-$0.620 &\checkmark & \checkmark & \checkmark & \checkmark & (2)\\
Mg I & 5183.604 & 2.712 & $-$0.180 &\checkmark & \checkmark & \checkmark & \checkmark & (2)\\
Mg I & 5172.684 & 2.712 & $-$0.402 &\checkmark & \checkmark & \checkmark & \checkmark & (2)\\
Mg I & 4702.991 & 4.346 & $-$0.666 &\checkmark & \checkmark & \checkmark & \checkmark & (2)\\
Mg I & 4571.096 & 0.000 & $-$5.691 &\checkmark & \checkmark & \checkmark & \checkmark & (2)\\
Mg I & 4167.271 & 4.346 & $-$1.004 &\checkmark & \checkmark & \checkmark & \checkmark & (2)\\
Al I & 3944.0 & 0.000 & $-$0.640 &\checkmark & \checkmark & \checkmark & \checkmark & (3)\\
Al I & 3961.520 & 0.010 & $-$0.340 &\checkmark & \checkmark & \checkmark & \checkmark & (3)\\
Si I & 4102.936 & 1.909 & $-$3.140 &\checkmark & \checkmark & \checkmark & \checkmark & (4)\\
Si I & 6155.134 & 5.619 & $-$0.400 &\checkmark & \checkmark & \checkmark & \checkmark & (4)\\
Ca I & 6493.781 & 2.521 & $-$0.109 &  \checkmark & \checkmark & \checkmark & \checkmark & (5)\\
Ca I & 6449.808 & 2.521 & $-$1.015 & \checkmark & \checkmark & \checkmark & \checkmark & (5)\\
Ca I & 6439.075 & 2.526 & 0.394 &  \checkmark & \checkmark & \checkmark & \checkmark & (5)\\
Ca I & 6102.723 & 1.879 & $-$0.862 & \checkmark & \checkmark & \checkmark & \checkmark & (5)\\
Ca I & 5601.277 & 2.526 & $-$0.552 & \checkmark & \checkmark & \checkmark & \checkmark & (5)\\
Ca I & 5588.749 & 2.526 & 0.313 &  \checkmark & \checkmark & \checkmark & \checkmark & (5)\\
Ca I & 5590.114 & 2.521 & $-$0.596 & \checkmark & \checkmark & \checkmark & \checkmark & (5)\\
Ca I & 5594.462 & 2.523 & 0.051 &  \checkmark & \checkmark & \checkmark & \checkmark & (5)\\
Ca I & 5598.480 & 2.521 & $-$0.134 & \checkmark & \checkmark & \checkmark & \checkmark & (5)\\
Ca I & 4578.551 & 2.521 & $-$0.170  & \checkmark & \checkmark & \checkmark & \checkmark & (5)\\
Ca I & 4425.437 & 1.879 & $-$0.286 &  \checkmark & \checkmark & \checkmark & \checkmark & (5)\\
Sc I* & 4023.677 & 0.021 & $-$ &\checkmark & \checkmark & \checkmark & \checkmark & (6)\\
Sc II* & 6245.637 & 1.507 & $-$ &\checkmark & \checkmark & \checkmark & \checkmark & (6)\\
Sc II* & 5526.790 & 1.768 & $-$ &\checkmark & \checkmark & \checkmark & \checkmark & (6)\\
Sc II* & 5239.813 & 1.455 & $-$ &\checkmark & \checkmark & \checkmark & \checkmark & (6)\\
Sc II* & 5031.021 & 1.357 & $-$ &\checkmark & \checkmark & \checkmark & \checkmark & (6)\\
Sc II* & 4670.407 & 1.357 & $-$ &\checkmark & \checkmark & \checkmark & \checkmark & (6)\\
Sc II* & 4431.352 & 0.605 & $-$ &\checkmark & \checkmark & \checkmark & \checkmark & (6)\\
Sc II* & 4324.996 & 0.595 & $-$ &\checkmark & \checkmark & \checkmark & \checkmark & (6)\\
Sc II* & 4354.598 & 0.605 & $-$ &\checkmark & \checkmark & \checkmark & \checkmark & (6)\\
Sc II* & 4415.557 & 0.595 & $-$ &\checkmark & \checkmark & \checkmark & \checkmark & (6)\\
Sc II* & 4246.822 & 0.315 & $-$ &\checkmark & \checkmark & \checkmark & \checkmark & (6)\\
Sc II* & 5641.001 & 1.500 & $-$ &\checkmark & \checkmark & \checkmark & \checkmark & (6)\\
Sc II* & 5657.896 & 1.507 & $-$ &\checkmark & \checkmark & \checkmark & \checkmark & (6)\\
Sc II* & 5657.896 & 1.507 & $-$ &\checkmark & \checkmark & \checkmark & \checkmark & (6)\\
Ti I & 5336.848 & 1.582 & $-$1.630 & \checkmark & \checkmark & \checkmark & \checkmark & (7)\\
Ti I & 5210.385 & 0.048 & $-$0.884 &\checkmark & \checkmark & \checkmark & \checkmark & (7)\\
Ti I & 5193.881 & 2.345 & $-$0.942 &  \checkmark & \checkmark & \checkmark & \checkmark & (7)\\
Ti I & 5173.743 & 0.000 & $-$1.118 &  \checkmark & \checkmark & \checkmark & \checkmark & (7)\\
Ti I & 5036.464 & 1.443 & 0.130 &  \checkmark & \checkmark & \checkmark & \checkmark & (7)\\
Ti I & 5038.397 & 1.430 & 0.013 &  \checkmark & \checkmark & \checkmark & \checkmark & (7)\\
Ti I & 5040.613 & 0.826 & $-$1.787 &\checkmark & \checkmark & \checkmark & \checkmark & (7)\\
Ti I & 5016.161 & 0.848 & $-$0.574 &\checkmark & \checkmark & \checkmark & \checkmark & (7)\\
Ti I & 5020.026 & 0.836 & $-$0.414 &\checkmark & \checkmark & \checkmark & \checkmark & (7)\\
Ti I & 5022.868 & 0.826 & $-$0.434 &\checkmark & \checkmark & \checkmark & \checkmark & (7)\\
Ti I & 5024.844 & 0.818 & $-$0.602 &\checkmark & \checkmark & \checkmark & \checkmark & (7)\\
Ti I & 4999.503 & 0.826 & 0.250 & \checkmark & \checkmark & \checkmark & \checkmark & (7)\\
Ti I & 4981.731 & 0.848 & 0.504 & \checkmark & \checkmark & \checkmark & \checkmark & (7)\\
Ti I & 4656.469 & 0.000 & $-$1.345 &\checkmark & \checkmark & \checkmark & \checkmark & (7)\\
Ti I & 4617.269 & 1.749 & 0.389 & \checkmark & \checkmark & \checkmark & \checkmark & (7)\\
Ti I & 4623.097 & 1.739 &0.110  & \checkmark & \checkmark & \checkmark & \checkmark & (7)\\
Ti I & 4534.776 & 0.836 & 0.280 & \checkmark & \checkmark & \checkmark & \checkmark & (7)\\
Ti I & 4533.241 & 0.848 & 0.476 & \checkmark & \checkmark & \checkmark & \checkmark & (7)\\
Ti I & 4512.734 & 0.836 & $-$0.480 &\checkmark & \checkmark & \checkmark & \checkmark & (7)\\
Ti I & 4441.439 & 3.186 & $-$1.792 &\checkmark & \checkmark & \checkmark & \checkmark & (7)\\
Ti II & 6491.561 & 2.061 & $-$1.793 &\checkmark & \checkmark & \checkmark & \checkmark & (7)\\
Ti II & 5418.751 & 1.582 & $-$2.110 &\checkmark & \checkmark & \checkmark & \checkmark & (7)\\
Ti II & 5490.690 & 1.566 & $-$2.650 &\checkmark & \checkmark & \checkmark & \checkmark & (7)\\
Ti II & 5381.015 & 1.566 & $-$1.970 & \checkmark & \checkmark & \checkmark & \checkmark & (7)\\
Ti II & 5211.536 & 2.590 & $-$1.356 & \checkmark & \checkmark & \checkmark & \checkmark & (7)\\
Ti II & 5185.913 & 1.893 & $-$1.370 &\checkmark & \checkmark & \checkmark & \checkmark & (7)\\
Ti II & 5188.680 & 1.582 & $-$1.050 & \checkmark & \checkmark & \checkmark & \checkmark & (7)\\
Ti II & 5154.070 & 1.566 & $-$1.920 & \checkmark & \checkmark & \checkmark & \checkmark & (7)\\
Ti II & 5129.152 & 1.892 & $-$1.300 & \checkmark & \checkmark & \checkmark & \checkmark & (7)\\
Ti II & 5013.677 & 1.582 & $-$1.990 & \checkmark & \checkmark & \checkmark & \checkmark & (7)\\
Ti II & 4911.193 & 3.124 & $-$0.650 & \checkmark & \checkmark & \checkmark & \checkmark & (7)\\
Ti II & 4865.612 & 1.116 & $-$2.810 & \checkmark & \checkmark & \checkmark & \checkmark & (7)\\
Ti II & 4805.085 & 2.061 & $-$0.960 &\checkmark & \checkmark & \checkmark & \checkmark & (7)\\
Ti II & 4763.881 & 1.221 & $-$2.360 & \checkmark & \checkmark & \checkmark & \checkmark & (7)\\
Ti II & 4779.985 & 2.048 & $-$1.370 & \checkmark & \checkmark & \checkmark & \checkmark & (7)\\
Ti II & 4589.985 & 1.237 & $-$1.620 & \checkmark & \checkmark & \checkmark & \checkmark & (7)\\
Ti II & 4568.314 & 1.224 & $-$2.650 & \checkmark & \checkmark & \checkmark & \checkmark & (7)\\
Ti II & 4563.761 & 1.221 & $-$0.790 & \checkmark & \checkmark & \checkmark & \checkmark & (7)\\
Ti II & 4544.028 & 1.243 & $-$2.530 & \checkmark & \checkmark & \checkmark & \checkmark & (7)\\
Ti II & 4529.474 & 1.572 & $-$1.650 & \checkmark & \checkmark & \checkmark & \checkmark & (7)\\
Ti II & 4493.513 & 1.080 & $-$2.830 & \checkmark & \checkmark & \checkmark & \checkmark & (7)\\
Ti II & 4501.270 & 1.116 & $-$0.760 & \checkmark & \checkmark & \checkmark & \checkmark & (7)\\
Ti II & 4443.801 & 1.080 & $-$0.700 & \checkmark & \checkmark & \checkmark & \checkmark & (7)\\
Ti II & 4444.558 & 1.116 & $-$2.210 & \checkmark & \checkmark & \checkmark & \checkmark & (7)\\
Ti II & 4411.074 & 3.095 & $-$0.670 &\checkmark & \checkmark & \checkmark & \checkmark & (7)\\
Ti II & 4411.925 & 1.224 & $-$2.550 &\checkmark & \checkmark & \checkmark & \checkmark & (7)\\
Ti II & 4394.051 & 1.221 & $-$1.770 &\checkmark & \checkmark & \checkmark & \checkmark & (7)\\
Ti II & 4395.850 & 1.243 & $-$1.970 &\checkmark & \checkmark & \checkmark & \checkmark & (7)\\
V I* & 4406.633 & 0.301 & $-$ &\checkmark & \checkmark & \checkmark & \checkmark & (8)\\
V I & 4390.099 & 2.616 & $-$1.504 & \checkmark & \checkmark & \checkmark & \checkmark & (8)\\
V I* & 4379.230 & 0.301 & $-$ & \checkmark & \checkmark & \checkmark & \checkmark & (8)\\
V I* & 4111.774 & 0.301 & $-$ & \checkmark & \checkmark & \checkmark & \checkmark & (8)\\
V I* & 4099.812 & 2.211 & $-$ & \checkmark & \checkmark & \checkmark & \checkmark & (8)\\
V II & 4023.378 & 1.805 & $-$0.689 &\checkmark & \checkmark & \checkmark & \checkmark & (8)\\
V II & 4036.777 & 1.476 & $-$1.594 &\checkmark & \checkmark & \checkmark & \checkmark & (8)\\
V II & 4183.428 & 2.050 & $-$1.112 &\checkmark & \checkmark & \checkmark & \checkmark & (8)\\
Cr I & 4646.148 & 1.030 & $-$0.700 &\checkmark & \checkmark & \checkmark & \checkmark & (9)\\
Cr I & 4651.282 & 0.983 & $-$1.460 &\checkmark & \checkmark & \checkmark & \checkmark & (9)\\
Cr I & 4652.152 & 1.004 & $-$1.030 &\checkmark & \checkmark & \checkmark & \checkmark & (9)\\
Cr I & 4626.174 & 0.968 & $-$1.320 &\checkmark & \checkmark & \checkmark & \checkmark & (9)\\
Cr I & 4616.120 & 0.983 & $-$1.190 &\checkmark & \checkmark & \checkmark & \checkmark & (9)\\
Cr I & 5409.772 & 1.030 & $-$0.720 &\checkmark & \checkmark & \checkmark & \checkmark & (9)\\
Cr I & 5345.801 & 1.004 & $-$0.980 &\checkmark & \checkmark & \checkmark & \checkmark & (9)\\
Cr I & 5348.312 & 1.004 & $-$1.290 &\checkmark & \checkmark & \checkmark & \checkmark & (9)\\
Cr I & 5296.691 & 0.983 & $-$1.400 &\checkmark & \checkmark & \checkmark & \checkmark & (9)\\
Cr I & 5298.277 & 0.983 & $-$1.150 &\checkmark & \checkmark & \checkmark & \checkmark & (9)\\
Cr I & 4591.457 & 3.422 & $-$1.888 &\checkmark & \checkmark & \checkmark & \checkmark & (9)\\
Cr I & 4545.945 & 0.941 & $-$1.370 &\checkmark & \checkmark & \checkmark & \checkmark & (9)\\
Cr I & 4254.332 & 0.000 & $-$0.114 &  \checkmark & \checkmark & \checkmark & \checkmark & (9)\\
Cr II & 4824.127 & 3.871 & $-$0.970 &\checkmark & \checkmark & \checkmark & \checkmark & (9)\\
Cr II & 4634.070 & 4.072 & $-$0.990 &\checkmark & \checkmark & \checkmark & \checkmark & (9)\\
Cr II & 4618.803 & 4.074 & $-$0.840 &\checkmark & \checkmark & \checkmark & \checkmark & (9)\\
Cr II & 4588.199 & 4.071 & $-$0.627 &\checkmark & \checkmark & \checkmark & \checkmark & (9)\\
Cr II & 4558.650 & 4.073 & $-$0.449 &\checkmark & \checkmark & \checkmark & \checkmark & (9)\\
Mn I* & 4823.524 & 2.319 & $-$ & \checkmark & \checkmark & \checkmark & \checkmark & (10)\\
Mn I* & 4783.427 & 2.298 & $-$ & \checkmark & \checkmark & \checkmark & \checkmark & (10)\\
Mn I* & 4766.418 & 2.920 & $-$ &  \checkmark & \checkmark & \checkmark & \checkmark & (10)\\
Mn I* & 4754.042 & 2.282 & $-$ &\checkmark & \checkmark & \checkmark & \checkmark & (10)\\
Mn I* & 4034.483 & 0.000 & $-$ &\checkmark & \checkmark & \checkmark & \checkmark & (10)\\
Mn I* & 4033.062 & 0.000 & $-$ &\checkmark & \checkmark & \checkmark & \checkmark & (10)\\
Mn I* & 4235.295 & 2.888 & $-$ &  \checkmark & \checkmark & \checkmark & \checkmark & (10)\\
Mn I* & 4041.355 & 2.114 & $-$ &  \checkmark & \checkmark & \checkmark & \checkmark & (10)\\
Co I & 4020.828 & 3.665 & $-$0.961 &\checkmark & \checkmark & \checkmark & \checkmark & (11)\\
Co I & 4110.530 & 1.049 & $-$1.080 &\checkmark & \checkmark & \checkmark & \checkmark & (11)\\
Co I & 4121.311 & 0.923 & $-$0.320 &\checkmark & \checkmark & \checkmark & \checkmark & (11)\\
Ni I & 6643.629 & 1.676 & $-$2.300 &\checkmark & \checkmark & \checkmark & \checkmark & (12)\\	
Ni I & 6108.107 & 1.676 & $-$2.450 &\checkmark & \checkmark & \checkmark & \checkmark & (12)\\
Ni I & 5155.762 & 3.898 & 0.011 &  \checkmark & \checkmark & \checkmark & \checkmark & (12)\\
Ni I & 5115.389 & 3.834 & $-$0.110 &\checkmark & \checkmark & \checkmark & \checkmark & (12)\\
Ni I & 5099.927 & 3.679 & $-$0.100 &\checkmark & \checkmark & \checkmark & \checkmark & (12)\\
Ni I & 5080.528 & 3.655 & 0.330 &  \checkmark & \checkmark & \checkmark & \checkmark & (12)\\
Ni I & 5081.107 & 3.847 & 0.300 &  \checkmark & \checkmark & \checkmark & \checkmark & (12)\\
Ni I & 5035.357 & 3.635 & 0.290 &  \checkmark & \checkmark & \checkmark & \checkmark & (12)\\
Ni I & 5017.568 & 3.539 & $-$0.020 &\checkmark & \checkmark & \checkmark & \checkmark & (12)\\
Ni I & 4904.407 & 3.542 & $-$0.170 &\checkmark & \checkmark & \checkmark & \checkmark & (12)\\
Ni I & 4866.262 & 3.539 & $-$0.210 &\checkmark & \checkmark & \checkmark & \checkmark & (12)\\
Ni I & 4714.408 & 3.380 & 0.260 & \checkmark & \checkmark & \checkmark & \checkmark & (12)\\
Ni I & 4715.757 & 3.543 & $-$0.320 &\checkmark & \checkmark & \checkmark & \checkmark & (12)\\
Cu I* & 5105.537 & $-$ & $-$ &\checkmark & \checkmark & \checkmark & \checkmark & (13)\\
Zn I & 4722.153 & 4.030 & $-$0.338 &\checkmark & \checkmark & \checkmark & \checkmark & (14)\\
Zn I & 4810.528 & 4.078 & $-$0.137 &\checkmark & \checkmark & \checkmark & \checkmark & (14)\\
Sr II & 4077.709 & 0.000 & 0.167 &\checkmark & \checkmark & \checkmark & \checkmark & (15)\\
Sr II & 4215.519 & 0.000 & $-$0.145 &\checkmark & \checkmark & \checkmark & \checkmark & (15)\\
Y II & 4124.907 & 0.409 & $-$1.500 &\checkmark & \checkmark & \checkmark & \checkmark & (16)\\
Y II & 4682.324 & 0.409 & $-$1.510  & \checkmark & \checkmark & \checkmark & \checkmark & (16)\\
Y II & 4854.863 & 0.992 & $-$0.11  & \checkmark & \checkmark & \checkmark & \checkmark & (16)\\
Y II & 4883.684 & 1.084 & 0.265  & \checkmark & \checkmark & \checkmark & \checkmark & (16)\\
Y II & 5087.416 & 1.084 & $-$0.170  & \checkmark & \checkmark & \checkmark & \checkmark & (16)\\
Y II & 5123.211 & 0.992 & $-$1.219  & \checkmark & \checkmark & \checkmark & \checkmark & (16)\\
Y II & 5200.406 & 0.992 & $-$0.570   & \checkmark & \checkmark & \checkmark & \checkmark & (16)\\
Y II & 5662.925 & 1.944 & 0.38  & \checkmark & \checkmark & \checkmark & \checkmark & (16)\\
Zr II & 4029.684 & 0.713 & $-$0.78  & \checkmark & \checkmark & \checkmark & \checkmark & (17)\\
Zr II & 4050.316 & 0.713 & $-$1.000  & \checkmark & \checkmark & \checkmark & \checkmark & (17)\\
Zr II & 4090.535 & 0.758 & $-$1.009  & \checkmark & \checkmark & \checkmark & \checkmark & (17)\\
Zr II & 4161.213 & 0.713 & $-$0.59  & \checkmark & \checkmark & \checkmark & \checkmark & (17)\\
Zr II & 4150.986 & 0.802 & $-$0.992  & \checkmark & \checkmark & \checkmark & \checkmark & (17)\\
Zr II & 4208.977 & 0.713 & $-$0.460  & \checkmark & \checkmark & \checkmark & \checkmark & (17)\\
Zr II & 4211.907 & 0.527 & $-$1.083  & \checkmark & \checkmark & \checkmark & \checkmark & (17)\\
Zr II & 4317.299 & 0.713 & $-$1.450  & \checkmark & \checkmark & \checkmark & \checkmark & (17)\\
Ba II & 6496.897 & 0.604 & $-$0.377 & \checkmark & \checkmark & \checkmark & \checkmark & (18)\\
Ba II* & 5853.700 & $-$ & $-$  & \checkmark & \checkmark & \checkmark & \checkmark & (18)\\
Ba II* & 6141.713 & $-$ & $-$  & \checkmark & \checkmark & \checkmark & \checkmark & (18)\\
Ba II* & 4554.033 & $-$ & $-$  & \checkmark & \checkmark & \checkmark & \checkmark & (18)\\
Ba II* & 4934.077 & $-$ & $-$  & \checkmark & \checkmark & \checkmark & \checkmark & (18)\\
La II* & 4333.753 & $-$ & $-$ & \checkmark & \checkmark & \checkmark & \checkmark & (19)\\
La II* & 4086.709  & $-$ & $-$ & \checkmark & \checkmark & \checkmark & \checkmark & (19)\\
La II* & 4123.218  & $-$ & $-$ & \checkmark & \checkmark & \checkmark & \checkmark & (19)\\
La II & 4921.776  & 0.244 & $-$0.45 & \checkmark & \checkmark & \checkmark & \checkmark & (19)\\
La II* & 4322.503  & $-$ & $-$ & \checkmark & \checkmark & \checkmark & \checkmark & (19)\\
Ce II & 4186.594 & 0.864 & 0.74  & \checkmark & \checkmark & \checkmark & \checkmark & (20)\\
Ce II & 4562.359 & 0.478 & 0.23  & \checkmark & \checkmark & \checkmark & \checkmark & (20)\\
Ce II & 4486.909 & 0.295 & $-$0.090  & \checkmark & \checkmark & \checkmark & \checkmark & (20)\\
Ce II & 4628.161 & 0.516 & 0.220  & \checkmark & \checkmark & \checkmark & \checkmark & (20)\\
Pr II* & 4143.11  & $-$ & $-$	& \checkmark & \checkmark & $-$ & \checkmark & (21)\\
Pr II* & 4179.39  & $-$ & $-$	& \checkmark & \checkmark & $-$ & \checkmark & (21)\\
Nd II & 4061.080 & 0.471 & 0.347  & \checkmark & \checkmark & \checkmark & \checkmark & (22)\\
Nd II & 4109.448 & 0.321 & 0.184  & \checkmark & \checkmark & \checkmark & \checkmark & (22)\\
Nd II* & 4232.374 & $-$ & $-$  & \checkmark & \checkmark & \checkmark & \checkmark & (22)\\
Nd II & 4135.321 & 0.631 & $-$0.400  & \checkmark & \checkmark & \checkmark & \checkmark & (22)\\
Nd II & 4327.932 & 0.559 & $-$0.430  & \checkmark & \checkmark & \checkmark & \checkmark & (22)\\
Nd II & 4338.690 & 0.742 & $-$0.290  & \checkmark & \checkmark & \checkmark & \checkmark & (22)\\
Nd II* & 4358.161 & $-$ & $-$  & \checkmark & \checkmark & \checkmark & \checkmark & (22)\\
Nd II & 4825.478 & 0.182 & $-$0.776  & \checkmark & \checkmark & \checkmark & \checkmark & (22)\\
Sm II* & 4424.337 & $-$ & $-$  & \checkmark & \checkmark & \checkmark & \checkmark & (23)\\
Sm II & 4434.318 & 0.378 & $-$0.204  & \checkmark & \checkmark & \checkmark & \checkmark & (23)\\
Sm II* & 4467.341 & $-$ & $-$  & \checkmark & \checkmark & \checkmark & \checkmark & (23)\\
Eu II* & 4129.801  & $-$ & $-$ & \checkmark &  \checkmark &  \checkmark &  \checkmark & (24)\\
Eu II* & 4205.05  & $-$ & $-$ & \checkmark &  \checkmark &  \checkmark &  \checkmark & (24)\\
Gd II & 3850.97 & 0.000 & $-$0.094  & \checkmark & \checkmark & \checkmark & $-$ & (25)\\
Gd II & 4037.33 & 0.662 & $-$0.020  & \checkmark & \checkmark & \checkmark & $-$ & (25)\\
Gd II & 4085.6 & 0.731 & $-$0.07 &  \checkmark & \checkmark & \checkmark & $-$ & (25)\\
Gd II & 4098.61 & 0.819 & 0.312 & \checkmark & \checkmark & \checkmark & $-$ & (25)\\
Tb II & 3976.84 & 0.000 & 0.08 & \checkmark & \checkmark & $-$ & $-$ & (26)\\
Dy II & 4103.306 & 0.103 & $-$0.390 & \checkmark & \checkmark & \checkmark & \checkmark & (27)\\
Dy II & 4111.343 & 0.000 & $-$0.850 & \checkmark & \checkmark & \checkmark & \checkmark & (27)\\
Dy II & 4449.704 & 0.000 & $-$1.030 & \checkmark & \checkmark & \checkmark & \checkmark & (27)\\
Ho II & 4045.44 & 0.000 & $-$0.05 & \checkmark & \checkmark & $-$ & $-$ & (28)\\
Ho II & 4152.58 & 0.079 & $-$0.930 & \checkmark & \checkmark & $-$ & $-$ & (28)\\
Er II & 3830.48 & 0.000 & $-$0.22  & \checkmark & \checkmark & \checkmark & $-$ & (29)\\
Er II & 3896.23 & 0.055 & $-$0.12  & \checkmark & \checkmark & \checkmark & $-$ & (29)\\
Tm II & 3848.02 & 0.000 & $-$0.140  & $-$ & \checkmark & $-$ & $-$ & (30)\\
Lu II & 5476.69 & 1.760 & $-$0.276 & $-$ & $-$ & $-$ & $-$  & (31)\\
Hf II & 3918.09 & 0.452 & $-$1.14  & \checkmark & \checkmark & $-$ & \checkmark  & (32)\\
Hf II & 4093.15 & 0.452 & $-$1.15  & \checkmark & \checkmark & $-$ & \checkmark & (32)\\
Os I & 4260.85 & 0.000 & $-$1.440 & \checkmark & \checkmark & \checkmark & \checkmark & (33)\\
Os I & 4420.47 & 0.000 & $-$1.20 & \checkmark & \checkmark & \checkmark & \checkmark & (33)\\	
Pb I* & 4057.8  & $-$ & $-$ & $-$ & $-$ & $-$ & $-$ & (34)\\
Th II & 4019.12 & 0.000 & $-$0.228 & \checkmark & $-$ & $-$ & \checkmark & (35)\\
	\hline
\end{longtable}
References for $\log gf$: (1) \cite{Sneden.etal.2014_Na5889_Na5895_Mg5528_Mg4702_Mg4167}, (2) \cite{Sneden.etal.2014_Na5889_Na5895_Mg5528_Mg4702_Mg4167, Yong.etal.2014_Mg5711, Heiter.etal.2015_Mg4571}, (3) \cite{Mendoza.etal.1995_Al3944_Al3961}, (4) \cite{Shi.etal.2008_Si4102_Si6155}, (5) \cite{Ryabchikova.etal.2000_Ca6493_Ca6449_Ca6439_Ca6102}, (6) \cite{Lawler.etal.2019_Sc_hfs}, (7) \cite{VALD_linelist.1999}[VALD]; \cite{NIST_database}[NIST] (8) \cite{Lawler.etal.2014_V-I, Wood.etal.2014_V-II}, (9) \cite{VALD_linelist.1999}[VALD]; \cite{NIST_database}[NIST], (10) \cite{Den_Hartog.etal.2011_Mn}, (11) \cite{Nitz.etal.1999.Co-I, Cardon.etal.1982.Co-I}, (12) \cite{Siqueira.etal.2015.Ni-I, Vejar.etal.2021.Ni-I} (13) \href{http://kurucz.harvard.edu/linelists.html}{Kurucz Linelist [http://kurucz.harvard.edu/linelists.html]}, (14) \cite{Takeda.etal.2002.Zn-I}, (15) \cite{Mishenina.etal.2017.Sr-II}, (16) \cite{Biemont.etal.2011_Y-II} , (17) \cite{Ljung.etal.2006.Zr-II, Malcheva.etal.2006.Zr-II}, (18) \citep{McWilliam.1998, Cui.etal.2013} , (19) \cite{Lawler.etal.2001_La-II} , (20) \cite{Palmer.etal.2000_Ce-II, Zhang.etal.2001_Ce-II}, (21) \cite{Sneden.etal.2009_Pr-II}, (22) \cite{VALD_linelist.1999}[VALD] , (23) \cite{VALD_linelist.1999}[VALD] , (24) \cite{Lawler.etal.2001_Eu-II}, (25) \cite{VALD_linelist.1999}[VALD] , (26) \cite{Lawler.etal.2001.Tb-II}, (27) \cite{Sneden.etal.2009.Dy-II}, (28) \cite{Lawler.etal.2004.Ho-II}, (29) \cite{Lawler.etal.2008.Er-II}, (30) \cite{Sneden.etal.2009.Tm-II}, (31) \cite{Bord.etal.1997.Lu-II}, (32) \cite{Lawler.etal.2006.Hf-II}, (33) \cite{Quinet.etal.2006.Os-II}, (34) \cite{Roederer.etal.2012.Pb-I}, (35) \cite{Nilsson.etal.2002.Th-II}. 

Note: * with the name of element represents hyperfine splitting.

\section{Evolution of heavy elements with metallicity}
\label{sec:heavy_element_evolution_with_FeH}
\begin{figure*}
	\centering
	\includegraphics[width=\textwidth]{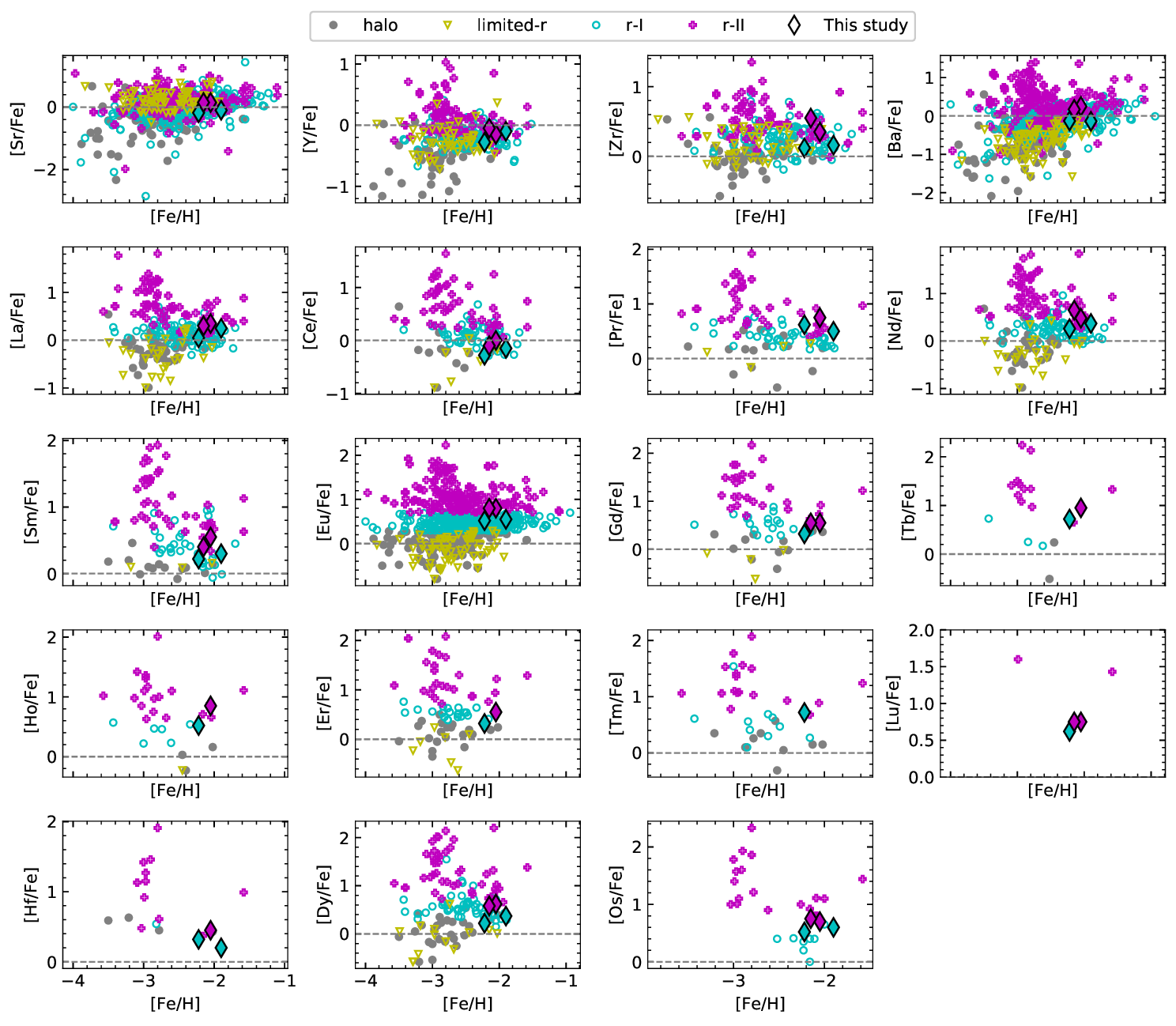}
    \caption{Abundance of heavy elements as a function of metallicity. The limited-$r$ stars are shown in yellow colours, the $r$-I stars in cyan colours, and the $r$-II stars in magenta colours. The four black diamonds indicate the $r$-I (cyan) and $r$-II (magenta) program stars in this study. References for the $r$-process stars are given in the caption of Fig.~\ref{fig:Ba_Eu_vs_Eu_Fe.eps}. Normal metal-poor halo stars, shown with grey filled, are taken from \protect\cite{SAGAbase.2008, Ryan.etal.1991, McWilliam.etal.1995, Burris.etal.2000, Fulbright.2000, Aoki.etal.2002, Johnson.2002, Cayrel.etal.2004, Honda.etal.2004, Aoki.etal.2005, Barklem.etal.2005, Jonsell.etal.2005, Preston.etal.2006, Lai.etal.2007, Zhang.etal.2009, Ishigaki.etal.2010, Roederer.etal.2010mar, Hollek.etal.2011, Allen.etal.2012, Aoki.etal.2013jan, Cohen.etal.2013, Ishigaki.etal.2013, Yong.etal.2013, Placco.etal.2014jan, Roederer.etal.2014jun, Hansen.etal.2015, Jacobson.etal.2015, Li.etal.2015jan}, and \protect\cite{JINAbase2018}.}
    \label{fig:Fe_H__X_Fe_all_heavy_elements}
\end{figure*}

\begin{figure*}
	\centering
	\includegraphics[width=\textwidth]{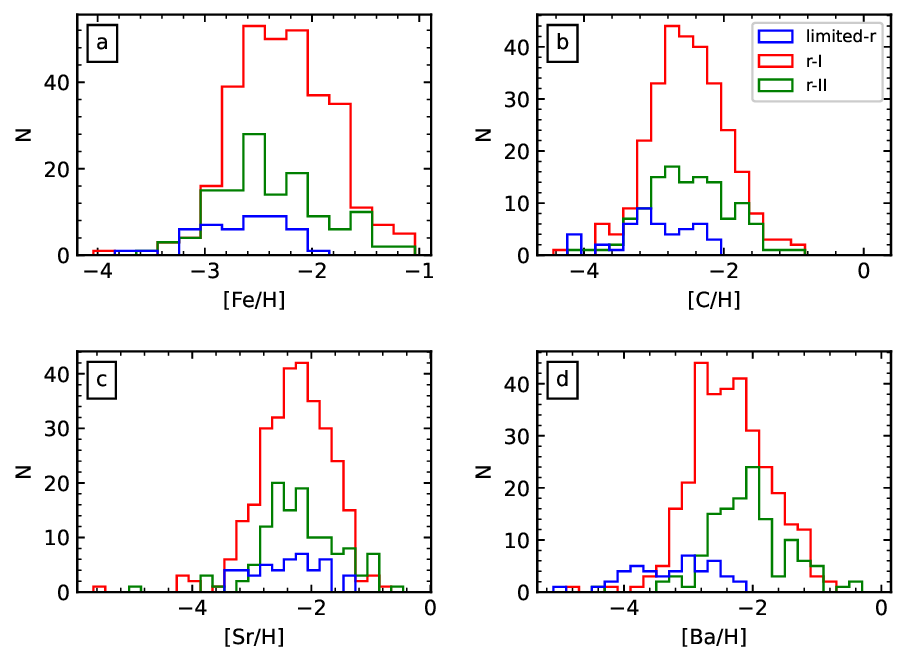}
    \caption{Comparison of metallicity, C, Sr, and Ba abundance distribution in RPE stars.The distribution of limited-$r$ stars is shown with the blue histogram, the red histogram shows the distribution of $r$-I stars, and the green histogram displays the distribution of $r$-II stars. References for the RPE stars are given in the caption of Fig.~\ref{fig:Ba_Eu_vs_Eu_Fe.eps}.}
    \label{fig:r_process_FeH_dist}
\end{figure*}

Fig.~\ref{fig:Fe_H__X_Fe_all_heavy_elements} shows the evolution of all the heavy elements from Sr to Os as a function of metallicity. Similar to the main text, limited-$r$ stars are shown in yellow colours, the $r$-I stars in cyan colours, and the $r$-II stars in magenta colours. The four black diamonds indicate the $r$-I (cyan) and $r$-II (magenta) program stars in this study.Fig.~\ref{fig:r_process_FeH_dist} shows the distribution of metallicity, [C/Fe], [Sr/H], and [Ba/H].

\bsp	
\label{lastpage}
\end{document}